\documentclass[nonacm, sigconf, screen]{acmart}
\PassOptionsToPackage{table}{xcolor}
\usepackage[utf8]{inputenc} % allow utf-8 input
\usepackage[T1]{fontenc}    % use 8-bit T1 fonts
\usepackage{hyperref}       % hyperlinks
\usepackage{url}            % simple URL typesetting
\usepackage{threeparttable}
\usepackage{booktabs}       % professional-quality tables
\usepackage{amsfonts}       % blackboard math symbols
\usepackage{microtype}      % microtypography
\usepackage{graphicx}
\usepackage{amsmath}
\usepackage{amsthm}
\usepackage{longtable}
\usepackage{array, makecell}
\usepackage{multicol,xparse,environ}
\usepackage{pifont}
\usepackage{bbm}
\usepackage[raggedrightboxes]{ragged2e}
\usepackage{amsfonts}
\usepackage{algorithmic}
\usepackage{textcomp}

\usepackage{enumitem}
\usepackage{verbatim}
\usepackage{subfig}
\usepackage[font=bf]{caption}
\usepackage{soul}
\usepackage{multirow,multicol}
\usepackage{balance}
\usepackage{ifthen}
\usepackage{float}
\usepackage{newtxmath}  % \vmathbb 
\usepackage{bbding}
\usepackage[linesnumbered,ruled,vlined]{algorithm2e}

\usepackage{makecell}

\SetKwInput{KwInput}{Input}                % Set the Input
\SetKwInput{KwOutput}{Output}              % set the Output

\AtBeginDocument{%
  }

\setcopyright{acmlicensed}
\copyrightyear{2018}
\acmYear{2018}
\acmDOI{XXXXXXX.XXXXXXX}

\acmConference[Conference acronym 'XX]{Make sure to enter the correct
  conference title from your rights confirmation emai}{June 03--05,
  2018}{Woodstock, NY}
\acmISBN{978-1-4503-XXXX-X/18/06}

\definecolor{mycolor}{RGB}{0, 0, 0}

\newcommand{\aonegtTW}{0.3}
\newcommand{\aonereTW}{0.42}

\newcommand{\atwogtTW}{0.14}
\newcommand{\atworeTW}{0.29}

\newcommand{\cellpad}{0.4mm}

\newcommand{\athreegtTW}{0.3}
\newcommand{\athreereTW}{0.43}

\newcommand{\ablationreTW}{0.5}

\newcommand{\attacktwoeggTW}{0.39}

\newcommand{\adv}{\mathcal{A}}
\newcommand{\advS}{{\mathcal{S}^{\adv}}}
\newcommand{\advT}{{\mathcal{T}^{\adv}}}
\newcommand{\advE}{{\mathcal{E}^{\adv}}}

\newcommand{\advSC}{{\mathcal{S}^c}}
\newcommand{\advTC}{{\mathcal{T}^{c}}}
\newcommand{\advEC}{{\mathcal{E}^{c}}}
\newcommand{\advESC}{{\hat{\mathcal{E}}^{c}}}

\newcommand{\advqS}{{{S}^{\adv}}}
\newcommand{\advqT}{{{T}^{\adv}}}
\newcommand{\advqE}{{{E}^{\adv}}}

\newcommand{\tgt}{{\text{tgt}}}
\newcommand{\tgtS}{{S^\tgt}}
\newcommand{\tgtT}{{{T}^{\tgt}}}
\newcommand{\tgtE}{{{E}^{\tgt}}}

\newcommand{\tgtSset}{{\mathcal{S}^{\tgt}}}
\newcommand{\tgtTset}{{\mathcal{T}^{\tgt}}}
\newcommand{\tgtEset}{{\mathcal{E}^{\tgt}}}

\newcommand{\be}{\boldsymbol{e}}

\newcommand{\qb}{{q_b}}

\newcommand{\tdict}{{T_{\text{all}}}}

\DeclareMathOperator*{\argmax}{arg\,max}
\DeclareMathOperator*{\argmin}{arg\,min}

\begin{document}

\clearpage
\pagenumbering{arabic}
\setcounter{page}{1}

\title{Prompt Inference Attack on Distributed Large Language Model Inference Frameworks}\thanks{This paper has been accepted to CCS 2025.}

\author{Xinjian Luo}
\email{xinjian.luo@mbzuai.ac.ae}
\affiliation{%
  \institution{Mohamed bin Zayed University of Artificial Intelligence}
  \country{United Arab Emirates}
}

\author{Ting Yu}
\email{ting.yu@mbzuai.ac.ae}
\affiliation{%
  \institution{Mohamed bin Zayed University of Artificial Intelligence}
  \country{United Arab Emirates}
}

\author{Xiaokui Xiao}
\email{xkxiao@nus.edu.sg}
\affiliation{%
  \institution{National University of Singapore}
  \country{Singapore}
}

\renewcommand{\shortauthors}{Luo et al.}

\begin{abstract}

The inference process of modern large language models (LLMs) demands prohibitive computational resources, rendering them infeasible for deployment on consumer-grade devices. To address this limitation, recent studies propose distributed LLM inference frameworks, which employ split learning principles to enable collaborative LLM inference on resource-constrained hardware. 
However, distributing LLM layers across participants requires the transmission of intermediate outputs, which may introduce privacy risks to the original input prompts --- a critical issue that has yet to be thoroughly explored in the literature.

In this paper, we rigorously examine the privacy vulnerabilities of distributed LLM inference frameworks by designing and evaluating three prompt inference attacks aimed at reconstructing input prompts from intermediate LLM outputs.
These attacks are developed under various query and data constraints to reflect diverse real-world LLM service scenarios. Specifically, the first attack assumes an unlimited query budget and access to an auxiliary dataset sharing the same distribution as the target prompts. The second attack also leverages unlimited queries but uses an auxiliary dataset with a distribution differing from the target prompts. The third attack operates under the most restrictive scenario, with limited query budgets and no auxiliary dataset available.
We evaluate these attacks on a range of LLMs, including state-of-the-art models such as Llama-3.2 and Phi-3.5, as well as widely-used models like GPT-2 and BERT for comparative analysis.
Our experiments show that the first two attacks achieve reconstruction accuracies exceeding $90\%$, while the third achieves accuracies typically above $50\%$, even under stringent constraints. 
These findings highlight substantial privacy risks in distributed LLM inference frameworks, issuing a strong alert on their deployment in real-world applications. 
Additionally, our analysis uncovers distinct distributional properties of intermediate embeddings across LLM layers, providing valuable insights into the LLM inference process and the development of effective defense mechanisms for distributed LLM frameworks.

\end{abstract}

%%
%% Keywords. Separate the keywords with commas.

\settopmatter{printfolios=true}
\maketitle

\sloppy

\section{Introduction}\label{sec-intro}
\begin{figure*}
    \vspace{-3mm}
    \includegraphics[width=.95\textwidth]{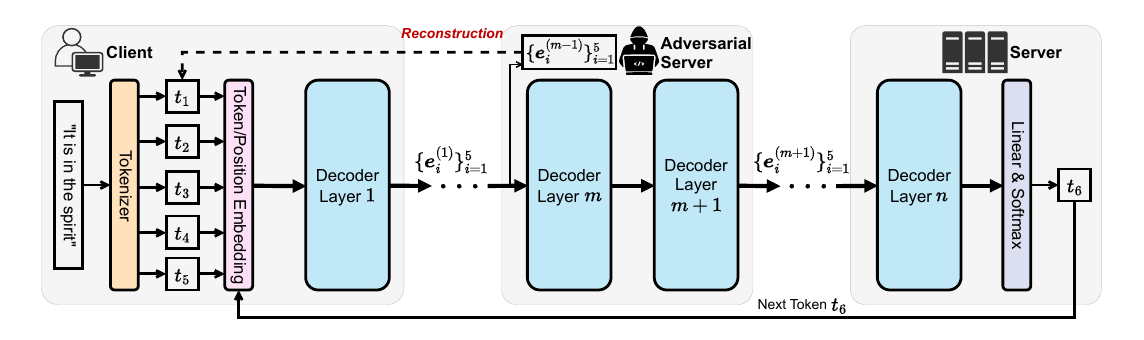}
  \vspace{-3mm}
  \caption{The overview of distributed LLM inference for next-token prediction.}
  \label{fig-overview}
  \vspace{-3mm}
\end{figure*}

Although demonstrating remarkable performance, recently released large language models (LLMs) require substantial GPU memory for inference, making them challenging to deploy on consumer-grade devices~\cite{Petals-acl}. To address this limitation, recent studies~\cite{Petals-acl,Petals-nips,cake,zhang2024edgeshard-llminference,icml-hexgen-distributedllm,LinguaLinked-distributedllm,sigcomm-pipellm-distributedllm} have proposed distributed LLM inference frameworks inspired by split learning~\cite{Splitlearning}, where LLM layers are distributed among multiple participants to enable collaborative inference. In this way, a portion of the inference process can be allowed to run locally on devices with limited computational resources. 
For example, Fig.~\ref{fig-overview} illustrates the Petals framework~\cite{Petals-acl}, which facilitates distributed LLM inference by introducing two types of modules: a client module that hosts the tokenizer, token embedding layer, and a few decoder layers, and a server module that manages intermediate decoder layers. 
%
% Participants in this framework can host either or both modules. 
%
During inference, intermediate sequence embeddings are exchanged between the client and server modules to enable a complete next-token prediction.

For these distributed LLM inference frameworks, a critical question arises: \textit{\textcolor{mycolor}{to what extent can the privacy of input prompts be compromised via intermediate outputs?}} 
This concern is particularly significant as users increasingly rely on LLMs for tasks involving confidential information, and the leakage of sensitive data to third parties could have serious consequences.
% 
% For example, LLMs are commonly employed to assist in academic paper reviews~\cite{paperreview}. However, distributed LLM frameworks could potentially expose private manuscript content in the prompts to third parties, breaching confidentiality and ethical standards. \ting{Some may question whether it is ethical to review papers using LLMs. I think it would be better to use some other examples, e.g., users tend to ask LLMs about medical/healthcare questions.}
%
For example, in digital healthcare, LLMs can analyze users' descriptions of symptoms and mobile sensor data for medical diagnosis~\cite{yang2024drhouse}. Any leakage of such data in the prompts can compromise user privacy and significantly undermine trust in related applications.
Similarly, LLMs are widely applied in business analytics~\cite{businessanalysis}, where input prompts often include proprietary information such as customer reviews and sales data. Leakage of such proprietary information could result in reputation damage and financial losses for companies.
\textcolor{mycolor}{While some developers of distributed LLM frameworks acknowledge potential privacy risks associated with the transmission of intermediate results~\cite{Petals-acl}, they fail to provide definitive answers regarding \textit{how and to what extent} private input prompts could be leaked}.
To bridge this critical gap, we propose inference attacks to systematically reconstruct private prompts from intermediate outputs in distributed LLM frameworks, offering an in-depth exploration of this under-explored vulnerability.

While conceptually related to input inference attacks in split learning~\cite{zhu2023passivesplitlearning,splitlearning-tiger}, our prompt inference task presents two critical differences that limit the applicability of existing split learning attack methods. First, split learning attacks focus on reconstructing image data, whereas attacks on LLMs target text data. Second, split learning attacks operate during the training phase and typically require training a shadow model to simulate layers outside the adversary’s control. In contrast, our attacks are designed for the inference phase, where only one-time intermediate outputs are available for reconstructing each prompt, and training a shadow model for LLMs is impractical due to the prohibitive computational demands posed by the much larger size of LLMs compared to image classification models.
Sentence embedding inversion attacks~\cite{li2023sentencepromptinference2,SongR20promptinference,PanZJY20promptinference,MorrisZCSR24promptinference1} also share some conceptual overlap with our task. However, these methods target relatively small sentence embedding models and rely on transformer-based attack models for reconstructing sentences from embeddings. This approach suffers from two limitations.
First, the length of input sentence embeddings for attack models is typically constrained (e.g., $<15$ in \cite{GuKRVM23embdinv3}) because training transformer-based models for long contexts requires substantial amounts of data and computational resources --- resources that most adversaries lack. Second, these reconstruction methods \textcolor{mycolor}{could be} context-dependent, where an error in reconstructing a word may propagate and severely degrade the reconstruction accuracy for subsequent words, limiting the overall effectiveness of reconstruction.
In this paper, we aim to develop novel attacks on LLMs that can overcome the aforementioned limitations. 
%
% \ting{This may sound like sentence embedding inversion attacks could serve as a baseline. And the reviewer may ask us to compare the effectiveness of this baseline attack with ours. Did we do that?}
%
% Simply adapting techniques from prior work~\cite{zhu2023passivesplitlearning,splitlearning-tiger,li2023sentencepromptinference2,SongR20promptinference,PanZJY20promptinference,MorrisZCSR24promptinference1} to the setting of distributed LLM inference would not help 
%
To achieve our attack targets, new insights into the workflow of LLM inference need to be developed.

To design effective attacks on distributed LLM frameworks, we analyze the distribution of sequence embeddings across different layers of LLMs. Our key observation reveals that intermediate embeddings output by different LLM layers typically form distinct clusters in the embedding space. Building on this insight, we propose three types of attacks by progressively tightening constraints on query budgets and auxiliary data, which can reflect diverse real-world application scenarios for distributed LLM inference and enable a comprehensive assessment of their privacy risks.
Specifically, the first adversary operates with unlimited query budgets and auxiliary data following the same distribution as the target prompts. Here, instead of taking the prompt reconstruction task as a sentence generation problem~\cite{MorrisZCSR24promptinference1, li2023sentencepromptinference2, MorrisKSR23promptinference}, we reformulate it as a classification problem and leverage a trained classifier for prompt reconstruction.
The second adversary, while also having unlimited query budgets, assumes auxiliary data with arbitrary distributions.
For this case, we analyze the influence of sequence length and word constitution on intermediate embeddings and introduce a sequence embedding synthesis approach to augment query data for classifier training.
The third adversary faces the most restrictive scenario: limited query budgets and no auxiliary data. In response, we design a three-phase reconstruction framework that first utilizes nearest neighbor search and semi-supervised learning for identifying key concepts of input prompts, and then employ a constrained beam search to reconstruct finer details.

Compared to related studies~\cite{zhu2023passivesplitlearning,splitlearning-tiger,li2023sentencepromptinference2,SongR20promptinference,PanZJY20promptinference,MorrisZCSR24promptinference1}, our attacks are lightweight (no shadow model training required), free from length constraints, and context-independent (errors in word reconstruction hardly propagate).
% \ting{Would beam search have error propagation problem?}. 
%
Moreover, our methods are adaptable across different LLM architectures, making them LLM-independent.
Experiments conducted on a variety of LLMs, including Llama-3.2~\cite{llama3.2}, Phi-3.5~\cite{phi3.5}, GPT-2~\cite{brown2020language-gpt3}, and BERT~\cite{DevlinCLT19Bert}, demonstrate reconstruction accuracies exceeding $90\%$ for the first two attacks and reaching larger than $50\%$ in most cases for the third attack.
Notably, our results show that the embeddings in early LLM layers are more vulnerable to privacy breaches than those in later layers. This difference is mainly attributed to the distributional variations of sequence embeddings across the LLM inference pipeline.
We believe these new findings will provide valuable insights into LLM inference mechanisms and the development of robust defense mechanisms for distributed LLM inference frameworks.
To the best of our knowledge, this is the first work that develops attacks targeting distributed LLMs.
Our contributions are summarized as follows:
\begin{itemize}
    \item We uncover clustering patterns in sequence embeddings across LLM layers and leverage this to design three prompt inference attacks under varying constraints. Our attacks are lightweight, free of length constraints, context-independent, and LLM-independent.
    \item We perform experiments on well-recognized LLMs including Llama-3.2, Phi-3.5, GPT-2, and BERT. The results demonstrate larger than $90\%$ reconstruction accuracies in most attack settings. Our results also reveal the differences in susceptibility along the LLM inference pipeline.
    \item We underscore the significant privacy risks in distributed LLM inference frameworks. We believe that our new findings on the LLM embedding distributions not only guide the development of future defenses but also offer valuable insights into LLM inference mechanisms.
\end{itemize}

\section{Preliminaries}
In this section, we first outline the process of LLM inference and then present the general workflow of distributed LLM frameworks.

\subsection{Decoder-Only LLMs}
The Transformer architecture~\cite{transformer} was originally proposed to increase training parallelization, which relies solely on attention mechanisms to capture the dependencies between input and output sequences. 
Due to its scalability and efficiency, Transformer has become the backbone of modern LLMs.
Note that the original Transformer was designed for machine translation with an encoder-decoder structure. 
Since the decoder component of Transformers is more suitable for text generation tasks, it has emerged as the primary architecture for state-of-the-art LLMs. Examples include GPT-2~\cite{brown2020language-gpt3}, Phi-3.5~\cite{phi3.5}, and Llama-3.2~\cite{llama3.2}, collectively known as \textit{decoder-only} transformer architectures. 
%
% This paper primarily focuses on such architectures.

The primary task for decoder-only LLMs is \textit{next-token prediction}, the basis for their pre-training. 
Other tasks, such as semantic similarity~\cite{acl24semanticsimilarity} and text classification~\cite{fields2024textclassification}, are typically implemented by modifying the final linear layer while retaining the decoder structure used for next-token prediction. 
Fig.~\ref{fig-overview} shows a typical forward pass in an LLM during next-token prediction.
An input prompt $S$ is tokenized into a sequence of tokens $T=\{t_i\}^l_{i=1}$ with $t_i\in \tdict=\{0,1, \cdots,|\tdict|-1\}$, where $\tdict$ represents the set of all unique token IDs, and $l$ is the sequence length. For example, in GPT-2, the input phrase ``It is in the spirit'' is tokenized as $T=\{1026, 318, 287, 262, 4437\}$ with $|\tdict|=50257$.
These tokens are then passed to an embedding layer, producing token embeddings $B=\{\boldsymbol{b}_i\}^l_{i=1}$ with $\boldsymbol{b}_i\in \mathbb{R}^{d}$, where $d$ denotes a pre-defined model dimension (e.g., $d=1280$ for GPT-2).
The token embedding layer essentially acts as a dictionary, mapping token IDs in $\tdict$ to their corresponding embeddings.

The embeddings $B$ are then processed through a sequence of decoder layers, generating intermediate embeddings $E^{(m)}=\{\be^{(m)}_i\}^l_{i=1}$, where $\be^{(m)}_i\in \mathbb{R}^{d}$ and $m$ denotes the decoder layer index. 
Specifically, each intermediate embedding $\be^{(m)}_i$ is computed as $\be^{(m)}_i=f_{1\cdots m}(\boldsymbol{b}_1,\cdots,\boldsymbol{b}_i)$ with $f_{1\cdots m}$ denoting the first $m$ decoder layers, i.e., the generation of the $i$-th intermediate embedding is conditioned on all tokens proceeding token $t_i$. The embedding $\be^{(m)}_i$ can also be viewed as a \textit{sequence embedding} for the token sequence $\{t_1,\cdots,t_i\}$ at layer $m$ .
After processing through all $n$ decoder layers, the last embedding $\be^{(n)}_l$ of the final embedding outputs $E^{(n)}$ is passed to a linear layer, which predicts the next token $t_{l+1}$. The token $t_{l+1}$ will then be appended to $T$ for subsequent predictions.
Thus, a complete forward pass of LLMs can be represented as:
\begin{equation*}
    S\xrightarrow{\text{tokenizer}} T \xrightarrow{\text{embedding}} B 
    \xrightarrow{f_1 \cdots f_{n}}  E^{(n)} \xrightarrow{}  e^{(n)}_l\xrightarrow{\text{linear layer}} t_{l+1}.
\end{equation*}

\subsection{Distributed LLM Inference}\label{subsec-dllmI}
The performance of LLMs is well known to scale with model size, leading state-of-the-art LLMs to consist of tens or even hundreds of billions of parameters.
Consequently, running such massive models requires high-end hardware resources, which are inaccessible to most researchers and practitioners. To address this limitation, distributed LLM inference frameworks~\cite{Petals-acl, Petals-nips,zhang2024edgeshard-llminference,cake} have emerged as a practical solution, enabling LLM inference on consumer-grade devices.
The core idea behind distributed LLM inference is to partition the model into smaller modules. Typically, the client module hosts the embedding layer and a few initial decoder layers, while the server module hosts the remaining intermediate decoder layers.
During inference, the client initiates a query, processes the input through its hosted layers, and then transmits the intermediate outputs to the server modules.
%
% , which complete the forward pass.
%
Fig.~\ref{fig-overview} illustrates this process in a distributed setup.
By splitting the model in this way, each participant only needs to host a portion of the model parameters, making it feasible to perform LLM inference on low-end hardware.
In addition, this collaborative framework allows participants to flexibly choose whether to host a client, a server, or both~\cite{Petals-acl}.

A significant concern in distributed LLM inference frameworks arises from the transmission of intermediate outputs between participants, which may expose sensitive information about the input prompt. 
\textcolor{mycolor}{While some developers~\cite{Petals-acl} acknowledge potential privacy risks associated with sharing intermediate results, they fail to provide conclusive evidence or a comprehensive analysis of \textit{how and to what extent} input prompt privacy may be compromised}.
%
% To address this critical gap, we propose three prompt inference attacks under varying constraints, aiming to evaluate and quantify the privacy risks associated with distributed LLM inference frameworks.
\section{Problem Statement}
In this section, we first introduce the system model illustrating how an adversary participates in distributed LLM inference and then discuss possible adversarial settings for prompt inference attacks.

\subsection{System Model}
We follow the settings of distributed LLM frameworks as described in~\cite{Petals-acl, Petals-nips}, where an adversary can host both a client module and a server module. Without loss of generality, we assume that the adversary’s server module begins at the $m$-th decoder layer, as illustrated in Fig.~\ref{fig-overview}.
During the inference phase, a participant initializes a query by inputting a prompt $\tgtS\sim P_{\tgt}$ into the client module.
%
% , which processes the prompt and transmits the resulting intermediate outputs to the next server module in the inference pipeline.
%
$P_{\tgt}$ denotes the distribution or domain of prompt $\tgtS$.
$\tgtS$ is also treated as the \textit{target prompt} to be reconstructed by the adversary.
The intermediate outputs will be continually processed to be $E^{\tgt(m-1)}$ and then forwarded to the adversary's server module starting from layer $m$. Consequently, the adversary always has access to the intermediate results $E^{\tgt(m-1)}$ generated for any query initiated by other participants.
Since we focus on developing generalized attacks that can work with embeddings from any intermediate decoder layer, we will omit the specific layer index $m-1$ and refer the embeddings received by the adversary as $\tgtE$ for simplicity.
Similarly, the adversary can generate its own queries from auxiliary data $\advqS\sim P_{\adv}$ using the client module and obtain the corresponding embeddings $\advqE$.
The maximum number of tokens allowed in total queries is referred to as the \textit{query budgets}.

\subsection{Attack Model}\label{subsec-attack-model}
In this paper, we focus on developing black-box, passive prompt inference attacks aiming at {reconstructing the target prompts $\tgtS$} initialized by other participants. The \textit{black-box} setting~\cite{LuoJWWXO24-tifs, luo2022feature-ccs} assumes that the adversary has no knowledge of components beyond the client and server modules it hosts. The \textit{passive} (or semi-honest) setting~\cite{jiang2024data-tkde, luo2021feature-icde} indicates that the adversary does not tamper with the LLM inference process but instead infers private prompts $\tgtS$ solely from the received sequence embeddings $\tgtE$.
In addition, we consider designing attacks under different constraints on query budgets and auxiliary data availability, as summarized in Table~\ref{tb-attack-setting}.

Specifically, $\adv_1$ has unlimited query budgets and access to an auxiliary dataset $\advS$ sampled from the same distribution as the target prompt $\tgtS$, i.e., $\advS \sim P_{\adv} \approx P_{\tgt}$. 
This scenario reflects cases where both auxiliary and target data originate from the same domain, such as Wikipedia. The assumption of unlimited query budgets aligns with current distributed LLM inference frameworks~\cite{Petals-acl, Petals-nips}.
The second adversary $\adv_2$ also has unlimited query budgets but operates under a mismatch in distributions between auxiliary data and target prompts, i.e., $P_{\adv} \not\approx P_{\tgt}$. 
This scenario indicates that the adversary has no knowledge of the domain of target prompts $\tgtS$.
For example, while $\advqS$ could be sampled from Wikipedia, $\tgtS$ may be sampled from Reddit~\footnote{\url{https://www.reddit.com/r/AskReddit/}}.
The third adversary $\adv_3$ operates under the most restrictive setting, with limited query budgets and no auxiliary data.
Although limited query budgets are not explicitly discussed in prior distributed LLM inference studies~\cite{Petals-acl, Petals-nips}, they are relevant in similar LLM-related commercial scenarios, such as pay-as-you-go API services provided by OpenAI~\cite{openai-pricing} and Google~\cite{google-pricing}. Under this setting, the objective is to maximize attack effectiveness with small query costs.

\begin{table}[!t]
\setlength{\tabcolsep}{4pt}
\caption{Attack settings of our prompt inference attacks.}\label{tb-attack-setting}
\vspace{-3mm}
\center
\small
\begin{tabular}{cccc}
\toprule
{Adversary} & {Query Budgets} & {Auxiliary Data $\advS$} & {Data Distribution}  \\
\midrule
% \hline
$\adv_1$ & Unlimited & \ding{51} &	$P_{\adv} \approx  P_{\tgt}$ \\ 
% \hline
$\adv_2$ & Unlimited & \ding{51} &	$P_{\adv} \not\approx  P_{\tgt}$ \\
% \hline
$\adv_3$ & Limited & \ding{55} & $/$  \\
\bottomrule
\end{tabular}
\vspace{-5mm}
\end{table}
\section{The Attack Methods}\label{sec-attack}
In this section, we introduce the proposed attack methods corresponding to the three adversaries introduced in Section~\ref{subsec-attack-model}.

\subsection{The First Adversary $\adv_1$}\label{subsec-adv1}
\begin{figure}[!t]
\centering
% \begin{small}
\begin{tabular}{c}
% \vspace{-2mm}
\subfloat[]{\includegraphics[width=0.45\columnwidth]{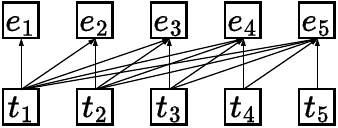}\label{fig-corr-orig}}
\hspace{6mm}
\subfloat[]{\includegraphics[width=0.45\columnwidth]{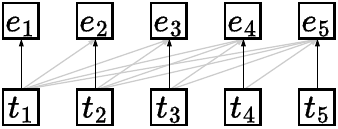}\label{fig-corr-ober}}
\end{tabular}
\vspace{-3mm}
\caption{The correspondence between input tokens and intermediate embeddings.}
\label{fig-corr}
% \end{small}
\vspace{-3mm}
\end{figure}

The first adversary operates with unlimited query budgets and an auxiliary dataset $\advS$, based on which $\adv_1$ can construct query tokens $\advT$.
By sending $\advT$ into the distributed inference framework, $\adv_1$ can obtain the corresponding embeddings $\advE$ from the locally hosted server module.
Using the pairs $(\advE, \advT)$, the adversary can train an attack model to reconstruct the target tokens $\tgtT$ from $\tgtE$. Ultimately, decoding $\tgtT$ via the client module's tokenizer yields the original prompt $\tgtS$.
The main challenge is designing an effective attack model for training on $(\advE, \advT)$.
Specifically, for a query sentence $\advqS$, the adversary can obtain a token sequence $\advqT=\{t_i\}_{i=1}^l$ and its intermediate embeddings $\advqE=\{\be_i\}^l_{i=1}$. Although $|\advqT|=|\advqE|$, the mapping between these two sets is not one-to-one.
In fact, each embedding $\be_i$ at position $i$ is conditioned on all tokens before position $i$, i.e., $\{t_1,\cdots,t_i\}$, as shown in Fig.~\ref{fig-corr-orig}.
Existing sentence reconstruction studies~\cite{MorrisZCSR24promptinference1, li2023sentencepromptinference2, GuKRVM23embdinv3, SongR20promptinference, MorrisKSR23promptinference} suggest that transformers, taking $\{\be_1,\cdots,\be_l\}$ as the input and producing $\{t_1,\cdots,t_l\}$ as the output, are well-suited for this inversion task, as they effectively capture dependencies among sequence elements. However, relying on transformers as the attack model has two significant drawbacks.
First, training a robust transformer-based model requires extensive data and computational resources, which may not be feasible in most real-world scenarios. Consequently, transformer-based models are often limited to reconstructing short sequences, e.g., $l \leq 15$~\cite{GuKRVM23embdinv3}.
Second, transformers reconstruct tokens sequentially, so errors in earlier tokens may negatively impact subsequent ones~\footnote{\color{mycolor}
To justify this, we first train three transformer-based attack models following \cite{GuKRVM23embdinv3}, whose training datasets and hyperparameters are the same but with different initial parameters. Then, we use a prompt ``ash thorp octane photo real organic line 3d glowing'' for reconstruction, where the first three tokens are fixed, and the following tokens need to be reconstructed by attack models. The results reconstructed by the three models are [``{ash thorp octane} \textit{picture Octane: The 3d Story}'', `{ash thorp octane} \textit{photo 3d photo light}'', ``{ash thorp octane} \textit{book is real at the University}''], indicating that the first reconstructed token may influence the following reconstructions, such as picture $\rightarrow$ Story, and book $\rightarrow$ University.
}.
%
% For example, an incorrect reconstruction of $t_3$ in Fig.~\ref{fig-corr-orig} can propagate errors to $t_4$ and $t_5$.
%
In this paper, we aim to design more efficient and generalized attacks that eliminate the reliance on transformer-based models, which require new observations on intermediate sequence embeddings of LLMs.

\begin{figure}
  \includegraphics[width=1.\columnwidth]{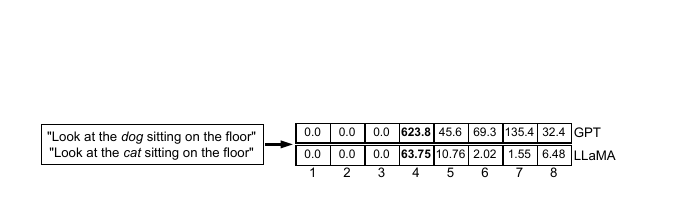}
  \vspace{-7mm}
  \caption{The $\ell_1$ norm of token-level embedding differences between $S^{\text{dog}}$ and $S^{\text{cat}}$ after layer 1 of GPT-2 and Llama-3.2.}
  \label{fig-embd-diff-norm}
  \vspace{-6mm}
\end{figure}

\begin{figure}[!t]
\centering
% \begin{small}
\begin{tabular}{c}
% \vspace{-2mm}
\subfloat[After Decoder Layer 6]{\includegraphics[width=0.48\columnwidth]{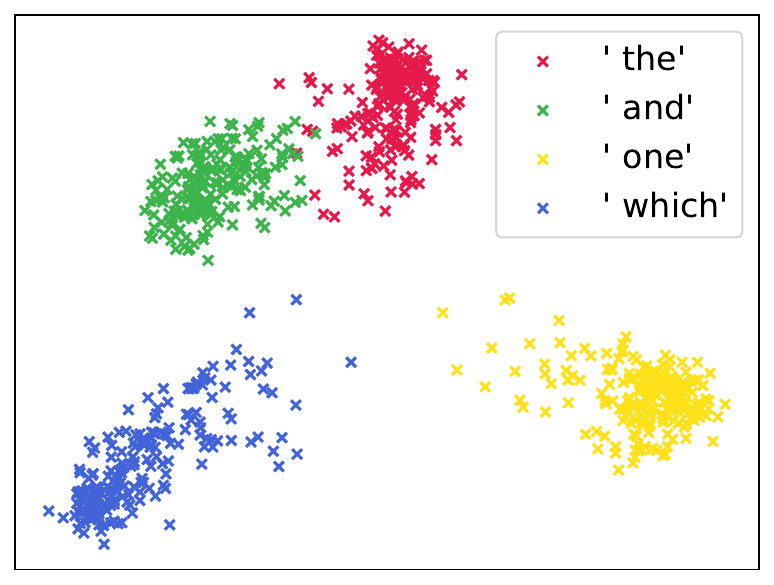}\label{fig-exp-dist-1}}
\hspace{0mm}
\subfloat[After Decoder Layer 16]{\includegraphics[width=0.48\columnwidth]{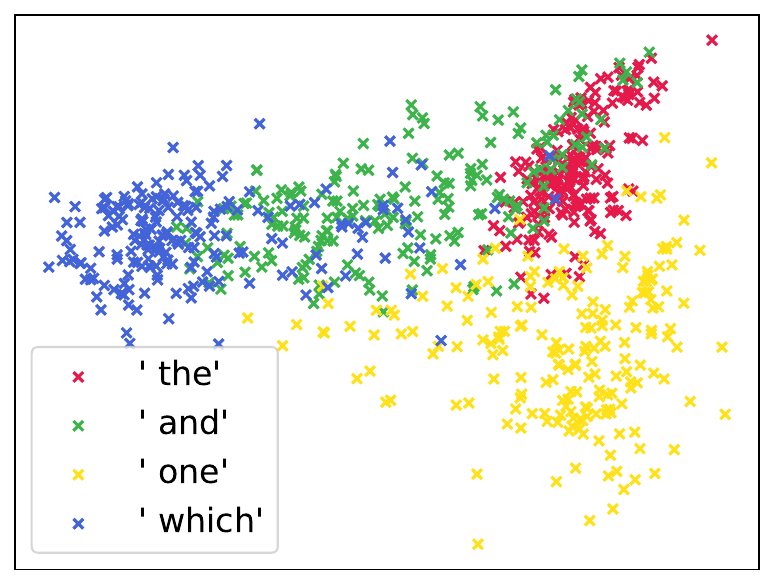}\label{fig-exp-dist-2}}
\end{tabular}
\vspace{-3mm}
\caption{The embedding distributions after different layers of Llama-3.2. ` the' denotes that the embedding corresponds to a token sequence ending with the token ` the'.}
\label{fig-exp-dist}
% \end{small}
\vspace{-4mm}
\end{figure}

% \subsection{The Key Observation}
\vspace{0.5mm}\noindent
\textbf{The Key Observation.}
Our investigation into the correspondence between $\advqT$ and $\advqE$ reveals a key observation: although $\be_i$ is conditioned on $\{t_1,\cdots,t_i\}$, the influence of $t_i$ significantly outweighs that of the preceding tokens $\{t_1,\cdots,t_{i-1}\}$, as demonstrated in Fig.~\ref{fig-corr-ober}.
To illustrate this, consider the example prompts $S^{\text{dog}}=\text{``Look at the dog sitting on the floor''}$ and $S^{\text{cat}}=\text{``Look at the cat sitting on the floor''}$. These prompts differ only in their fourth tokens, with $t_4^{\text{dog}}=\text{` dog'}\neq t_4^{\text{cat}}=\text{` cat'}$, while $t_i^{\text{dog}}=t_i^{\text{cat}}$ for all $i\neq 4$.
We first input these two prompts into an LLM and obtain the corresponding intermediate sequence embeddings after layer 1. 
After that, we compute the $\ell_1$ norm of differences between these two sequence embeddings at each position, $||\be_i^{\text{dog}} - \be_i^{\text{cat}}||_1$ for all $i$.
Fig.~\ref{fig-embd-diff-norm} presents the results tested on GPT-2~\cite{gpt2} and Llama-3.2~\cite{llama3.2}.

The main observation is that the embedding difference at position 4 is substantially larger than at subsequent positions.
This large difference at position 4 aligns with the distinct tokens $t_4^{\text{dog}}\neq t_4^{\text{dog}}$.
However, the small differences at positions 5 to 8 are counterintuitive, given that each embedding $\be_i$ is computed based on all preceding tokens. Intuitively, a significant change in $t_4$ should propagate to embeddings at positions 5 to 8.  
Considering that $t_i^{\text{dog}}= t_i^{\text{cat}}$ for $5\leq i\leq 8$, we conclude that \textit{the influence of the ending token $t_i$ in a sequence dominates the sequence embedding $\be_i$}, formalized as:
\begin{equation}\label{eq-embd-breakdown}
    \be_i = g(t_i) +  \epsilon_g (\{t_1,\cdots, t_{i-1}\}),
\end{equation}
where $g$ denotes the neural transformation between $\be_i$ and $\{t_1,\cdots,t_i\}$, and $\epsilon_g$ denotes a perturbation function conditioned on the preceding tokens $\{t_1,\cdots, t_{i-1}\}$.
To further validate this observation that $||g(t_i)||\gg || \epsilon_g (\{t_1,\cdots, t_{i-1}\})||$, we processed a Wikipedia dataset~\cite{wikitext} with Llama-3.2 to obtain embeddings after layers 6 and 16 and map each embedding $\be_i$ to its corresponding ending token $t_i$ for each sequence $\{t_1,\cdots,t_i\}$.
Using principal component analysis (PCA)~\footnote{\url{https://en.wikipedia.org/wiki/Principal_component_analysis}}, we reduced the embedding dimensions to two and visualized the embedding distributions for common tokens $t_i \in \{\text{` the'}, \text{` and'}, \text{` one'}, \text{` which'}\}$ in Fig.~\ref{fig-exp-dist}. 
These tokens appear most frequently in English, and their corresponding sequence embeddings can comprehensively capture the influence of preceding tokens on the embeddings of ending tokens.
It is worth noting that different points in Fig.~\ref{fig-exp-dist} correspond to different positions $i$, i.e., $\be_i$ is only mapped to the ending token $t_i$ of a sequence, while the sequence length $i$ can be arbitrary.
From Fig.~\ref{fig-exp-dist}, we observe that embeddings $\be_i$ in the embedding space form distinct clusters primarily determined by the ending token $t_i$, while the influence of preceding tokens $\{t_1, \cdots, t_{i-1}\}$ results in minimal intra-cluster shifts. These shifts are significantly smaller than the inter-cluster distances caused by differences in the ending tokens $t_i$.

\vspace{0.5mm}\noindent
\textbf{The Attack Method.}
Building on the observations from Fig.~\ref{fig-embd-diff-norm} and \ref{fig-exp-dist}, an adversary can first leverage the auxiliary dataset $\advS$ to obtain $(\advE, \advT)$, then construct mappings $(\be_i, t_i)$  from $(\advE, \advT)$ and train a classification model $\phi: \be_i \rightarrow t_i $ to learn the correspondence between sequence embedding $\be_i$ and their ending tokens $t_i$.
Once $\phi$ is trained, the adversary can use it to reconstruct the target sequence $\tgtS$ from $\tgtE$ on a token-by-token basis, on the condition that $\tgtS$ and $\advS$ are drawn from similar distributions.
Algorithm~\ref{alg-attack-1} presents detailed steps for $\adv_1$.
Since $\adv_1$ has unlimited query budgets, collecting a larger $\advS$ will result in better performance of $\phi$.   
However, our experiments indicate that having 200 occurrences of each token in $\advS$ suffices to train a high-performing attack model.

Unlike transformer-based methods~\cite{MorrisZCSR24promptinference1, li2023sentencepromptinference2, GuKRVM23embdinv3, SongR20promptinference, MorrisKSR23promptinference}, our approach does not rely on preceding tokens $\{t_1,\cdots,t_{i-1}\}$ for reconstructing $t_i$, making its reconstruction accuracy independent of sequence length or errors in earlier tokens.
Notably, Algorithm~\ref{alg-attack-1} can achieve accuracies higher than $90\%$ across all layers in state-of-the-art LLMs, as demonstrated in Section~\ref{subsec-exp-adv1}.

\begin{algorithm}[!tb]
\begin{small}
\caption{Adversary $\adv_1$ with Unlimited Query Budgets and an Auxiliary Dataset $\advS$ with $P_\adv \approx P_\tgt$}
\label{alg-attack-1}
\KwInput{Auxiliary dataset $\advS$, learning rate $\alpha$, target intermediate embeddings $\tgtE$}
\KwOutput{Reconstructed private prompt $\tgtS$}
$(\advE, \advT)\gets$ Send queries from $\advS$ for LLM inference \;
$\{(\be_i, t_i)\}_{i=1}^{|\advT|} \gets$ Construct training data from $(\advE, \advT)$ \; 
$\boldsymbol{\theta}_{\phi} \leftarrow \mathcal{N}(0,1)$ \tcp*{Initialize the attack model}
\ForEach{epoch} {
    \ForEach{batch}{
        $loss \gets 0$ \;
        Randomly select a batch of samples \;
        \ForEach{sample $i$}{
            $\hat{t}_{i} \leftarrow {\phi}( {\be}_{i}; \boldsymbol{\theta}_{\phi})$ \;

            $loss\gets loss + L(\hat{t}_{i}, t_{i})$ \;
        }
        $\boldsymbol{\theta}_{\phi} \leftarrow \boldsymbol{\theta}_{\phi} - \alpha \cdot \bigtriangledown_{\boldsymbol{\theta}_{\phi}} loss$ \tcp*{Update the attack model}
    }
}
$\tgtS\gets $ "" \;
\ForEach{$\be_i \in \tgtE$}{
    $\hat{t}_i\gets \phi(\be_i; \theta_{\phi})$ \;
    Decode $\hat{t}_i$ via the tokenizer and append the result to $\tgtS$ \;
}
\Return $\tgtS$\;
\end{small}
\end{algorithm}
% \vspace{-10mm}

\subsection{The Second Adversary $\adv_2$}\label{sec-attack-2}

Algorithm~\ref{alg-attack-1} performs effectively under the assumption that $\tgtS$ and $\advS$ follow a similar distribution.
However, in practical scenarios, the adversary may lack knowledge of the distribution of $\tgtS$. 
Consequently, a model $\phi$ trained on $\advS$ may not generalize well to $\tgtS$. For example, for embeddings output by layer 32 of GPT-2 trained on a Wikipedia dataset~\cite{wikitext}, Algorithm~\ref{alg-attack-1} can achieve a reconstruction accuracy of $98\%$ on Wikipedia-related prompts but experience a significant drop to $61\%$ for prompts sampled from a Q\&A dataset SQuAD2.0~\cite{squad2.0}.
This degradation occurs because approximately $25\%$ tokens in SQuAD2.0, denoted as $\advTC$, are not covered by the training set, i.e., $\advT \cap \advTC = \varnothing $.
To address this issue, it is necessary to generate embeddings $\advESC$ corresponding to $\advTC$ and augment the training dataset $(\advE, \advT)$ with the $(\advESC, \advTC)$ pairs.
In the case that $\advTC$ is unknown, we can simply use its superset $\advTC=\tdict \setminus\advT$ in the algorithm design.
Let $\advEC$ represent the unknown embeddings in the real context corresponding to $\advTC$.
To make the attack model trained on the synthetic embeddings $\advESC$ generalize well in real context, the main challenge is to generate $\advESC$ with a distribution similar to $\advEC$, specifically in how to generate valid sequences ${t_1,\cdots, t_{i}}$ for $t_i\in \advTC$ and $i\in\mathbb{N}^+$.
%

\begin{figure}[!tb]
\vspace{-3.8mm}
\centering
% \begin{small}
\begin{tabular}{c}
\subfloat[Fixed Synthetic Position $i=100$]{\includegraphics[width=0.49\columnwidth]{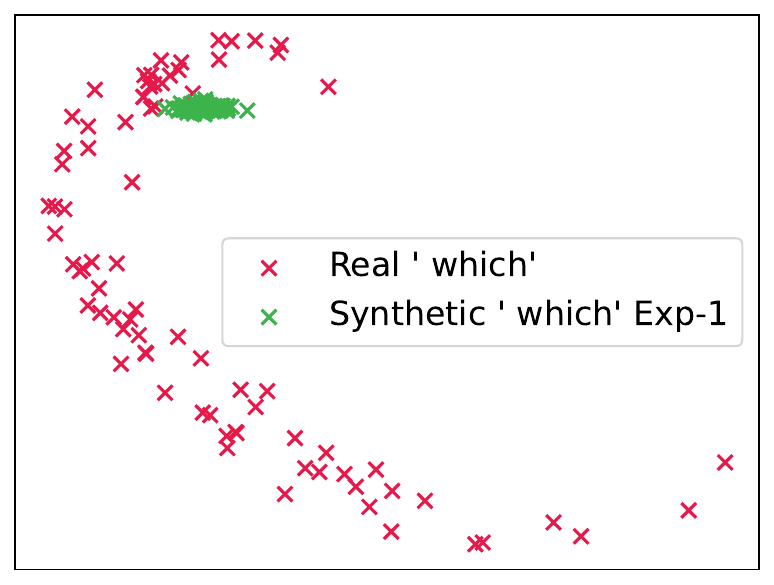}\label{fig-exp-fix-pos}}
\hspace{1.5mm}
\subfloat[Fixed Synthetic Sequence]{\includegraphics[width=0.49\columnwidth]{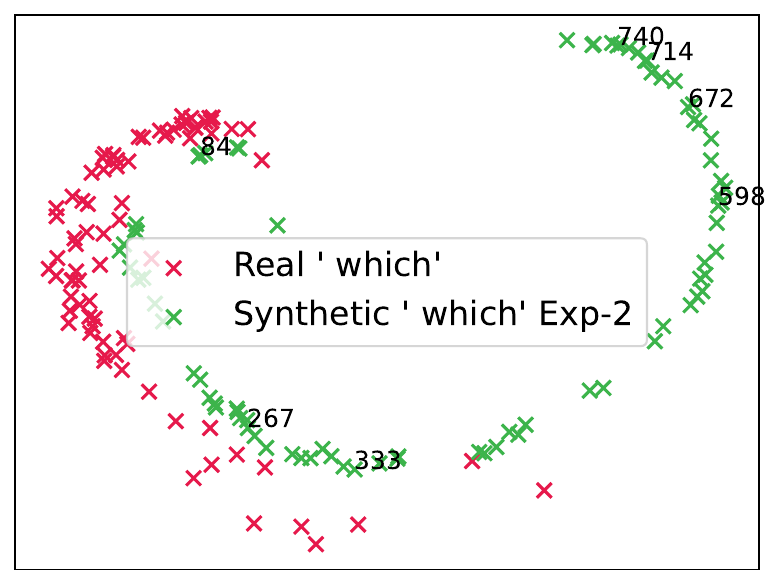}\label{fig-exp-fix-seq}}
\end{tabular}
\vspace{-3mm}
\caption{The real and synthetic embedding distributions for the token ` which' evaluated on layer 1 of GPT-2.}
\label{fig-exp-attack-2}
% \end{small}
\vspace{-2mm}
\end{figure}

% \subsection{The Key Observation}
\vspace{0.5mm}\noindent
\textbf{The Key Observation.}
Since $\advTC$ is absent from the auxiliary dataset $\advS$, a straightforward approach to generating real sequences ${t_1,\cdots, t_{i}}$ for $t_i\in \advTC$ involves searching the LLM's next-token outputs for $t_i\in \advTC$ when inputting $\advS$ as the prompts. However, this approach depends on LLM behaviors and does not allow precise control over the appearance of the target tokens $t_i$.
Alternatively, one might use GPT-4o~\cite{openai-pricing} to synthesize token sequences with a prompt like, ``Please generate a 10-token sentence with `test' as the third token''.
However, GPT-4o performs poorly in this case for two reasons: first, the target token `test' may be retokenized by GPT-4o's tokenizer into multiple tokens (e.g., `t' and `est'), causing GPT-4o to treat the target token as separate units; second, GPT-4o lacks an understanding of token composition in sentences and always output sentences with wrong token positions, such as ``He studied hard for the upcoming test''.
Therefore, instead of synthesizing valid sequences ${t_1,\cdots, t_{i}}$ for $t_i\in \advTC$,
we focus on directly generating the distribution of $\be_i$ as defined in  Eq.~\ref{eq-embd-breakdown}.
%
% based solely on $\advS$ is challenging when $\advTC\cap \advS =\varnothing$.
%
% we turn to directly generating the distribution of $\be_i$ in Eq.~\ref{eq-embd-breakdown}.
%
Notice that the generation of $\be_i$ for ${t_{i}}$ depends on two factors: the position $i$ of $t_i$ and the preceding tokens ${t_1,\cdots, t_{i-1}}$.
To assess the impact of these two factors on the distribution of $\be_i$, we design two experiments.
In the first, we fix $i = 100$ and generate 200 sequences by replacing ${t_1, \dots, t_{99}}$ with random real sequences sampled from $\advS$ and $t_{100}$ with a target token ` which'. 
In the second, we fix a sequence ${t_1, \dots, t_{l}}$ with $l = 800$ and generate 200 new sequences based on it by replacing the token at random positions $i$ with the target token ` which'.
We visualize the resulting distributions of $\be_i$ after PCA reduction in Fig.~\ref{fig-exp-attack-2}, where red points represent real-context embeddings sampled from SQuAD2.0, i.e., $\advEC$; and green points represent synthetic distributions, i.e., $\advESC$.
Position labels $i$ for the green points are provided in Fig.~\ref{fig-exp-fix-seq} for reference.
From Fig.~\ref{fig-exp-attack-2}, we observe that for a target token $t_i$, the preceding tokens ${t_1,\cdots, t_{i-1}}$ control the variance of its embedding $\be_i$ along certain dimensions, which are not visible in the figure, while the position $i$ influences the overall shape of $\be_i$ along the principal components.
Thus, to effectively imitate the distribution of $\advEC$, we need to perturb the embedding of target tokens along both the principal and minor components.

% \subsection{The Attack Method}
\vspace{0.5mm}\noindent
\textbf{The Attack Method.}
Building on the insights from Fig.~\ref{fig-exp-attack-2}, we propose to synthesize the embedding distribution of $t_i$ by constructing $\delta$ random sequences $\{t_1,\cdots, t_{i}\}$ that terminate with the target token $t_i\in \advTC$, where the position $i$ and preceding tokens $\{t_1,\cdots, t_{i-1}\}\subset \advT$ are randomized, and $\delta$ denotes the augmentation factor.
The detailed synthesis procedure is outlined in Algorithm~\ref{alg-attack-2}.
This randomized generation approach ensures that the synthetic embeddings can effectively cover the distribution of real sequence embeddings corresponding to $t_i$, as illustrated in Fig.~\ref{fig-embd-scatter-adv23} in the experiments. 

Note that this synthesis method also functions as a data augmentation strategy for training the classifier $\phi$. By exposing $\phi$ to the perturbed synthetic embedding distribution $\advESC$, which spans a broader range than the real embedding distribution $\advEC$, the robustness and generalization of $\phi$ are significantly enhanced for reconstructing $\tgtS$.
In our experiments, $\adv_2$ can considerably improve the reconstruction accuracies from $50\%$ to higher than $90\%$, closely matching the performance of $\adv_1$ under the ideal condition where $P_\adv \approx P_\tgt$.

\begin{algorithm}[!tb]
\begin{small}
\caption{Adversary $\adv_2$ with Unlimited Query Budgets and an Auxiliary Dataset $\advS$ with $P_\adv \not\approx P_\tgt$}
\label{alg-attack-2}
\KwInput{Auxiliary dataset $\advS$, token augmentation factor $\delta$, target intermediate embeddings $\tgtE$}
\KwOutput{Reconstructed private prompt $\tgtS$}
$\advT\gets$ Obtain the token sets of $\advS$ via the tokenizer\;
$\advTC\gets T_{\text{all} } \setminus \advT$ \tcp*{The token set not in $\advS$}
$\advSC\gets \varnothing$\;
\ForEach{$t^{c}\in\advTC$}{
    \ForEach{$j\in\{1,\cdots,\delta\}$}{
        $\{t_1,\cdots,t_n\} \gets$ Sampling a token sequence from $\advT$\;
        $i\gets$ Sampling a position from $\{1,\cdots,n\}$ \;
        $\advSC\gets \advSC\cup \{t_1,\cdots,t_{i-1}, t^{c}\}$\;
    }
}
$\tgtS \gets$ Execute Algorithm~\ref{alg-attack-1} with inputs $\advS \cup \advSC$ and $\tgtE$\;
\Return $\tgtS$\;
\end{small}
\end{algorithm}
% \vspace{-4mm}

\subsection{The Third Adversary $\adv_3$}
Both $\adv_1$ and $\adv_2$ rely on extensive queries to distributed LLM inference frameworks for constructing attack training datasets.
However, such queries may impose significant time costs, especially under unstable communication in distributed frameworks, and may incur financial costs when pay-as-you-go API services are used.
Therefore, developing a more efficient attack strategy that can work under limited query budgets would be necessary in these scenarios.
For the third adversary $\adv_3$, apart from the limitation on query budgets, we further remove the auxiliary data assumption, i.e., the attacker has no auxiliary data $\advS$ for query construction, making this setting the most restrictive one. 
Here, we follow the leading LLM service providers (e.g., OpenAI~\cite{openai-pricing}) and quantify the query budgets as the maximum number of tokens allowed in total queries.

\begin{figure}
\vspace{-2mm}
  \includegraphics[width=1.\columnwidth]{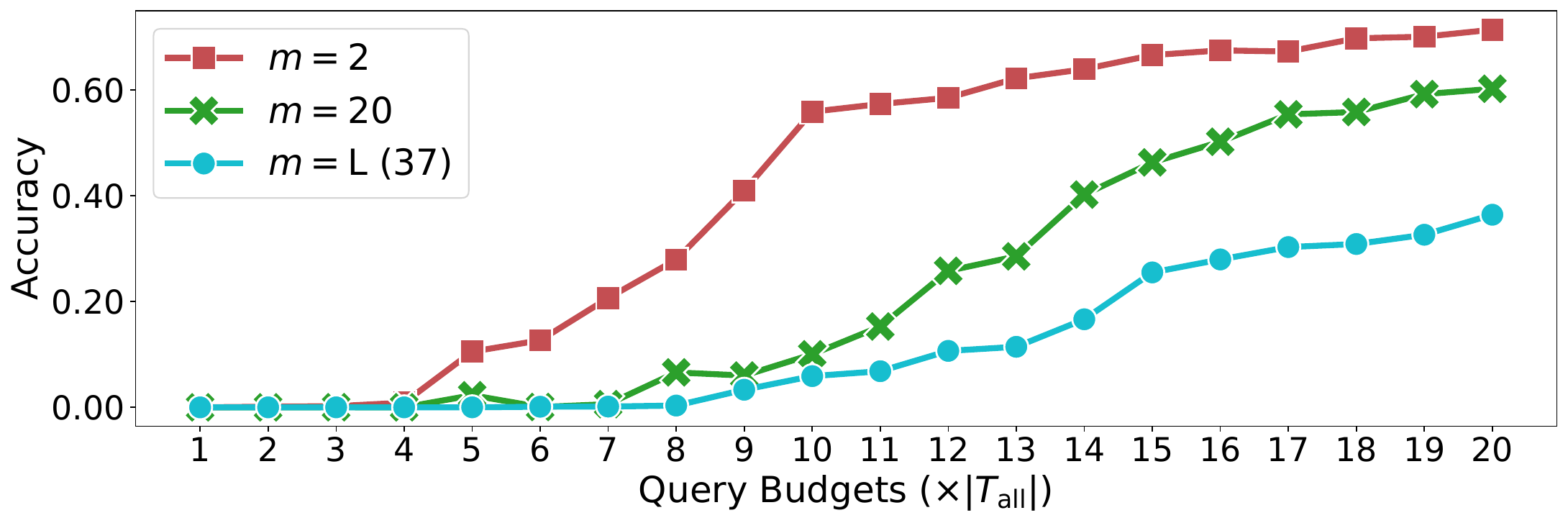}
  \vspace{-7mm}
  \caption{The reconstruction accuracy of $\adv_1$ with limited query budgets performed on GPT-2 and WikiText-2~\cite{wikitext}.}
  \label{fig-mlp-query-budgets}
  \vspace{-4mm}
\end{figure}

% \subsection{The Motivation}
\vspace{0.5mm}\noindent
\textbf{Challenge.}
Without any auxiliary dataset $\advS$, we can still generate the training data $(\advESC, \advTC)$ by setting $\advT=\varnothing$, $\advTC=\tdict$ (the set of all unique token IDs), and replacing preceding tokens $\{t_1,\cdots,t_{i-1}\}$ with random token ID combinations.
The main problem is about the augmentation factor $\delta$ per token ID, i.e., the number of embeddings an attacker can obtain for the same token. Under a limited query budget $\qb$, $\delta$ is constrained to $\lfloor \qb/|\tdict|\rfloor$, which may reduce to one, allowing only a single query per token ID.
Note that we consider the minimal query budget $\qb$ to be $|\tdict|$, as it represents a sufficiently small budget based on the estimates in Table~\ref{tb-price-estimation}, and setting $\qb < |\tdict|$ not only restricts queries to a subset of token IDs but also significantly increases the false positive rates for these tokens during reconstruction.
The minimal budget $\qb = |\tdict|$ can produce a small training dataset, leading to severely underfitted classifiers with poor generalization.
For example, Fig.~\ref{fig-mlp-query-budgets} presents the testing accuracies of $\adv_1$ under limited query budgets, where $m$ denotes the layer number. The results indicate that training a classifier directly with limited query budgets (e.g., $\qb<10\times|\tdict|$) yields poor performance, particularly for later layers.
Therefore, new attack methods need to be designed for this restrictive setting.

\begin{figure}[!t]
  \includegraphics[width=.93\columnwidth]{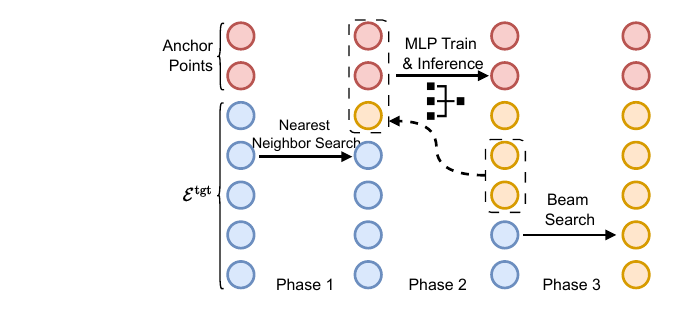}
  \vspace{-2mm}
  \caption{Overview of the semi-supervised three-phase reconstruction framework for $\adv_3$. Red points denote the embeddings with labels, i.e., $\advESC$; yellow points denote the target embeddings that have been assigned labels; blue points denote the target embeddings to be assigned labels.}
  \label{fig-overview-attack3}
  \vspace{-6mm}
\end{figure}

\vspace{0.5mm}\noindent
\textbf{Overview.}
For $\adv_3$, we propose a semi-supervised three-phase reconstruction framework. The first two phases involve classification tasks to reconstruct the backbone of sequences with high confidence, while the third phase focuses on semantic reconstruction to fill in unresolved details.
Before initializing this attack framework, the adversary needs to construct a sequential token set $\advTC$ with $|\advTC|=\qb$ in a way that the positions and preceding tokens for each token ID are randomized, as inspired by observations from $\adv_2$.
To put it simply, $\advTC$ can be built by first repeating all tokens in $\tdict$ at least $\lfloor \qb/|\tdict|\rfloor$ times and then randomly shuffling them. After that, the adversary queries the LLM using $\advTC$ and obtains $(\advESC, \advTC)$ pairs, where $\advESC$ are called the \textit{anchor points}. 
We will compare the effectiveness of this data generation method with the method used by $\adv_2$ in Section~\ref{subsec-exp-adv3}.
For effective attacks, \textit{$\qb$ must exceed $|\tdict|$} to ensure at least one anchor point per token ID.
After obtaining anchor points, the adversary can collect a set of target embeddings $\tgtEset$ and employ the semi-supervised framework to reconstruct $\tgtTset$ and $\tgtSset$.
The details of this attack framework are discussed in subsequent sections, with an overview provided in Fig.~\ref{fig-overview-attack3}.

\begin{figure}[!t]
\centering
% \begin{small}
\begin{tabular}{c}
\vspace{-2mm}
\subfloat[Layer 6]{\includegraphics[width=0.49\columnwidth]{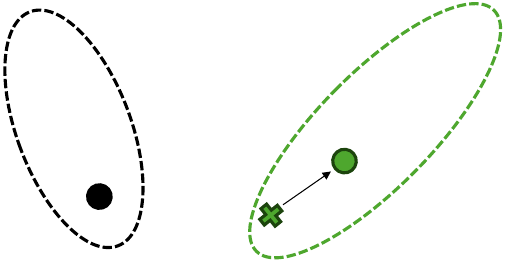}\label{fig-exp-nn-front}}
\hspace{6mm}
\subfloat[Layer 16]{\includegraphics[width=0.36\columnwidth]{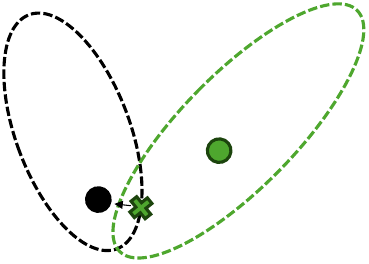}\label{fig-exp-nn-end}}
\end{tabular}
\vspace{-2mm}
\caption{Examples of the nearest neighbor search method. Solid circles denote the anchor points, and the cross denotes the target embedding to be assigned labels.}
\label{fig-exp-nn}
% \end{small}
\vspace{-5mm}
\end{figure}

\subsubsection{Phase 1: Nearest Neighbor Search}\label{subsec-nn-search}
As shown in Fig.~\ref{fig-exp-dist}, sequence embeddings with the same ending token typically form a dense cluster in the embedding space, and different clusters are sparsely distributed.
Leveraging this property, a straightforward method to assign a token label to a target embedding $\be^{\tgt}$ is to compute the average distance from $\be^{\tgt}$ to the anchor points in each cluster and assign the token label of the nearest cluster to $\be^{\tgt}$:
\begin{equation}\label{eq-nn-search-main}
    t^{\tgt} = \argmin_j d^{\tgt}_{\#j}, \text{ where } d^{\tgt}_{\#j}=\sum_{\be^{c}_{\#j} \in E^{c}_{\#j}} ||\be^{\tgt} - \be^{c}_{\#j} ||_p / | E^{c}_{\#j}|,
\end{equation}
where $\#j$ denotes the token ID $j$ associated with a cluster, and $E^{c}_{\#j}\subset \advESC$ denotes all anchor points in this cluster.

However, this method assumes that all embedding clusters exhibit isotropic distributions, which is generally invalid for real-world datasets, as demonstrated by the red points in Fig.~\ref{fig-exp-attack-2}.
As a result, it can easily lead to errors when clusters overlap, particularly in later layers of LLMs.
For example, in Fig.~\ref{fig-exp-nn-end}, the target embedding, represented by a cross, is mistakenly assigned to the black cluster by nearest neighbor search, although its true label is green.
Additionally, with a limited number of anchor points, it is challenging to accurately capture the overall shapes of embedding clusters, further reducing prediction accuracy.
To mitigate this issue, we propose incorporating a confidence score for the predictions generated by the nearest neighbor search. This allows us to focus on assigning labels only to target embeddings with high confidence.
The confidence score is calculated as follows. For a target embedding $\be^{\tgt}$, we first compute the $k$ smallest cluster distances $\{d_{\#j_1}^{\tgt},\cdots, d_{\#j_k}^{\tgt}\}$. Then, for each $o \in \{j_1, \cdots, j_k\}$,
\begin{equation}
    \hat{d}_{\#o}^{\tgt} = (c_1 - d_{\#o}^{\tgt})/c_2,  \label{eq-norm-d}
\end{equation}
\begin{equation}
     p_{\#o}^{\tgt} = \frac{\exp(\hat{d}_{\#o}^{\tgt})}{\sum_{r=j_1}^{j_k} \exp(\hat{d}_{\#r}^{\tgt})}, \label{eq-softmax-d}
\end{equation}
where Eq.~\ref{eq-norm-d} normalizes the distances with constants $c_1$ and $c_2$, and Eq.~\ref{eq-softmax-d} applies softmax to the normalized distances. Next, we compute the entropy of the normalized probabilities:
\begin{equation}\label{eq-confd-score}
    H(t^{\tgt})= -\sum_{o=j_1}^{j_k} p^{\tgt}_{\#o} \log p^{\tgt}_{\#o},
\end{equation}
The intuition behind Eq.~\ref{eq-confd-score} is that if $\be^{\tgt}$ is significantly closer to a cluster $j\in \{j_1, \cdots, j_k\}$ than to other $k-1$ clusters, the probability $p_{\#j}^{\tgt}$ of $\be^{\tgt}$ belonging to cluster $j$ will be higher, resulting in lower entropy $H(t^{\tgt})$, i.e., less uncertainty for the results $t^{\tgt}=j$. 
The confidence score for the result $t^{\tgt}$ from Eq.~\ref{eq-nn-search-main} is then computed as $1-H(t^{\tgt})$.
% The distance entropy is termed the confidence score.
%
Specifically, if $H(t^{\tgt})$ is below a small threshold $\tau_{\text{NN}}$ (the confidence is larger than $1-\tau_{\text{NN}}$), we assign $t^{\tgt}$ to $\be^{\tgt}$ and incorporate $(\be^{\tgt}, t^{\tgt})$ into the anchor points.
After processing all target embeddings in $\tgtEset$, the method proceeds to the next phase.

\subsubsection{Phase 2: Classifier Search}
The nearest neighbor search, limited by its assumption of isotropic cluster distributions, can only assign labels to a small fraction of $\tgtEset$, primarily when clusters are sparsely distributed.
However, by augmenting the anchor points $(\advESC, \advTC)$ during this process, we obtain additional samples for each cluster, which can enable the training of a non-linear classifier $\phi$ akin to the approach used in $\adv_1$. This classifier can learn the anisotropic distributions of different clusters more effectively.
It is important to note that the sample space $(\advESC, \advTC)$ may remain small under restricted query budgets after the first phase, thus applying a large gradient penalty during classifier training is essential to mitigate overfitting issues.
Once $\phi$ is trained, it can be employed to predict labels for target embeddings that remain unlabeled after the nearest neighbor search.

Despite this, $\phi$’s accuracy is likely constrained by the insufficient representation of real embedding distributions within $(\advESC, \advTC)$.
To enhance prediction reliability, we adopt a similar approach to Eq.~\ref{eq-confd-score}, computing logits entropy from the softmax outputs of $\phi$. Labels are assigned only to embeddings with logits entropy below a predefined threshold $\tau_{\phi}$.
This model training and inference phase can iterate multiple times to progressively expand the anchor points for training different $\phi$s.
For any embeddings that remain unlabeled after this phase, the framework proceeds to the third phase.

\subsubsection{Phase 3: Beam Search}
Following the first two phases, partially reconstructed sentences may resemble:  ``The[...] is a pioneer in[...] programming aimed at minorities[...] and[...]'', where ``[...]'' denotes a token failed to be reconstructed.
In this phase, we design a modified beam search to fill in the missing tokens.
Beam search is commonly used for generating plausible sentences in language models~\cite{beamsearch}, which operates by generating tokens sequentially while retaining only the $\beta$ (beam width) most probable sequences at each step.
The selection of the most probable sequences is guided by probabilities output by a language model. For this task, we propose to use a shadow LLM $f_{\text{shadow}}$ to fulfill the beam search, which can be any pre-trained open-source LLM, such as GPT-2.
Considering that our task is a token reconstruction task instead of a sentence generation task, we modify the traditional beam search method~\cite{beamsearch} to a context-aware and candidate-constrained search method. Algorithm~\ref{alg-beamsearch} presents details of the modified beam search method.

Specifically, the ``context-aware'' modification ensures that we determine the probabilities of possible missing tokens based on the reconstructed context, i.e., tokens already assigned labels (line~\ref{alg-line-beam-score}). Only those missing tokens that maximize the output probabilities of $f_{\text{shadow}}$ for the given context are retained (line~\ref{alg-line-reduce-beam-width}). 
This modification leverages the semantic knowledge embedded in $f_{\text{shadow}}$, ensuring that the reconstructed tokens are semantically consistent with the given context.
The ``candidate-constrained'' modification restricts the search space for each missing token to the top-5 candidates identified during the nearest neighbor search (line~\ref{alg-line-cand-t5}), rather than exploring the entire token dictionary.
Table~\ref{tb-top-k-comparison} presents the top-$k$ ($k\in\{1, 5, 10\}$) accuracies obtained by the nearest neighbor search under the minimum query budget, from which we observe that the top-5 accuracies are typically $30\%$ higher than the top-1 accuracies, while accuracy gains from top-5 to top-10 are minimal.
Consequently, we focus on reconstructing the missing tokens from the top-5 candidates via beam search, rather than generating sentences solely for plausible semantics.
In the final reconstructed sentences, tokens recovered through beam search can be flagged to highlight potential errors, thereby reducing the risk of transmitting misleading information to the adversary.
We summarize the semi-supervised framework of $\adv_3$ in Algorithm~\ref{alg-attack-3}.
% of the full version~\cite{fullversion}.

\begin{table}[t!]
\setlength{\tabcolsep}{12pt}
\center
\small
\caption{The accuracies for $\ell_1$-norm nearest neighbor search, tested on GPT-2 and Wikitext with $\qb=|\tdict|$.}\label{tb-top-k-comparison}
\vspace{-2mm}
\begin{tabular}{c|ccccc}
\hline
\multirow{2}{*}{Accuracy} & \multicolumn{4}{c}{Layers} \\
\cline{2-5}
   & 2  & 10 & 20 & 30   \\
\hline
\hline
 Top-1  & 0.7747  & 	0.4480  & 	0.2882  & 	0.1149  \\
 Top-5  &  0.8802  & 	0.5689  & 	0.3958  & 	0.1800\\
 Top-10  &  0.8984  & 	0.5930  & 	0.4387  & 	0.2108\\
\hline
\end{tabular}
\vspace{-4mm}
\end{table}

\begin{algorithm}[!tb]
\caption{Context-Aware Constrained Beam Search}\label{alg-beamsearch}
\KwIn{Beam width $\beta$, language model $f_{\text{shadow}}$, partially reconstructed token sequence $T^{\tgt}$, mask for missing tokens $M_{\text{miss}}$, top-5 candidate tokens $C_{\text{t5}}$}
\KwOut{Best sequence $S^*$}
$\text{Beam} \gets \{\langle \texttt{start}, 0 \rangle\}$ \tcp*[r]{Sequence and its score}
\For{$i = 1$ \textbf{to} $|T^{\tgt}|$}{
    $\text{Cands} \gets \emptyset$ \tcp*[r]{Candidate sequences}
    \ForEach{$\langle S, \text{score} \rangle \in \text{Beam}$}{
        \If{$M_{\text{miss}}[i]$ is true}{   
            \ForEach{$w \in C_{\text{t5}}[i]$}{\label{alg-line-cand-t5}
                $\text{new\_score} \gets \text{score} + \log f_{\text{shadow}}(w \mid S)$\; \label{alg-line-beam-score}
                $\text{Cands} \gets \text{Cands} \cup \langle S \| w, \text{new\_score} \rangle$\;
            }
        }
        \Else{
            $\text{Cands} \gets \text{Cands} \cup \langle S \| T^{\tgt}[i], \text{score} \rangle$ \;
        }
    }
    $\text{Beam} \gets \text{Top-}\beta(\text{Cands})$ \; \label{alg-line-reduce-beam-width}
}
$S^* \gets \arg\max_{\langle S, \text{score} \rangle \in \text{Beam}} \text{score}$ \tcp*[r]{Best sequence}
\Return $S^*$\;
\end{algorithm}

\begin{algorithm}[!tb]
\begin{small}
\caption{Adversary $\adv_3$ with Limited Budgets}
\label{alg-attack-3}
\KwInput{Budgets $\qb$, distance entropy threshold $\tau_{\text{NN}}$, logits entropy threshold $\tau_{\phi}$, beam width $\beta$, target embeddings $\tgtEset$}
\KwOutput{Reconstructed private prompt set $\tgtSset$}
$\delta \gets \lfloor \qb/|\tdict|\rfloor$ \;
$\advTC \gets$ Randomly repeat and combine all $t\in \tdict$ to form a $\qb$-token sequence \; \label{alg-line-attack3-datagen-s}
\If{$\delta>100$}{
    $\tgtSset \gets$ Execute Algorithm~\ref{alg-attack-1} with inputs $\advTC$ and $\tgtEset$\;
    \Return $\tgtSset$\;
}
$\advESC\gets$ Send queries from $\advTC$ for distributed LLM inference \; \label{alg-line-attack3-datagen-e}
$\{\langle E^{c}_{\#j}, \#j\rangle\}_{j=1}^{|\tdict|} \gets$ Construct anchor points from $(\advESC, \advTC)$ \; 
$\tgtSset \gets \varnothing$\;
$C_{\text{top-}k}\gets \varnothing$\; 
\ForEach(\tcp*[f]{Phase 1: nearest neighbor search}){$\be^{\tgt}_i\in \tgtEset$}{
    $\{t_{i\#j_1}^{\tgt},\cdots,t_{i\#j_k}^{\tgt}\}\gets $ Compute top-$k$ candidates\;
    $C_{\text{top-}k}\gets C_{\text{top-}k} \cup \{t_{i\#j_1}^{\tgt},\cdots,t_{i\#j_k}^{\tgt}\}$\;
    $H_{\text{NN}}\gets$ Compute distance entropy for $t_{i\#j}^{\tgt}$ using Eq.~\ref{eq-nn-search-main} and \ref{eq-confd-score}\;
    \If{$H_{\text{NN}}<\tau_{\text{NN}}$}{
        $\tgtSset\gets \tgtSset\cup \langle t_{i\#j}^{\tgt}, i \rangle $\;
        $E^{c}_{\#j}\gets E^{c}_{\#j}\cup \be^{\tgt}_i $ \;
    }
}
\ForEach(\tcp*[f]{Phase 2: classifier search}){iteration}{
    $\phi\gets$ Train a classifier on anchor points\;
    \ForEach{$\be^{\tgt}_i\in \tgtEset - \text{anchor points}$}{
        $\{p_{i\#j}^{\tgt}\}_{j=0}^{|\tdict|-1} \gets \phi(\be^{\tgt}_i)$ \;
        $H_{\phi}\gets$ Compute logits entropy\;
        \If{$H_{\phi}<\tau_{\phi}$}{
            $\tgtSset\gets \tgtSset\cup \langle \argmax_j p_{i\#j}^{\tgt}, i \rangle $\;
            $E^{c}_{\#j}\gets E^{c}_{\#j}\cup \be^{\tgt}_i $ \;
        }
    }
}
\If(\tcp*[f]{Phase 3: beam search}){$|\tgtSset|<|\tgtEset|$}{
    $T^{\tgt}, M_{\text{miss}}\gets$ Construct partial sequence and mask from $\tgtSset$\;
    $S^* \gets$ Execute Algorithm~\ref{alg-beamsearch} with inputs $\beta, T^{\tgt}, M_{\text{miss}}$ and $C_{\text{top-}k}$\;
    Update $\tgtSset$ based on $S^*$\;
}
\Return $\tgtSset$\;
\end{small}
\end{algorithm}
\section{Experiments}\label{sec-exp}
% In this section, we first introduce the experimental settings and then present the experimental results of the proposed attacks on state-of-the-art LLMs.

% \subsection{Experimental Setting}\label{subsec-exp-setting}
\vspace{0.5mm}\noindent
\textbf{Platform.}
All attack algorithms are implemented in Python using \textit{PyTorch}~\footnote{\url{https://pytorch.org/}}.
Experiments are conducted on a high-performance server equipped with AMD EPYC 7742 64-Core Processor $\times$ 256, NVIDIA A100-SXM4-40GB $\times$ 4, and 256GB RAM, running Ubuntu 22.04. 
We follow the settings of Petals~\cite{Petals-acl, Petals-nips} to set up a local distributed LLM inference framework~\footnote{\url{https://github.com/bigscience-workshop/petals}} and perform experiments on it.
%
% It is important to note that although the Petals community provides an online system, we found it unstable with high latency during LLM inference. Consequently, a private swarm was launched locally to ensure a reliable pipeline.

\vspace{0.5mm}\noindent
\textbf{LLMs.}
We evaluate the proposed attacks on three state-of-the-art decoder-only LLMs: Phi-3.5~\cite{phi3.5}, Llama-3.2~\cite{llama3.2}, and GPT-2~\cite{brown2020language-gpt3}, which are mainly designed for next-token prediction tasks.
Additionally, we assess the attacks on BERT~\cite{DevlinCLT19Bert}, an encoder-only model primarily used for sentence embedding generation.
The key architectural distinction is that decoder-only models condition the generation of embeddings $\be_i$ on the preceding $t-1$ tokens, as illustrated in Fig.~\ref{fig-corr-orig}, whereas encoder-only models generate embeddings $\be_i$ based on both preceding and following tokens within the input sequence. 
By including BERT in the evaluation, we aim to test the generalizability of the proposed attacks across different types of LLM architectures. The LLM details are shown in Table~\ref{tb-llm}.

\begin{table}[!t]
\setlength{\tabcolsep}{3pt}
\vspace{-2mm}
\caption{The used large language models.}\label{tb-llm}
\vspace{-2mm}
\centering
\small
\begin{tabular}{cccccc}
\toprule
{LLM} & {\#params} & \#layers & {Release Year} & {Type} & Developer \\
\midrule
% \hline
Phi-3.5 & 3.82B & 32 &	2024 & Decoder-only  & Microsoft\\ 
% \hline
Llama-3.2 & 1.24B & 16 & 	2024 & Decoder-only  & Meta \\ 
GPT-2 & 812M & 36 &	2019 & Decoder-only  & OpenAI \\
% \hline
BERT & 335M & 24 & 2018 & Encoder-only  & Google \\
\bottomrule
\end{tabular}
\vspace{-2mm}
\end{table}

\vspace{0.5mm}\noindent
\textbf{Datasets.}
We utilize four datasets in our experiments: WikiText-2~\cite{wikitext}, SQuAD 2.0~\cite{squad2.0}, Midjourney prompts~\cite{midjourney}, and \textcolor{mycolor}{PrivatePrompts}~\cite{MorrisZCSR24promptinference1}.
\textit{WikiText-2} comprises over 100 million tokens extracted from high-quality and featured Wikipedia articles.
\textit{SQuAD 2.0} includes over 150,000 crowd-sourced questions and is designed for evaluating the reading comprehension ability of LLMs.
\textit{Midjourney prompts} include 250,000 text-to-image messages collected from the Midjourney bot server over four weeks. 
%This dataset is preprocessed to retain only the text prompts written by users.
%
\textcolor{mycolor}{\textit{PrivatePrompts} consists of 251,270 manually-crafted sensitive strings with inserted PII entities, such as name and date. }
%
% Note that the focus of this study is on reconstructing the raw text content of these datasets; hence, the specific task types of these datasets are not closely relevant to the results in this paper.
%
For $\adv_1$, each dataset is randomly partitioned into training, testing, and evaluation splits in the ratio $(64\%, 16\%, 20\%)$.
The training and testing subsets are used for attack model development, while the evaluation subset is held out for attack evaluation.
For $\adv_2$, one dataset is used exclusively for training and testing, while another is designated for attack evaluation.
For $\adv_3$, all datasets are used only for attack evaluation. 
%
% The training data is randomly generated following the procedure outlined in Algorithm~\ref{alg-attack-3}.

\vspace{0.5mm}\noindent
\textbf{Attack Implementation.}
The classifier $\phi$ for all three attacks is implemented using a multi-layer perceptron (MLP) with six layers, where layer normalization and ReLU activation are used between consecutive layers. 
The input dimension matches the embedding size of the target LLM, while the output dimension corresponds to the tokenizer's vocabulary size, $|\tdict|$.
The MLP is trained for six epochs during $\adv_1$ and $\adv_2$ and for four epochs per iteration during $\adv_3$ of Algorithm~\ref{alg-attack-3}.
The distance and logits thresholds in Algorithm~\ref{alg-attack-3} are set to $\tau_{\text{NN}} = \tau_{\phi} = 0.05$. The beam width $\beta$ in the beam search phase is set to 6. 
%
% Notably, smaller thresholds reduce the risk of false positives in anchor point selection, thereby improving the accuracy of beam search in the third phase.

\vspace{0.5mm}\noindent
\textbf{Metric.}
The effectiveness of the proposed attacks is measured by two metrics: token Reconstruction Accuracy (RA) and \textcolor{mycolor}{Cosine Semantic Similarity (CSS)}. RA is calculated as: $\text{RA}=\sum_{t_i\in\tgtTset}\frac{\mathbbm{1}(\hat{t}_i=t_i)}{|\tgtTset|}$, where $\mathbbm{1}$ is the indicator function, and $\hat{t}_i$ denotes the reconstructed token corresponding to $t_i$. \textcolor{mycolor}{CSS is calculated by first extracting semantic embeddings of both reconstructed and ground truth prompts using Sentence-BERT~\footnote{\url{https://huggingface.co/efederici/sentence-bert-base}} and then measuring the cosine similarity between them.}
Based on our experiments, $\text{RA}>40\%$ is considered sufficient for plausible reconstruction, as key concepts of the ground truth prompts can already be reconstructed under this accuracy.
% , as shown in Table~\ref{tb-egg-adv3-in-paper}.

\vspace{0.5mm}\noindent
\textbf{Baselines.} \textcolor{mycolor}{
Three baselines are used in our experiments: two sentence embedding inversion attacks designed primarily for BERT-like LLMs, \textit{B-SEI}~\cite{GuKRVM23embdinv3} and \textit{Vec2Text}~\cite{MorrisKSR23promptinference}, and one LLM output inversion attack, \textit{Output2Prompt}~\cite{output2prompt}.
B-SEI and Vec2Text can be adapted to reconstruct prompts from different LLM layers, while Output2Prompt can only use LLM outputs for prompt reconstruction. 
Although Output2Prompt operates under different assumptions than embedding inversion attacks, comparing it with the others helps assess vulnerabilities across the entire LLM inference pipeline.
}

% Two existing sentence embedding inversion attacks~\cite{SongR20promptinference, GuKRVM23embdinv3} share the same attack settings as $\adv_1$, even though they are primarily developed for BERT-like sentence embedding models. 
% %
% Both methods train an inversion model to reconstruct prompts $\{t_i\}_{i=1}^l$ from intermediate embeddings $\{\be_i\}^l_{i=1}$; however, they differ in the choice of inversion model architectures. 
% %
% \citet{SongR20promptinference} employs a recurrent neural network (RNN), while \citet{GuKRVM23embdinv3} utilizes a transformer-based model. We adopt the transformer-based approach~\cite{GuKRVM23embdinv3} (referred to as \textit{B-SEI}) as the baseline, as it significantly outperforms the RNN-based method~\cite{SongR20promptinference}.

\subsection{Attack Performance of $\adv_1$}\label{subsec-exp-adv1}
We evaluate the performance of the first attack $\adv_1$ across different layers of various LLMs and datasets. Note that in the distributed LLM inference pipeline (Fig.~\ref{fig-overview}), $m$ denotes the first decoder layer hosted by the adversary. An attack on layer $m$ implies reconstructing prompts based on the embeddings $E^{(m-1)}$ fed into layer $m$. Typically, $m \geq 2$ in our experiments, as $m = 1$ corresponds to reconstructing prompts directly from initial token embeddings, which is trivial due to the one-to-one correspondence between these embeddings and input token IDs.
Considering that different LLMs have different numbers of layers ($n$), for each LLM, we evaluate the attack performance on five evenly selected layers spanning from the first to the last. Note that $m=\text{L}$ (i.e., $n+1$) implies reconstructing prompts from the embeddings fed into the linear layer. 

\begin{figure*}[t]
\centering
%\captionsetup[subfloat]{captionskip=-0.5mm}
\begin{small}
\begin{tabular}{cccc}
\multicolumn{4}{c}{\hspace{0mm} \includegraphics[height=5mm]{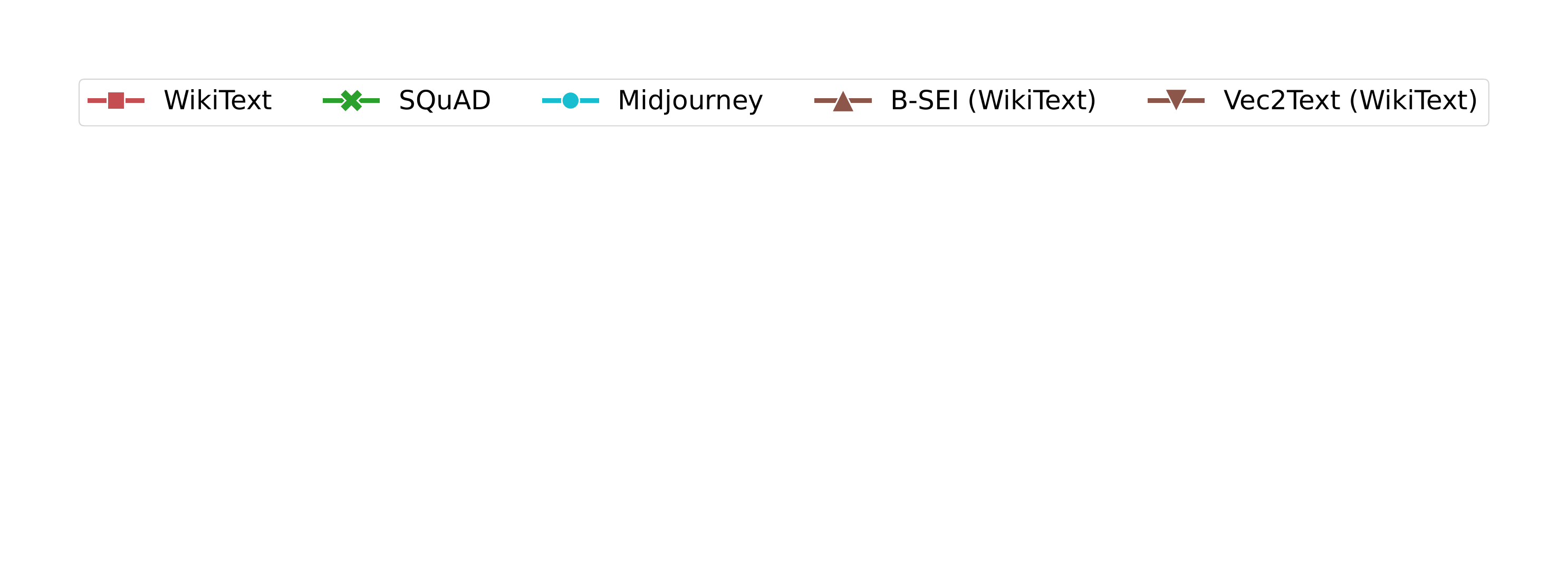}}
\vspace{-4mm}  \\
\hspace{-5mm}
\subfloat[{Phi-3.5}]{\includegraphics[width=0.25\textwidth]{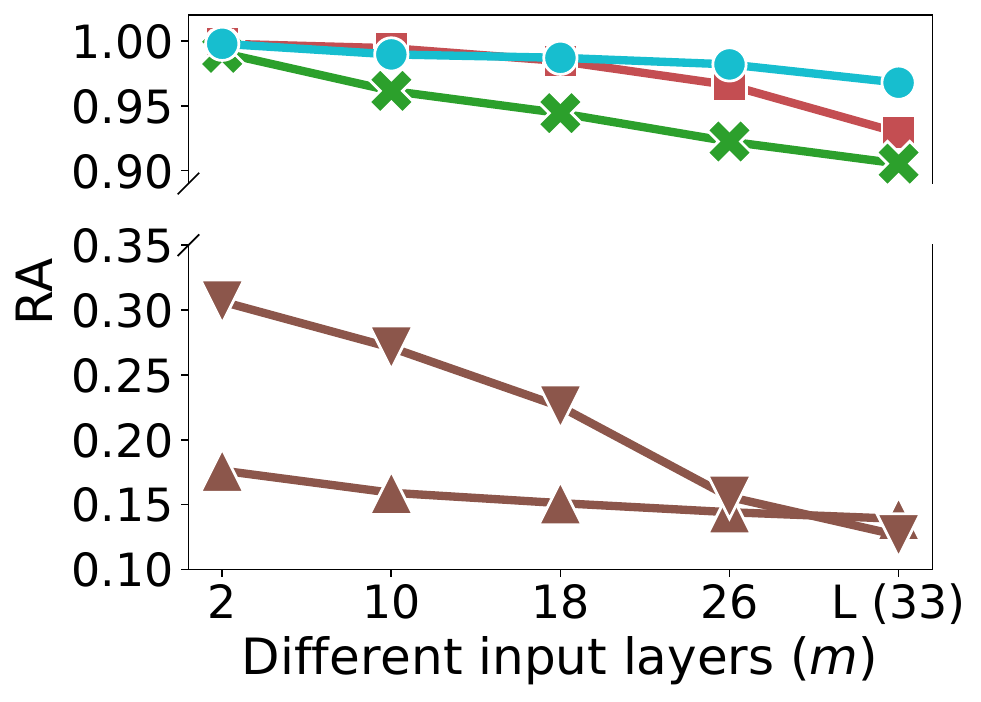}\label{subfig-attack1-l}}
&
\hspace{-5mm}
\subfloat[{Llama-3.2}]{\includegraphics[width=0.25\linewidth]{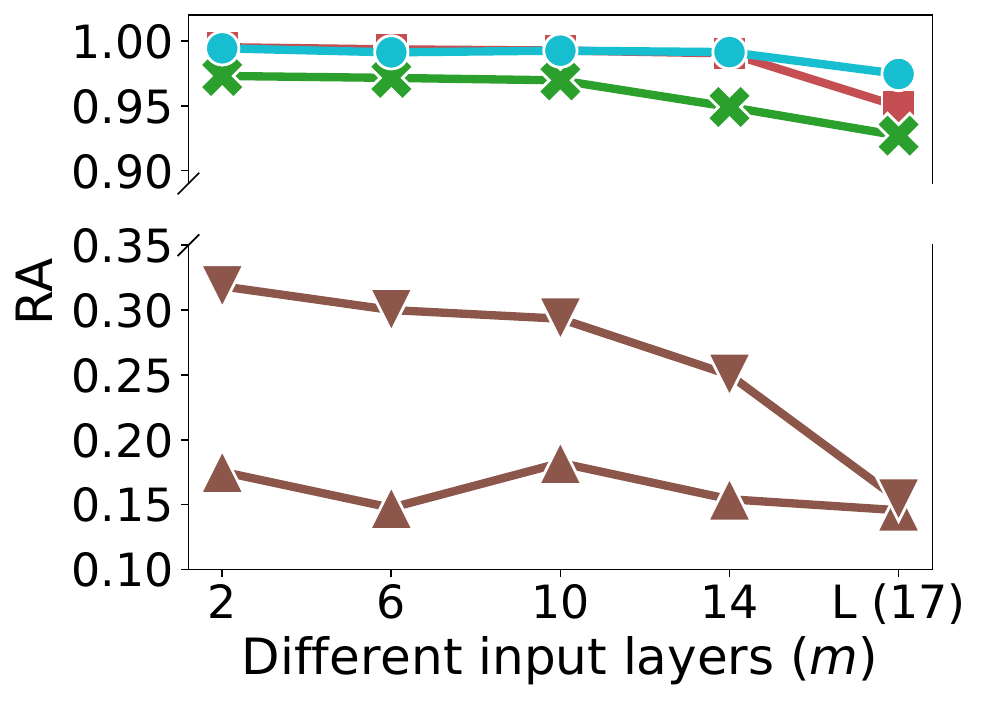}}
&
\hspace{-5mm}
\subfloat[{{GPT-2 }}]{\includegraphics[width=0.25\linewidth]{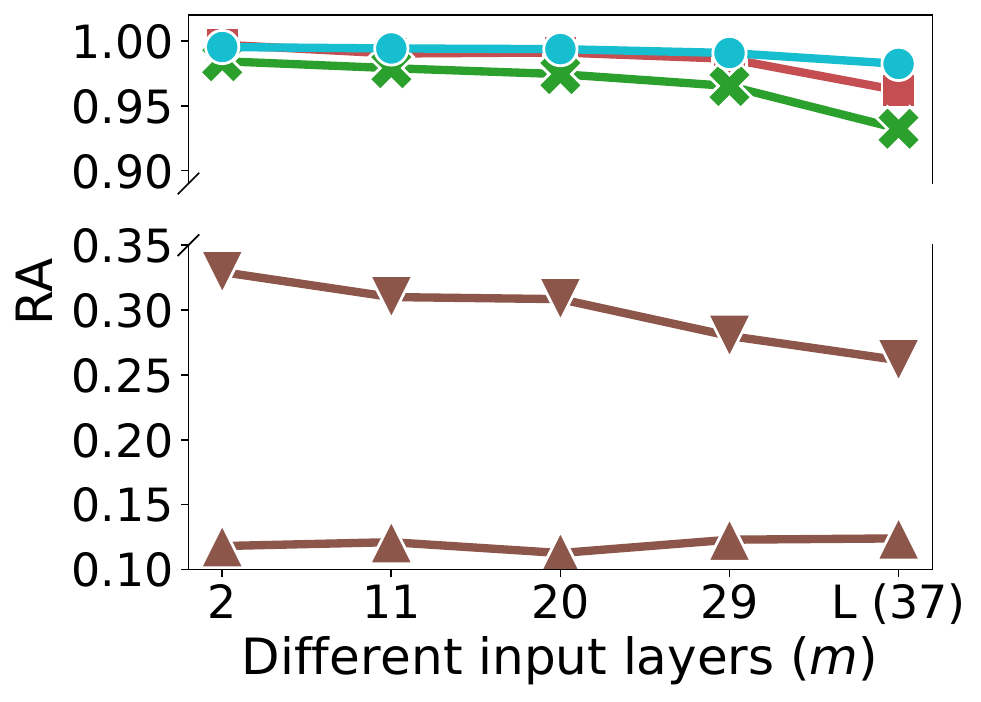}}
&
\hspace{-5mm}
\subfloat[{{BERT}}]{\includegraphics[width=0.25\linewidth]{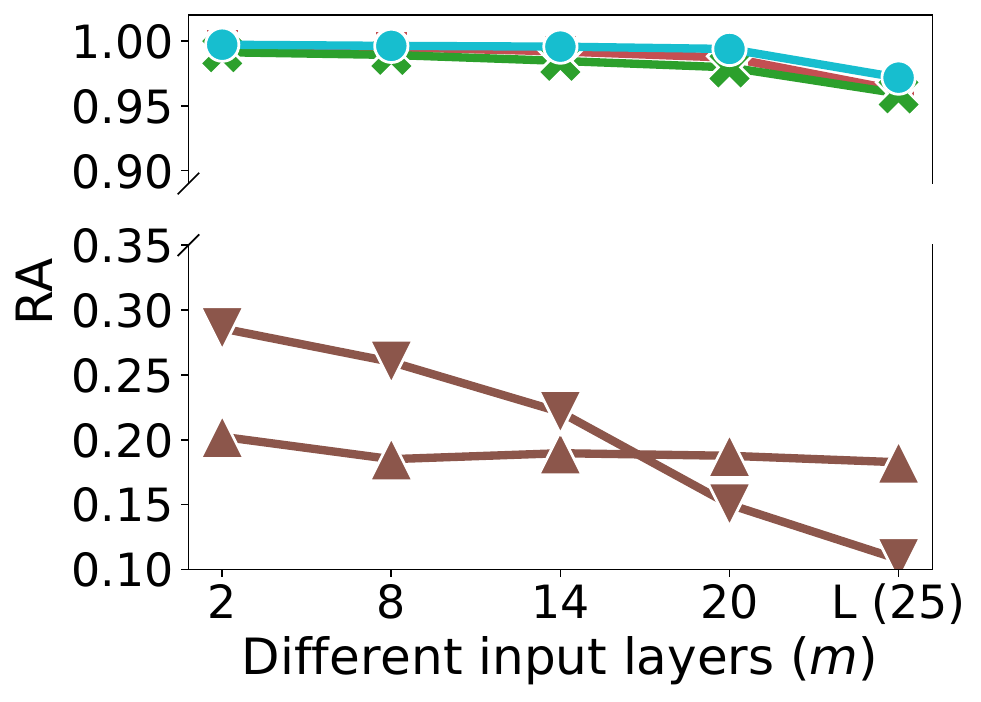}\label{subfig-attack1-r}} 
\vspace{-2mm}
\end{tabular}
\vspace{-1mm}
\caption{The reconstruction accuracies of $\adv_1$ performed on different models and datasets. Reconstruction examples are deferred to Table~\ref{tb-exp-adv1-append} of the full version~\cite{fullversion}.}
% \textcolor{mycolor}{(Revision: data for the baseline Vec2Text is added)}}
\label{fig-attack1-accur}
\end{small}
\vspace{-2mm}
\end{figure*}

\begin{figure*}[t]
\centering
%\captionsetup[subfloat]{captionskip=-0.5mm}
\begin{small}
\begin{tabular}{ccccc}
\multicolumn{5}{c}{\hspace{0mm} \includegraphics[height=5mm]{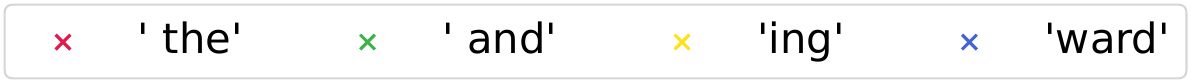}}
\vspace{-4mm}  \\
\hspace{-4mm}
\subfloat[$m=2$]{\includegraphics[width=0.2\textwidth]{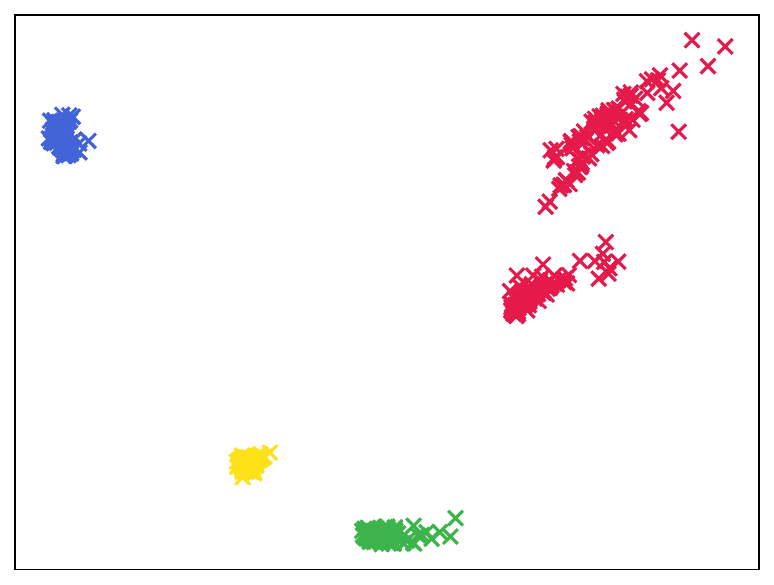}\label{subfig-embd-dist-phi3-l}}
&
\hspace{-5mm}
\subfloat[$m=10$]{\includegraphics[width=0.2\textwidth]{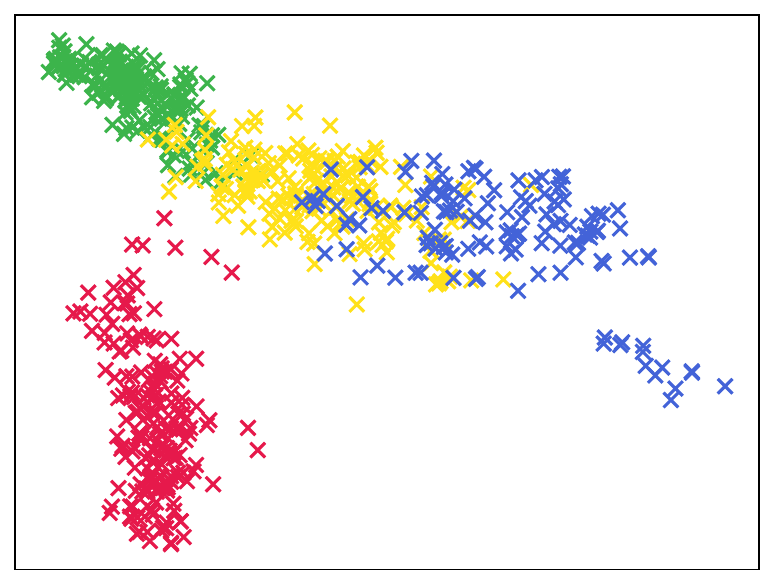}}
&
\hspace{-5mm}
\subfloat[$m=18$]{\includegraphics[width=0.2\textwidth]{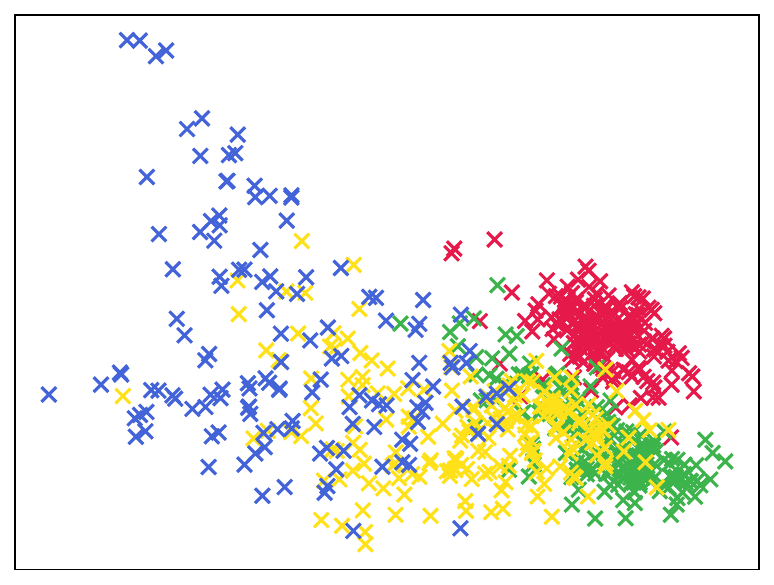}}
&
\hspace{-5mm}
\subfloat[$m=26$]{\includegraphics[width=0.2\textwidth]{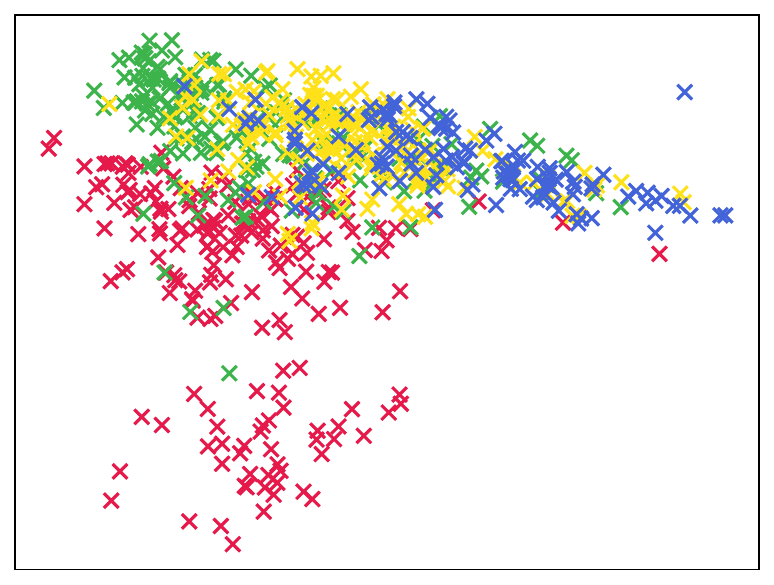}}
&
\hspace{-5mm}
\subfloat[$m=\text{L}$ (33)]{\includegraphics[width=0.2\textwidth]{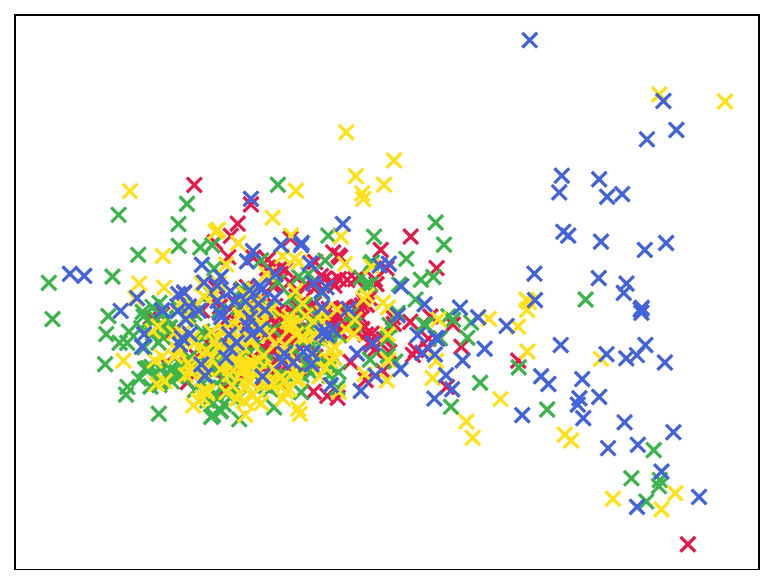}\label{subfig-embd-dist-phi3-r}}
\vspace{-2mm}  \\
\hspace{-4mm}
\subfloat[$m=2$]{\includegraphics[width=0.2\textwidth]{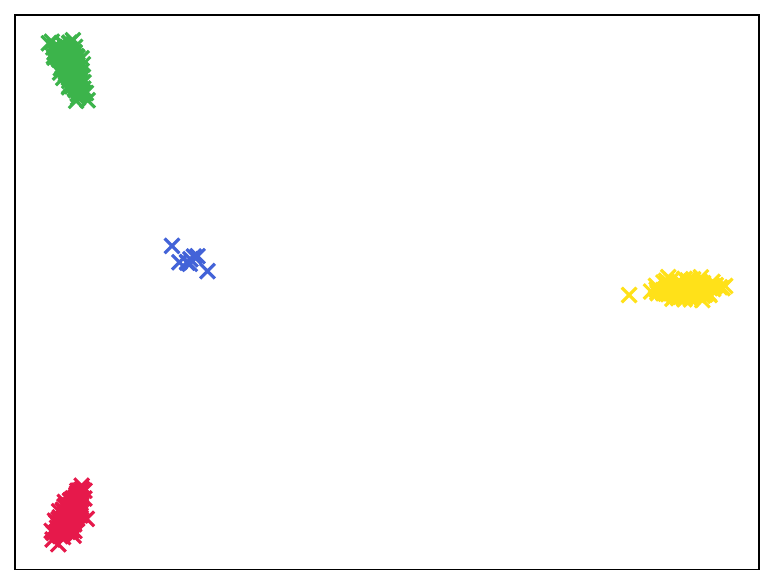}\label{subfig-embd-dist-bert-l}}
&
\hspace{-5mm}
\subfloat[$m=8$]{\includegraphics[width=0.2\textwidth]{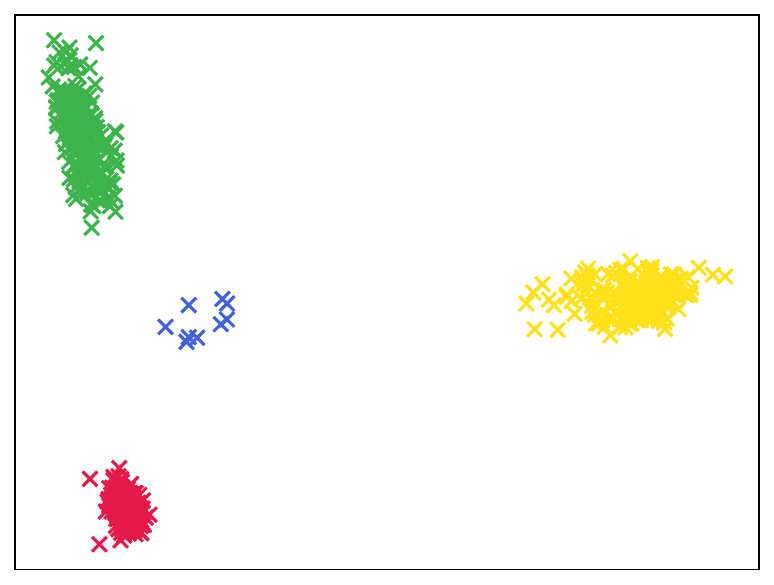}}
&
\hspace{-5mm}
\subfloat[$m=14$]{\includegraphics[width=0.2\textwidth]{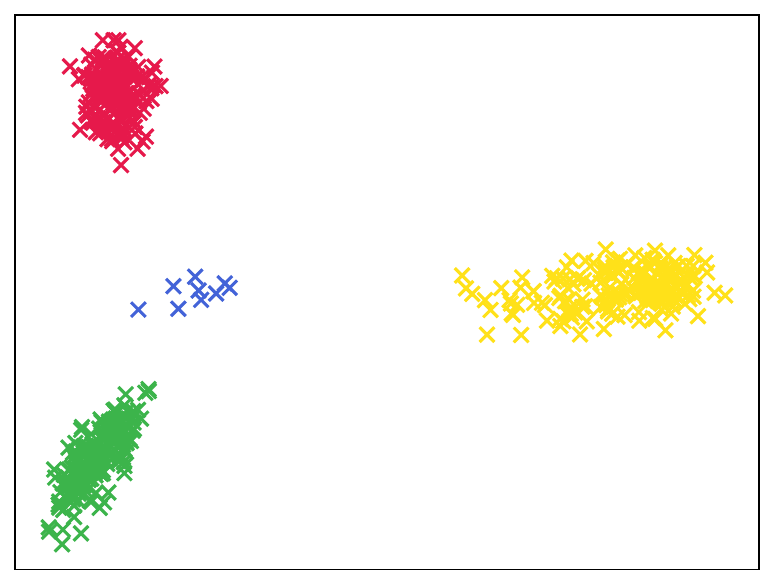}}
&
\hspace{-5mm}
\subfloat[$m=20$]{\includegraphics[width=0.2\textwidth]{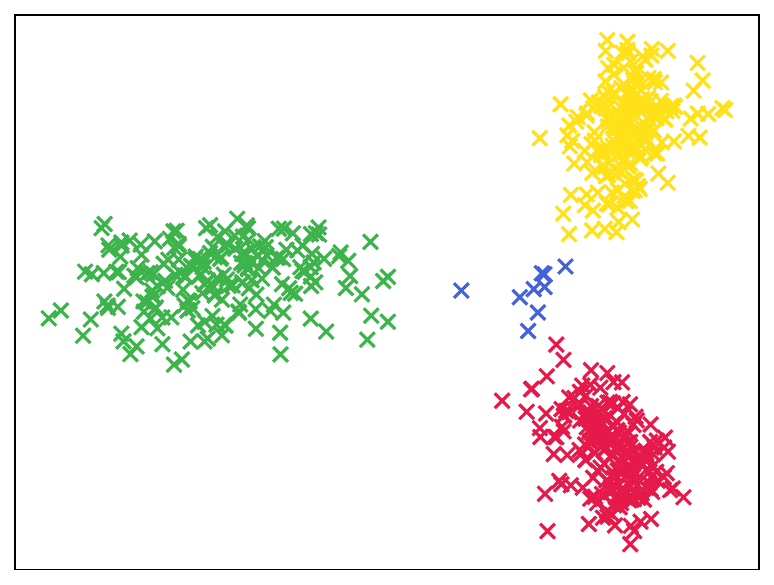}}
&
\hspace{-5mm}
\subfloat[$m=\text{L}$ (25)]{\includegraphics[width=0.2\textwidth]{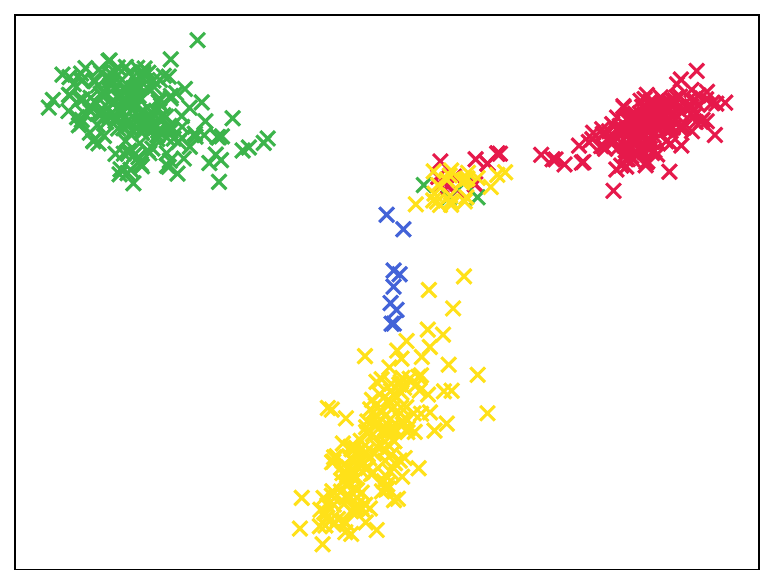}\label{subfig-embd-dist-bert-r}}
\end{tabular}
\vspace{-3mm}
\caption{The embedding distributions generated in WikiText-2 via (a)--(e) Phi-3.5 and (f)--(j) BERT for $\adv_1$. See Fig.~\ref{fig-embd-scatter-adv1-append} of the full version~\cite{fullversion} for more results. }
% \textcolor{mycolor}{(Revision: tokens ` one' and ` which' are replaced with `ing' and `ward')}}
\label{fig-embd-scatter-adv1}
\end{small}
\vspace{-2mm}
\end{figure*}

\vspace{0.5mm}\noindent
\textbf{{Results of $\adv_1$.}} 
The reconstruction accuracies of $\adv_1$ are shown in Fig.~\ref{fig-attack1-accur}, from which we draw three key observations.
\textit{First}, the RA of $\adv_1$ decreases as the layer number increases. \textit{Second}, $\adv_1$ achieves higher RA on encoder-based BERT compared to decoder-based LLMs. \textit{Third}, $\adv_1$ consistently achieves RA greater than $90\%$ across all layers in the evaluated models. 
To explain these observations, we visualize embedding distributions across different LLM layers by first using PCA to project the embeddings into two dimensions and then scatter these two-dimensional points in Fig.~\ref{fig-embd-scatter-adv1}.

\vspace{0.5mm}\noindent
\textbf{{Rationale.}} 
From Fig.~\ref{fig-embd-scatter-adv1}, we can see that embeddings corresponding to different tokens are sparsely distributed in the early layers. As the inference progresses, embedding clusters diffuse and overlap, reducing the separability between clusters and thus hindering the adversary's classifier. This diffusion explains the first observation that RA decreases in deeper layers. 
Additionally, Fig.~\ref{fig-embd-scatter-adv1} reveals that cluster diffusion in encoder-based BERT is less pronounced compared to decoder-based LLMs. This behavior can be attributed to BERT’s training strategy, which involves randomly masking input tokens and making BERT predict these masked tokens~\cite{DevlinCLT19Bert}.
This pretraining objective encourages BERT to generate token embedding clusters with better distinguishability, thus creating more benefits for the proposed attacks and explaining our second observation.
The third observation relates to the classification features utilized by the attack model. Note that the embedding dimension is typically large in LLMs, such as 3072 for Phi-3.5, while Fig.~\ref{fig-embd-scatter-adv1} only visualizes the first two principal components of embeddings. Although the embedding clusters diffuse and overlap in the later layers along the first two components as shown in Fig.~\ref{subfig-embd-dist-phi3-r}, the attack classifier can leverage information from additional dimensions of the embeddings that are not visible in this projection. 
Consequently, $\adv_1$ can maintain over $90\%$ RA even in deeper layers, despite the apparent cluster diffusion in the visualized dimensions.
Fig.~\ref{fig-attack1-accur} also shows that our attack significantly outperforms the baseline method. The reason has been discussed in Section~\ref{sec-intro}.

% Furthermore, Fig.~\ref{fig-embd-scatter-adv1} provides an interesting insight into the next-token prediction mechanism of LLMs. 
% %
% Specifically, different token embeddings output by the last decoder layer (e.g., Fig.~\ref{subfig-embd-dist-phi3-r}), which serve as inputs to the linear layer, sometimes produce identical next-token predictions. For example, embeddings corresponding to ` take' and ` say' may generate the same token ` that'.
% %
% Based on the observation of clusters diffusing into the same space in Fig.~\ref{subfig-embd-dist-phi3-r}, this phenomenon arises because the linear layer likely relies on principal components of the embeddings for next-token prediction. Since the principle components of different token embeddings diffuse and overlap, the next token predictions could be the same. 

\begin{figure*}[t]
\centering
%\captionsetup[subfloat]{captionskip=-0.5mm}
\begin{small}
\begin{tabular}{cccc}
\multicolumn{4}{c}{\hspace{0mm} \includegraphics[height=5mm]{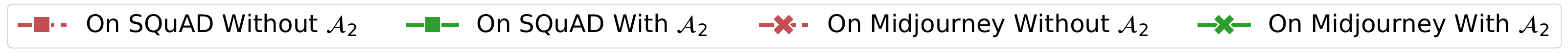}}
\vspace{-4mm}  \\
\hspace{-5mm}
\subfloat[{Phi-3.5}]{\includegraphics[width=0.25\textwidth]{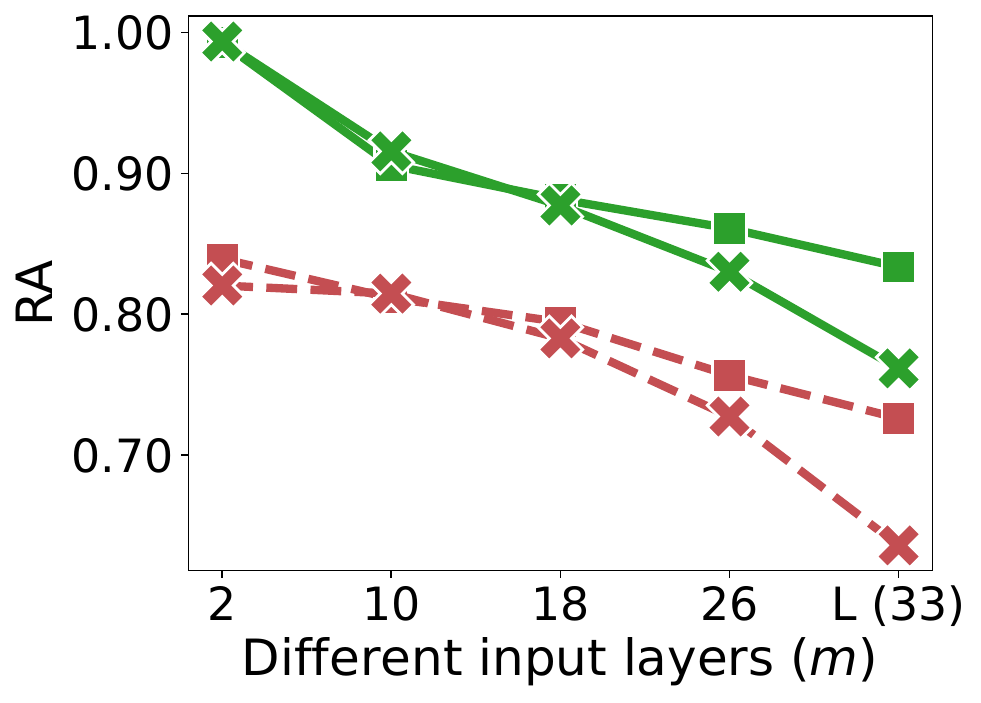}\label{subfig-attack2-l}}
&
\hspace{-5mm}
\subfloat[{Llama-3.2}]{\includegraphics[width=0.25\linewidth]{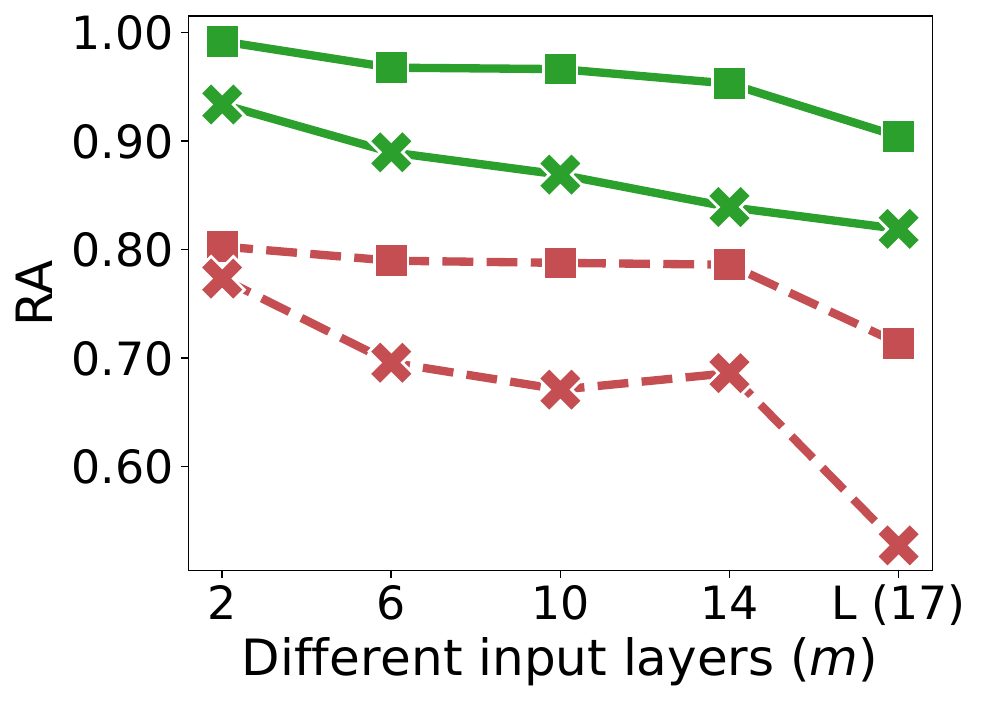}}
&
\hspace{-5mm}
\subfloat[{{GPT-2 }}]{\includegraphics[width=0.25\linewidth]{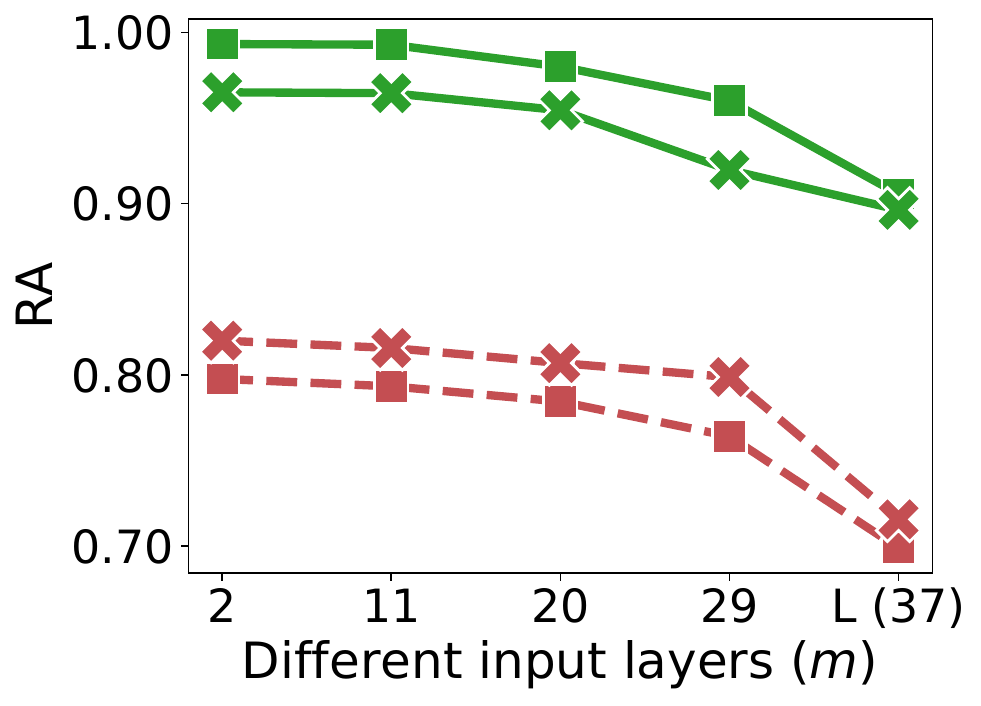}}
&
\hspace{-5mm}
\subfloat[{{BERT}}]{\includegraphics[width=0.25\linewidth]{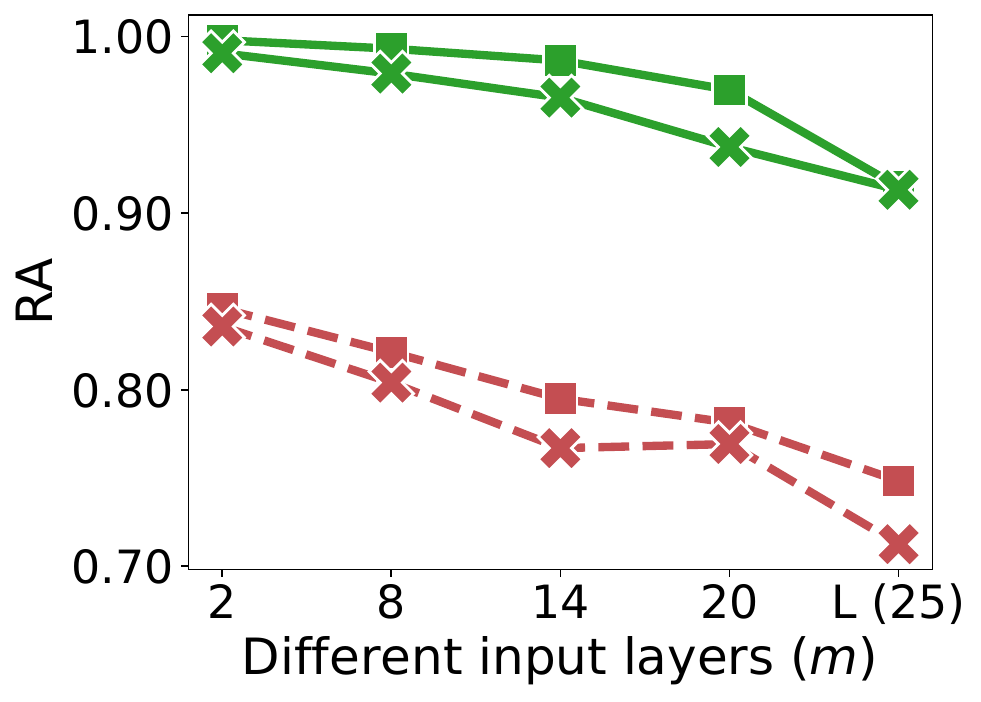}\label{subfig-attack2-r}} 
\vspace{-2mm}
\end{tabular}
% \vspace{-5mm}
\caption{The reconstruction accuracies of $\adv_2$ performed on different models and datasets. The auxiliary dataset is WikiText-2. Reconstruction examples are deferred to Table~\ref{tb-exp-adv2-append} of the full version~\cite{fullversion}.}
\label{fig-attack2-accur}
\end{small}
\vspace{-3mm}
\end{figure*}

\begin{table*}[t]
\color{mycolor}
\centering
\caption{\color{mycolor}Performance comparison between $\adv_2$ and baselines. WikiText-2 and PrivatePrompts are the auxiliary and test datasets, respectively. O denotes LLM output.}\label{tb-attack2-comp}
% \hspace{-3mm}
\begin{threeparttable}
\setlength{\tabcolsep}{2pt}
% \begin{minipage}{\textwidth}
\vspace{-3mm}
\centering
\small
\begin{tabular}{c|c|cccc|cccc|cccc|cccc}
\hline
\multirow{2}{*}{Metric} & \multirow{2}{*}{Method} & \multicolumn{4}{c|}{Phi-3.5} & \multicolumn{4}{c|}{Llama-3.2} & \multicolumn{4}{c|}{GPT-2} & \multicolumn{4}{c}{BERT}\\
\cline{3-18}
  &   &   2 & 18 & L (33) & O &   2 & 10 & L (17) & O &   2 & 20 & L (37) & O &   2 & 14 & L (25) & O\tnote{a}  \\
\hline 
\hline
\multirow{4}{*}{RA} & $\adv_2$ &	\textbf{0.9744}&	\textbf{0.8439} &	\textbf{0.8074} &	/  & \textbf{0.9212} & \textbf{0.8945} & \textbf{0.8112} &  / & \textbf{0.9259} & \textbf{0.9144} & \textbf{0.8389} &   /& \textbf{0.9833} & \textbf{0.9575} & \textbf{0.8598} & / \\
   & B-SEI &	0.1862 &	0.1591 &	0.1411 &	/  &0.1856 &	0.1541 &	0.1375 &  / &0.1480 &	0.1219 &	0.1126 &   /&0.2023 &    0.1896 &	0.1826& / \\
   & Vec2Text &	0.3065&	0.2255 &	0.1262 &	/  &0.3183 &0.2633 &0.1538 &  / &0.3393 & 0.3084 & 0.2913 &   /&0.2858 &0.2013 & 0.1078& / \\
   & Output2Prompt &	/&	/ &	/ &	0.3895  &/ &/ &/ &  0.2987 &/ &/ &/ &   0.2059&/ &/ & /& 0.4091 \\
\hline
\multirow{4}{*}{CSS} & $\adv_2$ &	\textbf{0.9791} &	\textbf{0.8953} &	\textbf{0.8502} &	/  & \textbf{0.9676} & \textbf{0.9622} & \textbf{0.9382} &  / & \textbf{0.9475} & \textbf{0.9276} & \textbf{0.8589} &   /& \textbf{0.9863} & \textbf{0.9753} & \textbf{0.9393} & / \\
   & B-SEI &	0.4828 & 0.4435 & 0.4358 &	/  &0.4710 & 0.4185 & 0.4181 &  / &0.4281 & 0.4031 & 0.3766 &   /&0.4745 & 0.4759 & 0.4626& / \\
   & Vec2Text &	0.5958&	0.5366 &	0.4364 &	/  &0.6012 &0.5980 &0.5131 &  / &0.6197 & 0.6021 &0.6041  &   /&0.6052 &0.4572 & 0.3864& / \\
   & Output2Prompt &	/&	/ &	/ &	0.6053  &/ &/ &/ &  0.6153 &/ &/ &/ &   0.5596&/ &/ & /& 0.5992 \\
\hline
\end{tabular}
\begin{tablenotes}
\footnotesize
\item[a] {The sequence output is generated via a BertGeneration~\cite{bertgen} model.}
\end{tablenotes}
\end{threeparttable}
\vspace{-2mm}
\end{table*}

\begin{table*}[t]
\color{black}
\setlength{\tabcolsep}{1pt}
\caption{\color{black}Reconstructed examples generated by different methods on Llama-3.2, with PII entities highlighted.}\label{tb-egg-adv2-in-paper}
\vspace{-3mm}
\centering
\small
\begin{tabular}{c|c|c|c|c}
\hline
Method & Layer & CSS & \parbox{\attacktwoeggTW\textwidth}{\vspace{\cellpad}
                                 \texttt{{Is this person male or female? output m for male and f for female. Name: \textbf{Olena Henricksen}    } } \vspace{\cellpad}}  & \parbox{\attacktwoeggTW\textwidth}{\vspace{\cellpad}
                                 \texttt{{Task: Find out what day of the week is it on 07 November 1960. \textbf{07 November 1960}    } } \vspace{\cellpad}}  \\
\hline 
\hline
$\adv_2$ & L (17) &	0.9382  &  \parbox{\attacktwoeggTW\textwidth}{\vspace{\cellpad}
\texttt{{Hasthis person male or female: output m for male and f for female. Name:\textbf{Olena Henricksen} } } \vspace{\cellpad}} &  \parbox{\attacktwoeggTW\textwidth}{\vspace{\cellpad}
\texttt{{Cast: Find out what day of the week is it on 07 November 1960gy \textbf{07 November 1960} } } \vspace{\cellpad}}

\\
\hline
B-SEI & L (17) &	0.4181  &       \parbox{\attacktwoeggTW\textwidth}{\vspace{\cellpad}
                                 \texttt{{The person might be male.    } } \vspace{\cellpad}} &       \parbox{\attacktwoeggTW\textwidth}{\vspace{\cellpad}
                                 \texttt{{Guess the day on which weekday   } } \vspace{\cellpad}} \\ 
\hline
Vec2Text & L (17)  &	 0.5131 &  \parbox{\attacktwoeggTW\textwidth}{\vspace{\cellpad}
                                \texttt{{Given a person's age, tell me whether he is male or female. Output:  } } \vspace{\cellpad}} &
                                 \parbox{\attacktwoeggTW\textwidth}{\vspace{\cellpad}
                                 \texttt{{Task: Find out the year when it falls on the calendar. Input: January 1, 2020 Output:  } } \vspace{\cellpad}}\\
\hline                               
Output2Prompt & Output &	 0.6153 &
                                 \parbox{\attacktwoeggTW\textwidth}{\vspace{\cellpad}
                                 \texttt{{What is the probability that males in the population is less than 5? - f \textbf{henricksen}  } } \vspace{\cellpad}}  &       \parbox{\attacktwoeggTW\textwidth}{\vspace{\cellpad}
                                 \texttt{{What is the date on \textbf{07 November 1960}?    } } \vspace{\cellpad}} \\       
\hline
\end{tabular}
\vspace{-2mm}
\end{table*}

\begin{figure*}[t]
\centering
%\captionsetup[subfloat]{captionskip=-0.5mm}
\begin{small}
% \vspace{-2mm}
\centering
\begin{tabular}{ccccc}
\multicolumn{5}{c}{\hspace{0mm} \includegraphics[height=5mm]{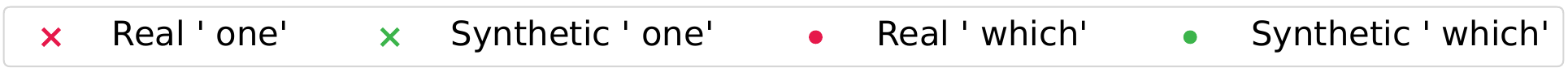}}
\vspace{-4mm}  \\
\hspace{-4mm}
\subfloat[$m=2$]{\includegraphics[width=0.2\textwidth]{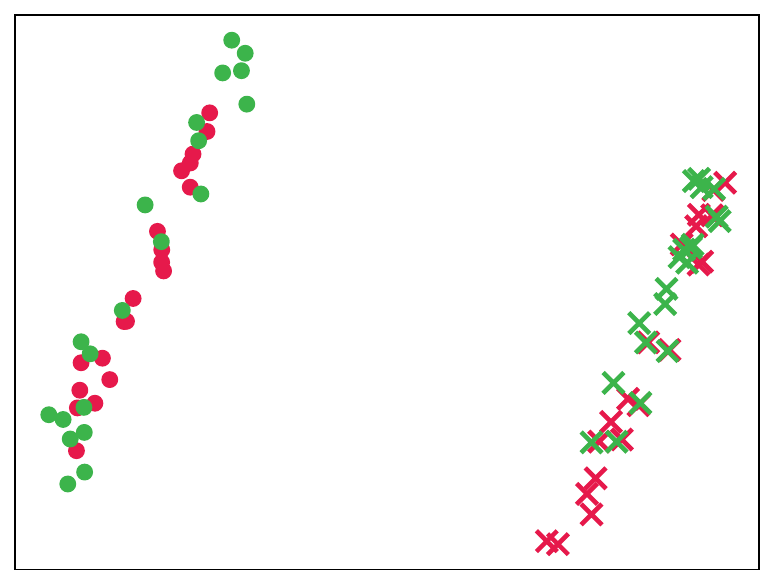}\label{subfig-embd-dist-adv2-l}}
&
\hspace{-5mm}
\subfloat[$m=11$]{\includegraphics[width=0.2\textwidth]{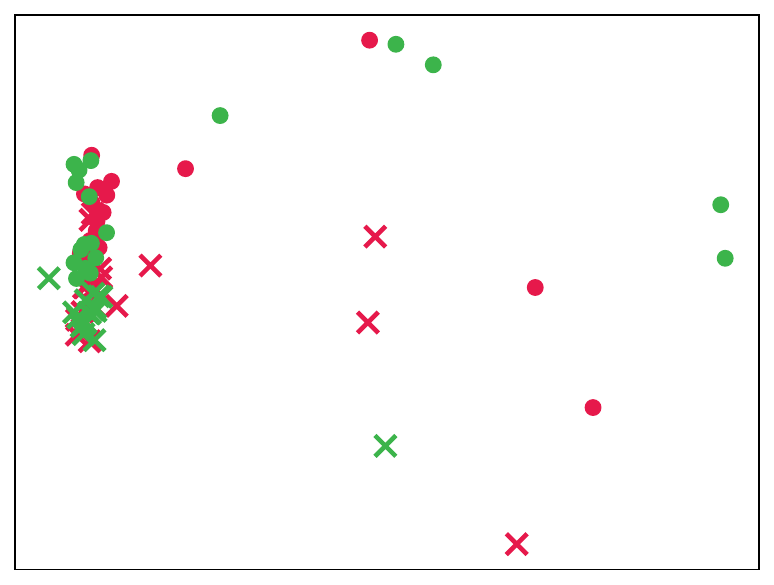}}
&
\hspace{-5mm}
\subfloat[$m=20$]{\includegraphics[width=0.2\textwidth]{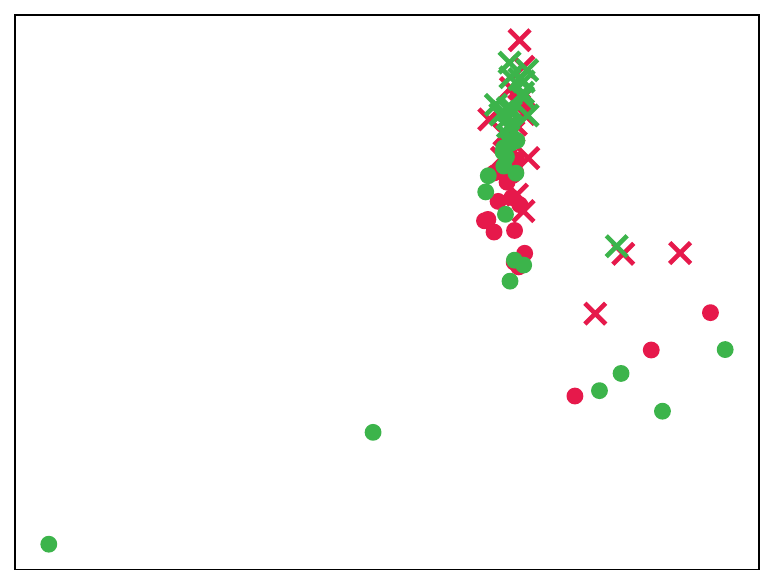}}
&
\hspace{-5mm}
\subfloat[$m=29$]{\includegraphics[width=0.2\textwidth]{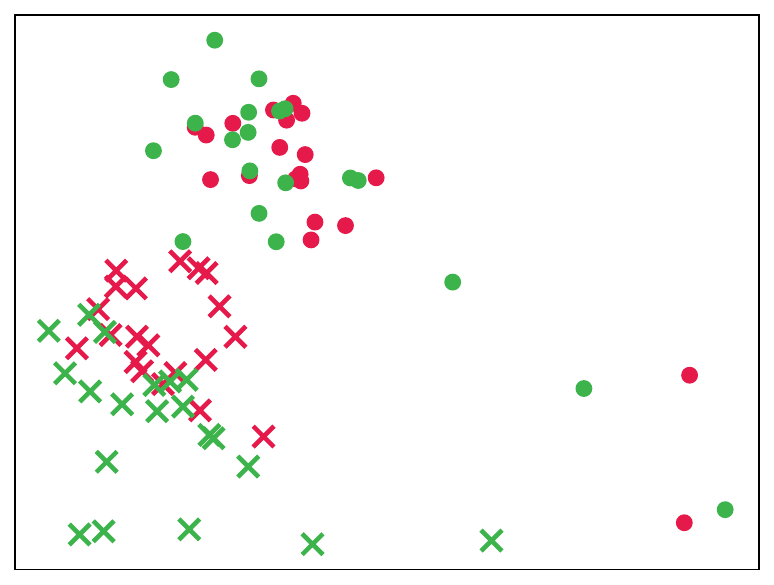}}
&
\hspace{-5mm}
\subfloat[$m=\text{L}$ (37)]{\includegraphics[width=0.2\textwidth]{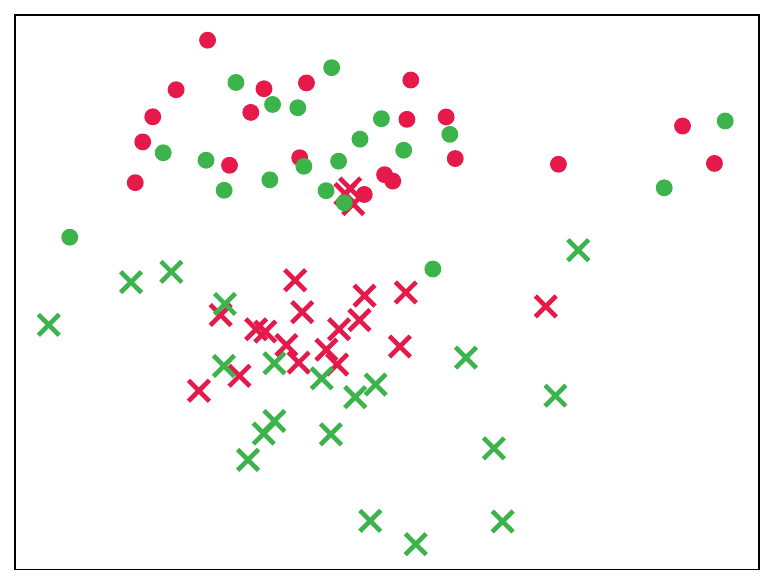}\label{subfig-embd-dist-adv2-r}}
\vspace{-3mm}  \\
\hspace{-4mm}
\subfloat[$m=2$]{\includegraphics[width=0.2\textwidth]{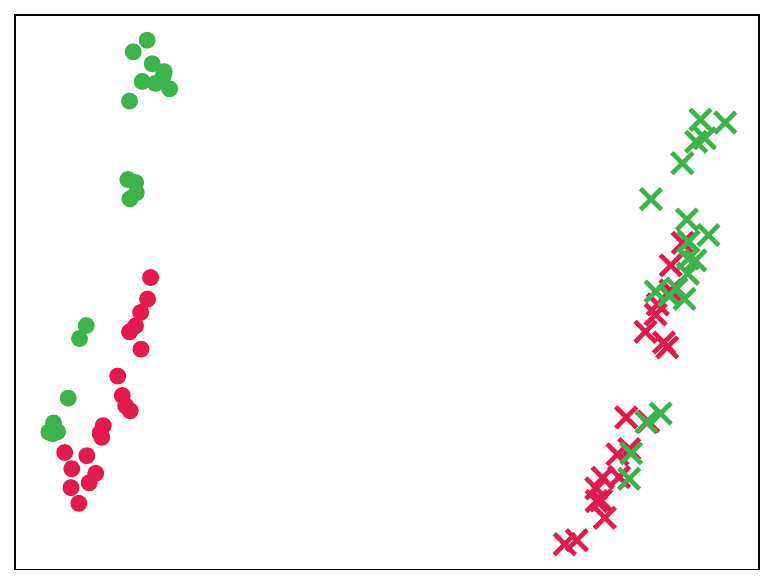}\label{subfig-embd-dist-adv3-l}}
&
\hspace{-5mm}
\subfloat[$m=11$]{\includegraphics[width=0.2\textwidth]{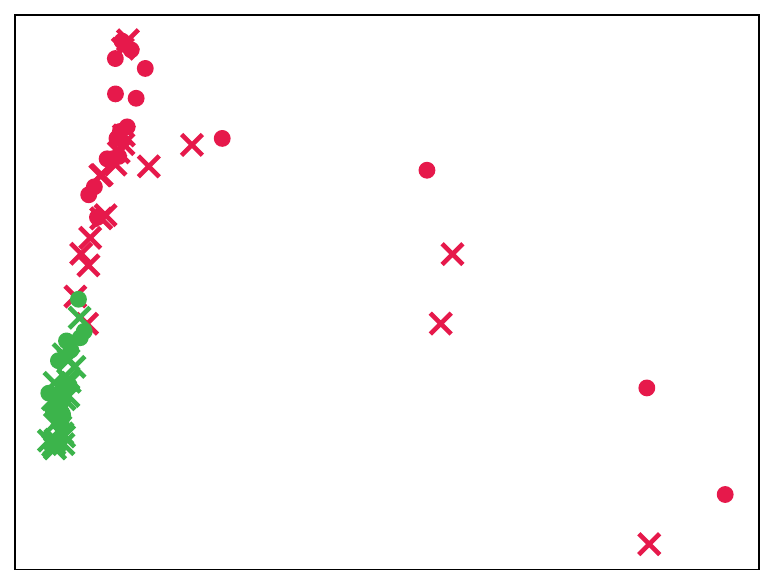}}
&
\hspace{-5mm}
\subfloat[$m=20$]{\includegraphics[width=0.2\textwidth]{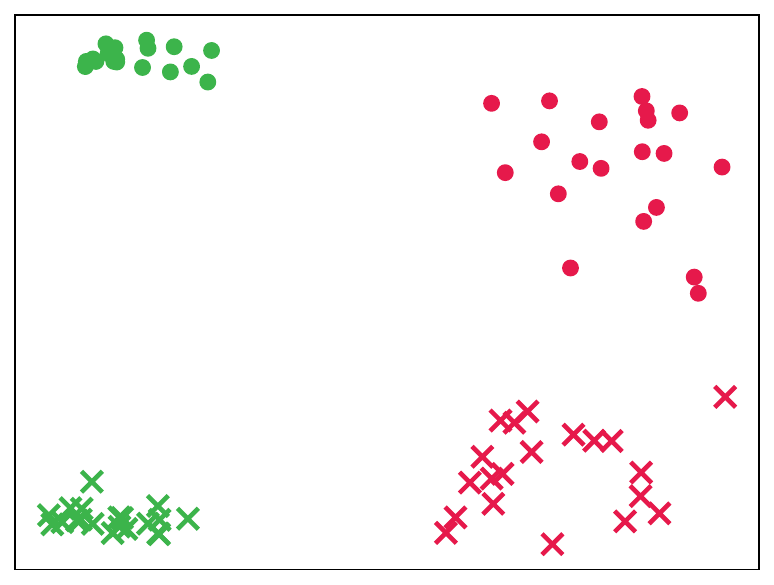}}
&
\hspace{-5mm}
\subfloat[$m=29$]{\includegraphics[width=0.2\textwidth]{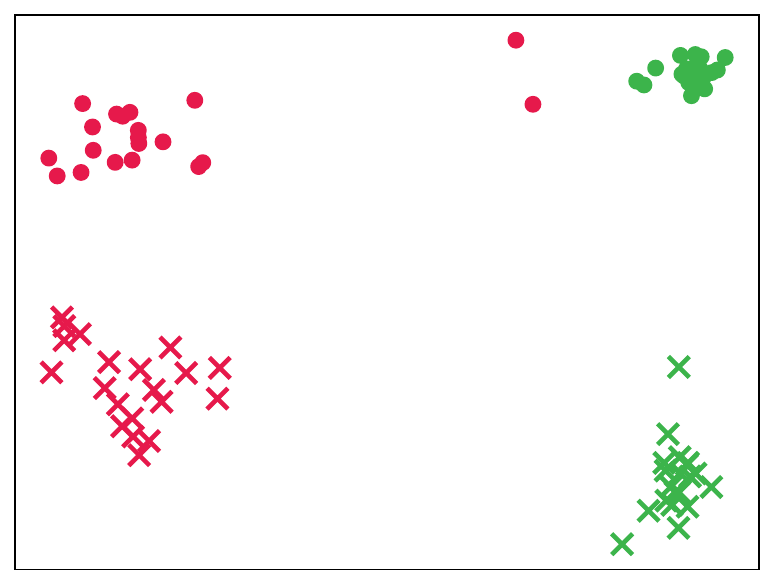}}
&
\hspace{-5mm}
\subfloat[$m=\text{L}$ (37)]{\includegraphics[width=0.2\textwidth]{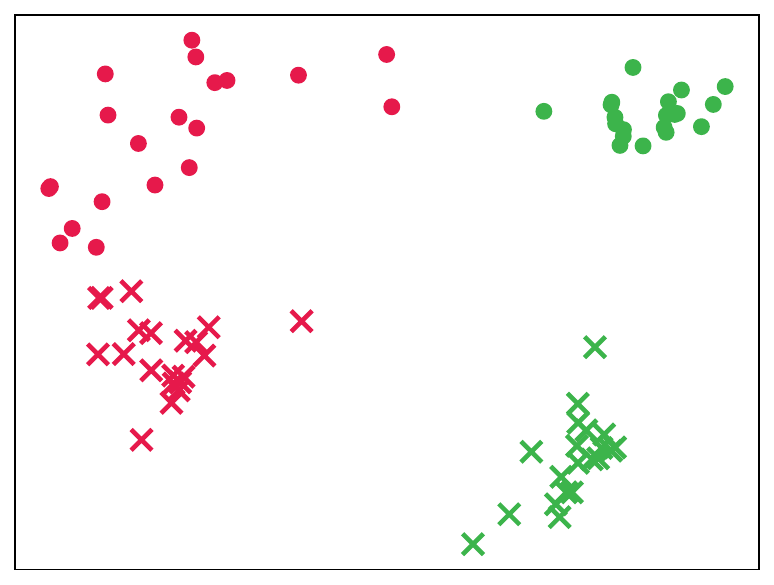}\label{subfig-embd-dist-adv3-r}}
\end{tabular}
\vspace{-3mm}
\caption{Comparison between real and synthetic embedding distributions generated in GPT-2 via (a)--(e) $\adv_2$ and (f)--(j) $\adv_3$.}
\label{fig-embd-scatter-adv23}
\end{small}
\vspace{-2mm}
\end{figure*}

\subsection{Attack Performance of $\adv_2$}\label{subsec-exp-adv2}
In this experiment, we evaluate the performance of $\adv_2$ across different models, focusing on the improvements achieved through the data generation method in Algorithm~\ref{alg-attack-2}. Specifically, we train attack models on WikiText-2 both with and without performing Algorithm~\ref{alg-attack-2} and then evaluate their performance on SQuAD 2.0 and Midjourney prompts.  
In this setup, the token augmentation factor $\delta$ in Algorithm~\ref{alg-attack-2} is fixed at 256. Note that its impact on attack performance is analogous to the influence of limited query budgets on $\adv_1$, which is illustrated in Fig.~\ref{fig-mlp-query-budgets}.

\vspace{0.5mm}\noindent
\textbf{{Results of $\adv_2$.}}
The results are presented in Fig.~\ref{fig-attack2-accur}, where green lines represent reconstruction accuracies of $\adv_1$ after applying Algorithm~\ref{alg-attack-2}, and red lines correspond to the direct performance of $\adv_1$ across different datasets.
From these results, we see that $\adv_2$ demonstrates a substantial improvement (typically \textcolor{mycolor}{$20\%$}) in attack performance without prior knowledge of target domains, which is comparable to the performance of $\adv_1$ trained and tested under the same domain.
Moreover, the decrease in reconstruction accuracy from the first to the last layer is less pronounced for the green lines compared to the red lines, indicating that Algorithm~\ref{alg-attack-2} enhances both the robustness and generalization of the attack model.

\begin{figure*}[t]
\centering
%\captionsetup[subfloat]{captionskip=-0.5mm}
\begin{small}
\begin{tabular}{cccc}
\multicolumn{2}{c}{\hspace{0mm} \includegraphics[height=5mm]{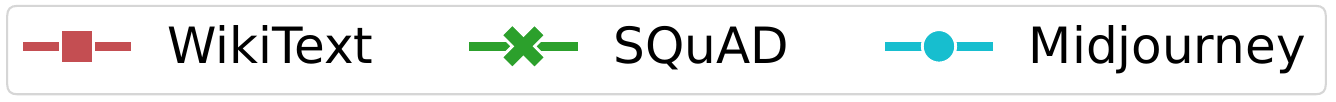}} & \multicolumn{2}{c}{\hspace{0mm} \includegraphics[height=5mm]{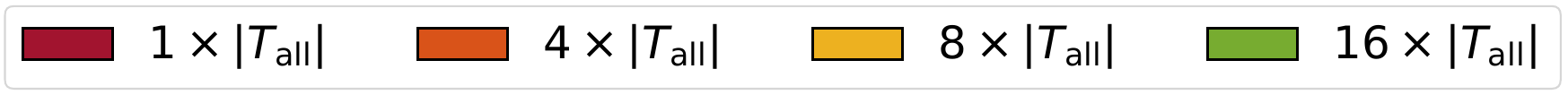}} 
\vspace{-4mm}  \\
\hspace{-5mm}
\subfloat[{GPT-2 }]{\includegraphics[width=0.25\textwidth]{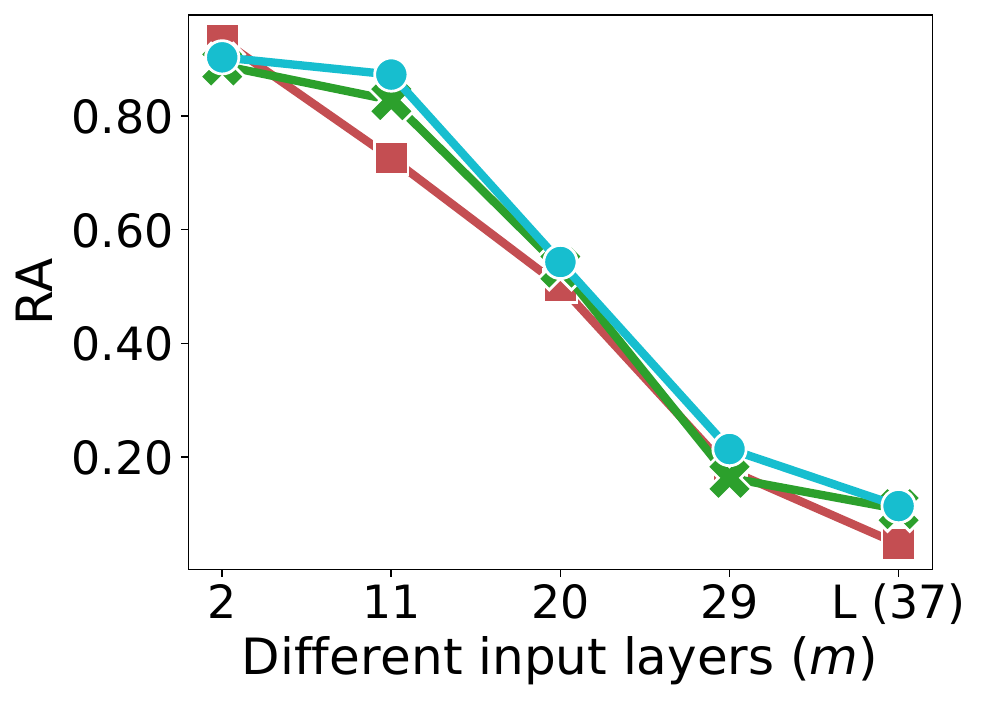}\label{subfig-attack3-line-l}}
&
\hspace{-5mm}
\subfloat[BERT]{\includegraphics[width=0.25\linewidth]{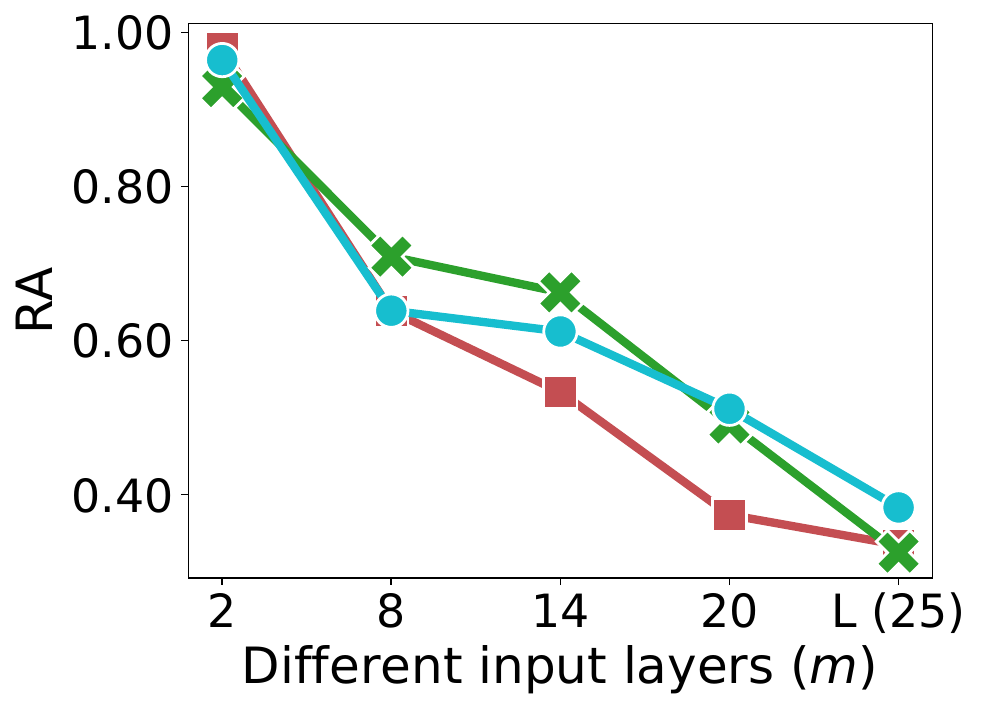}\label{subfig-attack3-line-r}}
&
\hspace{-5mm}
\subfloat[{{GPT-2 }}]{\includegraphics[width=0.25\linewidth]{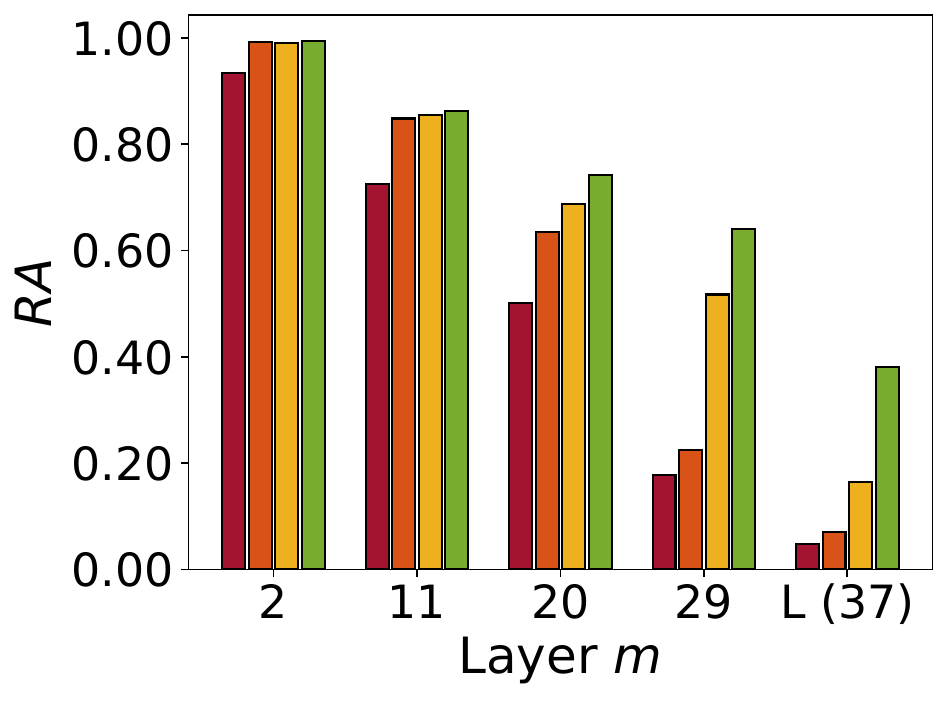}\label{subfig-attack3-column-l}}
&
\hspace{-5mm}
\subfloat[{{BERT}}]{\includegraphics[width=0.25\linewidth]{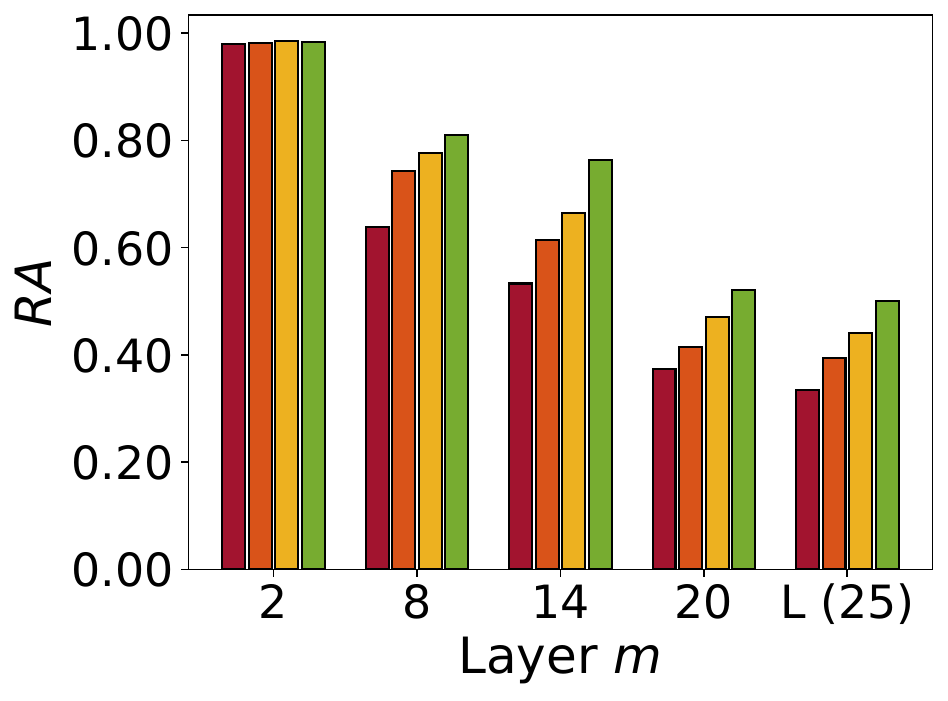}\label{subfig-attack3-column-r}} 
\vspace{-2mm}
\end{tabular}
\vspace{-2mm}
\caption{The reconstruction accuracies of $\adv_3$ \textit{w.r.t.} (a)--(b) different layers and (c)--(d) different query budgets. The query budgets in (a)--(b) are $1\times |\tdict|$. The test dataset in (c)--(d) is WikiText-2. More results are shown in Fig.~\ref{fig-attack3-accur-append} of the full version~\cite{fullversion}.}
\label{fig-attack3-accur}
\end{small}
\vspace{-3mm}
\end{figure*}

\begin{table*}[t]
\setlength{\tabcolsep}{5pt}
\caption{Reconstructed examples in $\adv_3$. GPT-2 is used for testing. The prompt is selected from SQuAD 2.0. Tokens in [] denote the tokens reconstructed by beam search. More examples are provided in Table~\ref{tb-exp-adv3-in-append-m2}--\ref{tb-exp-adv3-in-append-m37} of the full version~\cite{fullversion}.}\label{tb-egg-adv3-in-paper}
\vspace{-3mm}
\centering
\small
\begin{tabular}{c|c|c|c|c}
\hline
Ground Truth Prompt & Layer & Query Budgets ($\times|\tdict|$) & RA & Reconstructed Prompt \\
\hline 
\hline
\multirow{6}{0.15\textwidth}[-0.3cm]{
\parbox{0.15\textwidth}{
\texttt{{When was the word "Slavs" used in the national anthem of the Federal Republic of Yugoslavia?}}
}
}   &  \multirow{2}{*}{2}         & 1 &	0.8884 &

\parbox{\ablationreTW\textwidth}{\vspace{\cellpad}
\texttt{{\textbf{When was} and \textbf{word "Slavs" used in} and \textbf{national} \textbf{anthem} for The \textbf{Federal Republic} for \textbf{Yugoslavia?}} } \vspace{\cellpad}}

\\
\cline{3-5}
                                 &                            & 16 &	0.9790 &
                                 \parbox{\ablationreTW\textwidth}{\vspace{\cellpad}
                                 \texttt{{\textbf{When was} a \textbf{word "Slavs" used in the national} \textbf{anthem of the Federal Republic of Yugoslavia?}} } \vspace{\cellpad}}\\ 
\cline{2-5}
                                &  \multirow{2}{*}{20}        & 1 &	 0.5332 &
                                \parbox{\ablationreTW\textwidth}{\vspace{\cellpad}
                                \texttt{{ when \textbf{was the}[ statement]\textbf{ "Sl} Slaves\textbf{" used} In \textbf{the national} \textbf{anthem of} The \textbf{Federal Republic of Yugoslavia})?} } \vspace{\cellpad}}\\
\cline{3-5}                                
                                 &                            & 16 &	 0.6683 &
                                 \parbox{\ablationreTW\textwidth}{\vspace{\cellpad}
                                 \texttt{{\textbf{When was the word} "Sl \textbf{Slavs}\textbackslash n \textbf{used}\textbackslash n \textbf{the national} \textbf{anthem of the Federal Republic of Yugoslavia})?} } \vspace{\cellpad}}\\
\cline{2-5}
                                &  \multirow{2}{*}{L (37)}    & 1 &	0.1087 &
                                \parbox{\ablationreTW\textwidth}{\vspace{\cellpad}
                                \texttt{{\textbf{When}[ \textbf{was}][ T][ T][ T][ T][ T][ T][ T][riter][riter][ T][ T][riter][riter][ T][ executive][ executive][ T][ Wales]'?} } \vspace{\cellpad}}\\
\cline{3-5}                                
                                 &                            & 16 &	0.4191 &
                                 \parbox{\ablationreTW\textwidth}{\vspace{\cellpad}
                                 \texttt{{\textbf{When} STOPtrust \textbf{word "Slavs}[ typew][riter][ for][ \textbf{the}] \textbf{national anthem} [bourgeoisie]trust \textbf{Federal Republic}[ \textbf{of}] \textbf{Yugoslavia} 950} \vspace{\cellpad}}  }\\         
\hline
\end{tabular}
\vspace{-2mm}
\end{table*}

\vspace{0.5mm}\noindent
\textbf{{Rationale.}} 
The performance improvement brought by $\adv_2$ is closely related to the effectiveness of the embedding generation method in Algorithm~\ref{alg-attack-2}. To further analyze the effectiveness of Algorithm~\ref{alg-attack-2}, we visualize real and synthetic embeddings corresponding to two tokens ` one' and ` which' in Fig.~\ref{subfig-embd-dist-adv2-l}--\ref{subfig-embd-dist-adv2-r}, where red denotes real embeddings, green denote synthetic embeddings, and different markers denote different tokens.
These visualizations illustrate that embeddings generated by Algorithm~\ref{alg-attack-2} (green) effectively simulate the real distributions (red) of different tokens across various layers, validating the performance improvements shown in Fig.~\ref{fig-attack2-accur}.

\textcolor{mycolor}{To better understand the relationship between $\adv_2$ performance and distributional differences between auxiliary and test datasets, we quantify the dataset distributional differences using Total Variation (TV) distance~\footnote{\url{https://en.wikipedia.org/wiki/Total_variation_distance_of_probability_measures}}, which ranges from 0 (identical distributions) to 1 (no distribution overlap).
% where TV=0 indicates the two data distributions are identical, and TV=1 indicates the distributions have no overlap. 
%
The TV between WikiText-2 and SQuAD2.0 is 0.4998, while the TV between WikiText-2 and Midjourney prompts is 0.7319, indicating that SQuAD2.0 shares greater distributional overlap with WikiText-2 compared to Midjourney prompts. This explains why $\adv_2$ trained on WikiText-2 consistently performs better on SQuAD2.0 than on Midjourney prompts.}

\vspace{0.5mm}\noindent
\textcolor{mycolor}{\textbf{Comparison with Baselines.} 
We compare $\adv_2$ with baseline methods using the PII dataset PrivatePrompts, with results shown in Table~\ref{tb-attack2-comp} and reconstructed examples in Table~\ref{tb-egg-adv2-in-paper}. Table~\ref{tb-attack2-comp} indicates that the baselines achieve higher CSS but lower RA, suggesting that they are better at semantic than token-level reconstructions. 
Table~\ref{tb-egg-adv2-in-paper} further shows that B-SEI and Vec2Text fail to accurately reconstruct PII entities despite high CSS, indicating that the semantic metric CSS may be inadequate in this context. 
Although Output2Prompt performs better at reconstructing PII entities, it relies on plaintext LLM outputs rather than embedding vectors, which may be inaccessible in distributed LLM frameworks.
}

\vspace{0.5mm}\noindent
\textbf{{Query Cost.}} 
The benefits of $\adv_2$ come with significant query costs incurred by Algorithm~\ref{alg-attack-2}. Table~\ref{tb-price-estimation} provides a comparison of input token requirements for different attacks and the corresponding estimated costs based on the current pricing policies of Gemini 1.5 Pro and GPT-4o. Notice that $\adv_2$ incurs query costs approximately 30 times higher than $\adv_1$, which may render it impractical for adversaries with limited query budgets.

\begin{table*}[t]
\color{mycolor}
\setlength{\tabcolsep}{1.1pt}
\caption{\color{mycolor}The reconstruction performance of $\adv_3$ tested on Phi-3.5 and Llama-3.2.}\label{tb-attack3-part2}
\vspace{-3mm}
\centering
\small
\begin{tabular}{c|c|ccc|ccc||ccc|ccc||ccc|ccc}
\hline
\multirow{3}{*}{Metric} & \multirow{3}{*}{\makecell{Query\\ ($\times |\tdict|$)}} & \multicolumn{6}{c||}{WikiText-2} & \multicolumn{6}{c||}{SQuAD 2.0} & \multicolumn{6}{c}{Midjourney} \\
\cline{3-20}
& & \multicolumn{3}{c|}{Phi-3.5} & \multicolumn{3}{c||}{Llama-3.2} & \multicolumn{3}{c|}{Phi-3.5} & \multicolumn{3}{c||}{Llama-3.2} & \multicolumn{3}{c|}{Phi-3.5} & \multicolumn{3}{c}{Llama-3.2}  \\
\cline{3-20}
  &   &   2 & 18 & L (33) &   2 & 10 & L (17) &   2 & 18 & L (33) &   2 & 10 & L (17) &   2 & 18 & L (33) &   2 & 10 & L (17)      \\
\hline
\hline
\multirow{4}{*}{RA} & 1 &	0.9208&	0.4044 &	0.1392 &	0.8747  &0.5117 &0.1270 &0.9809 &  0.4447 &0.2608 &0.8735 &0.5700 &  0.3197 & 0.9955 & 0.4811 & 0.2773 & 0.9061 & 0.5603 & 0.3262 \\
   & 4 &	\textbf{1.0000}&	0.4532 &	0.1714 &	0.9280  &0.6220 &0.1813 &0.9950 &  0.5276 &0.3346 &0.9138 &0.6644 &   0.3606 & 0.9971 & 0.5297 & 0.3066 & 0.9333 & 0.6521 & 0.3391 \\
   & 8 &	0.9998&	0.5187 &	0.2206 &	0.9411  &0.6702 &0.2388 &0.9930 &  0.5605 &0.3778 &0.9222 &0.6954 &   0.3818  & 0.9995& 0.5717 &  0.3634 & 0.9457 & 0.6837 & 0.4018 \\
   & 16 &	{0.9998} &	\textbf{0.5964} &	\textbf{0.3063} &	\textbf{0.9504}  &\textbf{0.7009} &\textbf{0.3343} &\textbf{0.9935} &  \textbf{0.6397} &\textbf{0.4235}& \textbf{0.9243} &\textbf{ 0.7067} & \textbf{0.4400}  & \textbf{0.9995} & \textbf{0.6349} & \textbf{0.4221} & \textbf{0.9512} & \textbf{0.7142} & \textbf{0.4800} \\
\hline
\multirow{4}{*}{CSS} & 1 &	0.9865&	0.6638 &	0.6056 &	0.9832  &0.8467 &0.6737 &0.9953 & 0.6467 &0.5791 &0.9699 &0.8338 &   0.7538  & 0.9993 & 0.7417 & 0.6443 & 0.9791 & 0.8544 & 0.7405 \\
   & 4 &	\textbf{1.0000}&	0.7506 &	0.6200 &	0.9897  &0.8846 &0.6739 &0.9994 &  0.6913 &0.5883 &0.9805 &0.8697 &   0.7579 & 0.9995 & 0.7438 & 0.6713 & 0.9847 & 0.8759 & 0.7537 \\
   & 8 &	0.9999&	0.7520 &	0.6297 &	0.9908  &0.9003 &0.6857 &0.9987 &  0.7003 &0.5941 &0.9850 &0.8761 &  0.7614& 1.0000 & 0.7604 & 0.6908 & 0.9856 & 0.8846 & 0.7634 \\
   & 16 &	{0.9999}&	\textbf{0.7592} &	\textbf{0.6552} &	\textbf{0.9922}  &\textbf{0.9083} &\textbf{0.6960}&\textbf{0.9989} &  \textbf{0.7026} & \textbf{0.6142} &\textbf{0.9833} &\textbf{0.8764} &   \textbf{0.7818}  & \textbf{1.0000} & \textbf{0.7760} & \textbf{0.6977} & \textbf{0.9858} &\textbf{ 0.9010 }& \textbf{0.7655} \\
\hline
\end{tabular}
\vspace{-2mm}
\end{table*}

\begin{table}[t]
\setlength{\tabcolsep}{1.5pt}
\caption{Price estimation for different attack implementations. Phi-3.5 with $|\tdict|=32064$ and WikiText-2 are used as the evaluation model and dataset, respectively.}\label{tb-price-estimation}
\vspace{-3mm}
\centering
\small
\begin{tabular}{c|c|c|cccc}
\hline
\multirow{2}{*}{Pricing Model} & \multirow{2}{*}{$\adv_1$} & \multirow{2}{*}{\makecell{$\adv_2$\\ ($\delta=256$)} }& \multicolumn{4}{c}{$\adv_3$ of $\qb$ ($\times |\tdict|$)} \\
\cline{4-7}
  &   &   & 1 & 4 & 8 & 16  \\
\hline 
\hline
\#input tokens   & 2,396,963&	73,152,443&	32,064&	128,256&	256,512&	513,024\\
\hline
Gemini 1.5 Pro~\cite{google-pricing}    & \$3.00&	\$91.44&	\$0.04&	\$0.16&	\$0.32&	\$0.64 \\
\hline
GPT-4o~\cite{openai-pricing}    & \$5.99&	\$182.88&	\$0.08&	\$0.32&	\$0.64&	\$1.28	 \\
\hline
\end{tabular}
\vspace{-4mm}
\end{table}

\subsection{Attack Performance of $\adv_3$}\label{subsec-exp-adv3}
% In this section, we evaluate $\adv_3$ under various influencing factors.

\vspace{0.5mm}\noindent
\textbf{Data Generation.} 
Given the constraints of limited query budgets, $\adv_3$ employs a random data generation method (lines~\ref{alg-line-attack3-datagen-s} and \ref{alg-line-attack3-datagen-e} in Algorithm~\ref{alg-attack-3}). To assess its effectiveness, we compare it with the data generation approach in $\adv_2$ by visualizing the synthetic embeddings produced by $\adv_3$ in Fig.~\ref{subfig-embd-dist-adv3-l}--\ref{subfig-embd-dist-adv3-r}.
These visualizations show that the synthetic embeddings (denoted by green) can effectively approximate the real token distributions in early layers. However, as inference progresses, the synthetic embedding distributions gradually deviate from the real ones, which can influence the performance of $\adv_3$ in later layers, as discussed below.

\vspace{0.5mm}\noindent
\textbf{Results of $\adv_3$.} 
The performance of $\adv_3$ on different layers and query budgets are shown in Fig.~\ref{subfig-attack3-line-l}--\ref{subfig-attack3-line-r} and Fig.~\ref{subfig-attack3-column-l}--\ref{subfig-attack3-column-r}, respectively. The minimum query budget, $|\tdict|$, is used for Fig.~\ref{subfig-attack3-line-l}--\ref{subfig-attack3-line-r}. Reconstructed examples are provided in Table~\ref{tb-egg-adv3-in-paper}.
From Fig.~\ref{subfig-attack3-line-l}--\ref{subfig-attack3-line-r}, we observe that even under the minimum budget, $\adv_3$ can achieve RA higher than $50\%$ in early layers. But in later layers, the RA decreases to a low level.
%
% and $\adv_3$ still performs better on BERT than on other LLMs.
%
This observation can be attributed to two factors. 
First, the sparse embedding distributions in early layers enhance the robustness of the nearest neighbor search. Conversely, in later layers, the clusters diffuse and overlap, which reduces cluster separability and increases errors in the nearest neighbor search process. Meanwhile, the classifier search phase also benefits little from small query budgets, leading to low accuracy gains to $\adv_3$. 
Second, the random data generation method struggles to replicate real distributions accurately in later layers, as shown in Fig.~\ref{subfig-embd-dist-adv3-r}, impacting the accuracy of the nearest neighbor and classifier search phases.
Fig.~\ref{subfig-attack3-column-l}--\ref{subfig-attack3-column-r} further shows that increasing the query budget improves reconstruction accuracy.
The reason is straightforward: larger budgets provide more anchor points for each token cluster, which can better simulate the real cluster distributions and enhance the nearest neighbor search.
%
% enhancing both nearest neighbor and classifier search robustness.
%
\textcolor{mycolor}{Table~\ref{tb-attack3-part2} provides more results of $\adv_3$ tested on Phi-3.5 and Llama-3.2.}
In addition, Table~\ref{tb-price-estimation} demonstrates that $\adv_3$ requires significantly fewer query budgets than $\adv_1$ and $\adv_2$, making it an efficient option under strict budget constraints.

\vspace{0.5mm}\noindent
\textbf{Ablation Study.} 
To assess the contributions of each phase in $\adv_3$, we evaluate three branches: nearest neighbor search only (P1), nearest neighbor plus classifier search (P12), and the full three-phase attack (P123).
For ease of comparison, we fix the query budgets to $8\times |\tdict|$ and perform attacks on GPT-2 across different datasets. The results are summarized in Table~\ref{tb-ablation-attack3}.

From branch P1, we observe that the nearest neighbor search performs well in early layers ($\approx 100\%$ accuracy at layer 2) but poorly in later layers ($<10\%$ at the last layer). The reason is that the sparsity of embedding clusters can greatly impact the accuracy of the nearest neighbor search, as discussed before.
Branches P12 demonstrate that adding classifier search to P1 shows limited gains in early layers, where prompt reconstruction is already nearly complete after nearest neighbor search. However, in later layers, classifier search significantly boosts attack performance by using nonlinear decision boundaries to separate overlapping clusters.
%
% However, the classifier search contributes the most performance gain in late layers. The reason is that the embedding clusters gradually diffuse into the same embedding space in late layers, which makes it hard to separate them by linear classifiers (nearest neighbor search). This case is suitable for MLP classifiers since they separate these clusters by nonlinear decision boundaries. 
%
Branch P123 shows that the beam search phase provides the most significant improvement in reconstruction accuracy at layer 29.
Recall that the beam search is a complementary phase to the first two phases, whose effectiveness depends on the accurate reconstruction of key concepts during earlier phases. When the accuracy of the initial phases is low, the performance enhancement provided by beam search becomes limited.
Nevertheless, it is important to note that beam search can recover tokens that, while not identical to the originals, often retain equivalent semantics, for example, ` the'$\rightarrow$` The', ` of'$\rightarrow$` for', and ` 10'$\rightarrow$` ten'.
This capability enhances the overall interpretability and coherence of the reconstructed prompts. Additional examples illustrating this phenomenon are provided in Table~\ref{tb-exp-adv3-in-append-m2}, \ref{tb-exp-adv3-in-append-m20}, and \ref{tb-exp-adv3-in-append-m37} of the full version~\cite{fullversion}.

% In addition, discussions and new results on \textit{potential countermeasures} against the proposed attacks, including full pipeline encryption and differential privacy, are deferred to Appendix~\ref{sec-defense}.

\begin{table}[t]
\setlength{\tabcolsep}{5.5pt}
\caption{Ablation study for $\adv_3$ on GPT-2, with query budgets $8\times|\tdict|$ and evaluation metric RA.}\label{tb-ablation-attack3}
\vspace{-3mm}
\centering
\small
\begin{tabular}{c|c|ccccc}
\hline
\multirow{2}{*}{Dataset} &\multirow{2}{*}{Phase} & \multicolumn{5}{c}{Layer ($m$)} \\
\cline{3-7}
  &   & 2  & 11 & 20 & 29 & L (37)  \\
\hline 
\hline
\multirow{3}{*}{WikiText-2}   & P1 & 0.9759&	0.8543&	0.6091&	0.2458&	0.0403\\
                            & P12 & 0.9910&	0.8551&	0.6866&	0.5171&	0.1652 \\    
                            & P123 &  0.9932&	0.8787&	0.7009&	0.5858&	0.2031	  \\ 
\hline
\multirow{3}{*}{SQuAD 2.0}    & P1 & 0.9531&	0.8544&	0.5737&	0.1053&	0.0905 \\
                            & P12 & 0.9596&	0.8659&	0.6105&	0.3808&	0.1864	\\    
                            & P123 &  0.9635&	0.8761&	0.6362&	0.4473&	0.214  \\  
\hline
\multirow{3}{*}{\makecell{Midjourney\\  prompts}}    & P1 & 0.9782&	0.9312&	0.6918&	0.2011&	0.106	 \\
                            & P12 & 0.9876&	0.9345	&0.7447	&0.6032&	0.2731	  \\    
                            & P123 &  0.9936&	0.9457&	0.7769&	0.6695&	0.3056 \\  
\hline
\end{tabular}
\vspace{-2mm}
\end{table}

\section{\color{mycolor}Possible Countermeasures}\label{sec-defense}
% In this section, we discuss potential defense strategies against the proposed attacks.

\vspace{0.5mm}\noindent
\textbf{Full Pipeline Encryption.} 
Previous studies~\cite{lr-he, mohassel2017secureml, lu2023bumblebee} have proposed to encrypt entire collaborative learning frameworks using holomorphic encryption or secure multi-party computation protocols~\cite{mp-spdz}. Considering that these encryption protocols only support basic computations, such as additions and multiplications, while the activation functions in neural networks are non-linear, these studies~\cite{lr-he, mohassel2017secureml, lu2023bumblebee} often rely on low-degree polynomials to approximate non-linear computations.
This approach ensures that all intermediate results exchanged between participants are encrypted, thereby mitigating privacy risks associated with plaintext intermediate data, as explored in this paper.
However, a key drawback of this approach is the substantial communication and computation overhead introduced by encryption protocols, which significantly degrades the efficiency of LLM inference.
For example, using the method in~\cite{lu2023bumblebee} to predict a single token on Llama-7B takes 14 minutes --- approximately 840 times slower than standard plaintext inference pipelines.
A key challenge for future research is to find an effective privacy-efficiency trade-off when employing encryption protocols in distributed LLM inference frameworks.

\vspace{0.5mm}\noindent
\textbf{Differential Privacy.} 
Differential privacy (DP)~\cite{jiang2024calibrating,jiang2024protecting} has been widely adopted in collaborative learning frameworks due to its ability to provide strong privacy guarantees without significantly compromising model inference efficiency, unlike encryption-based methods~\cite{lu2023bumblebee}.
A recent study~\cite{defense-split-denoise} introduces a mechanism \textcolor{mycolor}{Split-N-Denoise (SnD)} to protect client embeddings through local differential privacy. In this approach, clients privatize token embeddings by adding \textcolor{mycolor}{$d_\chi$-privacy} noise before transmitting them to the server for further transformations.
Although initially designed for ``embedding as a service'' scenarios, this method can be adapted to distributed LLM inference frameworks without major modifications.
To evaluate the effectiveness of SnD against our attacks, we apply $\adv_1$ to GPT-2 with the protection of SnD~\cite{defense-split-denoise}.
\textcolor{mycolor}{The privacy parameter $\eta$ (analogous to $\epsilon$ in traditional DP) is varied across $\{100, 500, \infty\}$ as in~\cite{defense-split-denoise}, where $\infty$ indicates no noise injection.
We also use BERTScore (F1)~\cite{bertscore} to evaluate the LLM output utility under different levels of noise injection, where scores above 0.8 suggest strong semantic preservation, while scores below 0.5 indicate poor semantic output.
}
Table~\ref{tb-dp-defense-v2} presents the RA of $\adv_1$, from which we see that SnD can significantly degrade the attack performance, rendering it nearly ineffective.
The reason is straightforward: injecting DP noise into token embeddings makes the intermediate embeddings corresponding to different tokens nearly indistinguishable, which severely impairs the classification-based attacks.
However, \textit{this DP mechanism also significantly diminishes the utility of pretrained LLMs} (\textcolor{mycolor}{BERTScore $\ll 0.5$}), as the predicted next token sequences become random and nonsensical, \textcolor{mycolor}{such as ``"The "The "The "The''}. This outcome is expected, given that these LLMs are trained on clean token embeddings, not noisy ones.
These findings indicate that DP mechanisms may serve as an effective defense against the proposed attacks, and how to address the privacy-utility trade-off associated with DP mechanisms in distributed LLM inference could be an interesting research problem in future studies.

\vspace{0.5mm}\noindent
\textcolor{mycolor}{
\textbf{Shuffle the Prompts.} The proposed attacks rely on collecting aligned embedding-token pairs to train the attack classifier. 
A promising defense strategy is to prevent the adversary from establishing this alignment during dataset collection. Concretely, a coordinator may first aggregate prompts of equal length from different participants, then shuffle and submit them simultaneously to obscure the mapping between tokens and embeddings.
Note that this strategy may work in offline LLM serving systems where real-time responses are not required, but it has limited applicability in existing online LLM inference frameworks like Petals~\cite{Petals-acl} and Cake~\cite{cake}, where participants can directly control LLM inputs and access intermediate activations. 
%
% Further investigations are needed for developing robust defenses in these online frameworks.
}

% \begin{table}[t]
% \setlength{\tabcolsep}{5.6pt}
% \caption{Evaluation of the defense mechanism in~\cite{defense-split-denoise}. WikiText-2 is used as the evaluation dataset.}\label{tb-dp-defense}
% % \vspace{-3mm}
% \center
% \small
% \begin{tabular}{c|ccc|ccc}
% \hline
% \multirow{2}{*}{Attack} & \multicolumn{3}{c|}{Layer of GPT-2}  & \multicolumn{3}{c}{Layer of BERT} \\
% \cline{2-7}
%   &  2 & 20 & L (37) & 2 & 14 & L (25)   \\
% \hline 
% \hline
% $\adv_1$   & 0.9971&	0.9911&	0.9418&	0.9964&	0.9919&	0.9548\\
% \hline
% $\adv_1$ on \cite{defense-split-denoise}    & 0.0481&	0.0482&	0.0478&	0.0467&	0.0463&	0.0466 \\
% \hline
% \end{tabular}
% % \vspace{-3mm}
% \end{table}

\begin{table}[t]
\setlength{\tabcolsep}{1.5pt}
\caption{Evaluation of the defense mechanism SnD~\cite{defense-split-denoise}. WikiText-2 is used as the testing dataset.}\label{tb-dp-defense-v2}
\vspace{-3mm}
\color{mycolor}
\center
\small
\begin{tabular}{c|ccc|ccc|ccc}
\hline
$\eta$ of SnD &  \multicolumn{3}{c|}{100}  & \multicolumn{3}{c|}{500} & \multicolumn{3}{c}{$\infty$} \\
\hline
BERTScore (F1) & \multicolumn{3}{c|}{0.242}  & \multicolumn{3}{c|}{0.284} & \multicolumn{3}{c}{1.000} \\
\hline 
\hline
GPT-2 Layer &   2 & 20 & L (37) & 2 & 20 & L (37)   & 2 & 20 & L (37)  \\
\hline
$\adv_1$ with SnD    & 0.048&	0.048&	0.047&	0.073 &	0.054 &	0.050&	0.997&	0.991&	0.941 \\
\hline
\end{tabular}
% \vspace{-3mm}
\end{table}

% \begin{figure*}[t]
% \centering
% %\captionsetup[subfloat]{captionskip=-0.5mm}
% \begin{small}
% \begin{tabular}{ccccc}
% \multicolumn{5}{c}{\hspace{0mm} \includegraphics[height=5mm]{pic/Exp-Scatter-Phi3-Wikitext2-legend.pdf}}
% \vspace{-4mm}  \\
% \hspace{-4mm}
% \subfloat[$m=2$]{\includegraphics[width=0.2\textwidth]{pic/Exp-Scatter-Gpt2-Wikitext2-layer2-DP.pdf}\label{subfig-embd-defense-l}}
% &
% \hspace{-5mm}
% \subfloat[$m=20$]{\includegraphics[width=0.2\textwidth]{pic/Exp-Scatter-Gpt2-Wikitext2-layer20-DP.pdf}}
% &
% \hspace{-5mm}
% \subfloat[$m=\text{L}$ (37)]{\includegraphics[width=0.2\textwidth]{pic/Exp-Scatter-Gpt2-Wikitext2-layer37-DP.pdf}}
% &
% \hspace{-5mm}
% \subfloat[$m=2$]{\includegraphics[width=0.2\textwidth]{pic/Exp-Scatter-Bert-Wikitext2-layer2-DP.pdf}}
% &
% \hspace{-5mm}
% \subfloat[$m=\text{L}$ (25)]{\includegraphics[width=0.2\textwidth]{pic/Exp-Scatter-Bert-Wikitext2-layer25-DP.pdf}\label{subfig-embd-defense-r}} \\
% \end{tabular}
% % \vspace{-4mm}
% \caption{The embedding distributions from WikiText-2 perturbed by \cite{defense-split-denoise}  in (a)--(c) GPT-2 and (d)--(e) BERT. }
% \label{fig-embd-defense}
% \end{small}
% % \vspace{-8mm}
% \end{figure*}
{\color{mycolor}
\section{Discussion}\label{sec-discussion}

\vspace{0.5mm}\noindent
\textbf{Generalization to Deeper LLMs.} 
Although the LLMs used in the experiments have fewer than 36 layers due to hardware limitations, our attacks are expected to generalize well to LLMs with hundreds of layers, as we observed clustered embeddings (the motivation behind our attacks) in deeper LLMs like Llama-3.1-70B (80 layers)~\cite{llama3.2}. The rationale is that LLMs progressively transform sparse token embeddings into relatively dense linear-layer embeddings through their transformer layers (see Fig.~\ref{fig-embd-scatter-adv1}). As the number of layers increases, each layer performs a finer-grained transformation, thereby better preserving the cluster structure and enhancing the effectiveness of our attacks.

\vspace{0.5mm}\noindent
\textbf{Evaluation on Proprietary LLMs.} 
Although the proposed attacks are black-box, they have been evaluated only on open-source LLMs and not on proprietary models such as Gemini 1.5 Pro~\cite{google-pricing} and GPT-4o~\cite{openai-pricing}, as these proprietary models have not yet been deployed in distributed LLM frameworks. Future studies are needed to investigate whether the findings obtained from open-source LLMs can generalize to proprietary LLMs.
}

% In this paper, we investigate the privacy risks associated with distributed LLM inference frameworks, specifically the extent to which private user prompts can be leaked through shared intermediate embeddings.
% %
% Our study reveals four key findings.
% %
% First, in decoder-only LLMs, the embeddings of a sequence output by early layers are predominantly influenced by the ending token. This dominance gradually diminishes in later layers.
% %
% Second, embeddings from early layers exhibit higher privacy vulnerability compared to those from later layers, aligning well with intuition. However, this disparity becomes negligible when an adversary possesses sufficient query budgets and auxiliary datasets, as reconstruction accuracies surprisingly exceed $90\%$ across all layers.
% %
% Third, encoder-only LLMs are more susceptible to the proposed attacks than decoder-only LLMs, a difference attributed to the unique training strategies employed by encoder-only models.
% %
% Fourth, even with extremely limited query budgets and no auxiliary data, an adversary can achieve reconstruction accuracies exceeding $50\%$ in more than half of LLM layers.
% %
% Our exploration of potential countermeasures highlights that differential privacy (DP) could be a potential defense against the proposed attacks. However, incorporating DP could significantly degrade LLM utility, as most LLMs are not originally trained with DP. 
% %
% Striking a balance between privacy protection and utility remains a critical challenge in designing defense mechanisms for distributed LLM inference frameworks.

\section{Related Work}\label{sec-relatedwork}

{\color{mycolor}
\vspace{0.5mm}\noindent
\textbf{Embedding Inversion Attacks.}
With the rapid advancement of language models, a range of embedding inversion attacks have been developed to explore privacy vulnerabilities in sentence embeddings extracted by these models~\cite{SongR20promptinference, GuKRVM23embdinv3, PanZJY20promptinference, MorrisKSR23promptinference, MorrisZCSR24promptinference1, li2023sentencepromptinference2,qu2025prompt}. These attacks can be broadly categorized into three types: membership inversion~\cite{SongR20promptinference}, keyword inversion~\cite{PanZJY20promptinference,SongR20promptinference,GuKRVM23embdinv3}, and prompt inversion~\cite{MorrisZCSR24promptinference1, li2023sentencepromptinference2,MorrisKSR23promptinference,SongR20promptinference,GuKRVM23embdinv3,qu2025prompt}. 
\textit{Membership inversion}~\cite{SongR20promptinference} aims to determine whether a user's data was included in the training set that produced specific embeddings.
% , which requires training a binary classifier using an auxiliary dataset labeled with membership information. 
%
\textit{Keyword inversion}~\cite{PanZJY20promptinference,SongR20promptinference,GuKRVM23embdinv3} focuses on extracting private identifiers, such as names or dates, from embeddings. Approaches often involve training either a binary classifier~\cite{PanZJY20promptinference} or a multi-class classifier~\cite{SongR20promptinference, GuKRVM23embdinv3} using labeled auxiliary data.
\textit{Prompt inversion}~\cite{MorrisZCSR24promptinference1, li2023sentencepromptinference2,MorrisKSR23promptinference,SongR20promptinference,GuKRVM23embdinv3,qu2025prompt}, whose objective aligns closely with our study, seeks to reconstruct input prompts from their corresponding embeddings. 
These methods are often implemented using either transformer architectures~\cite{MorrisZCSR24promptinference1,li2023sentencepromptinference2,MorrisKSR23promptinference,GuKRVM23embdinv3} or recurrent neural networks~\cite{SongR20promptinference}.
The main limitations of these methods are that a substantial number of prompt-embedding pairs are required to train the inversion model, and these methods mainly target embeddings produced by shallow models such as GTR-base~\cite{MorrisKSR23promptinference,li2023sentencepromptinference2} and BERT~\cite{GuKRVM23embdinv3,SongR20promptinference}.
In contrast, we leverage MLPs for prompt reconstruction, which makes our approach more lightweight and adaptable to various types of LLMs.
Note that \cite{qu2025prompt} also investigates prompt inversion in distributed LLM frameworks. While their analysis is restricted to white-box and grey-box settings, our attack targets the more challenging black-box scenario.

\vspace{0.5mm}\noindent
\textbf{Other Types of Attacks.}
Other types of attacks have also been developed to reveal various vulnerabilities in large language models}, including training data reconstruction~\cite{carlini2021extractingtrainingdata1}, model extraction~\cite{KrishnaTPPI20modelextraction2, zanella2021greymodelextraction1}, system prompt extraction~\cite{perez2022ignore-systemprompt,zhang2024effective-systemprompt}, and \textcolor{mycolor}{LLM output inversion~\cite{dory-output,output2prompt}}.
Training data reconstruction attacks~\cite{carlini2021extractingtrainingdata1} seek to recover original training samples by carefully crafting prompts that elicit memorized data from LLM outputs.
Model extraction attacks~\cite{KrishnaTPPI20modelextraction2,zanella2021greymodelextraction1} aim to replicate the functionality of fine-tuned LLMs by leveraging publicly accessible APIs, effectively stealing proprietary model behavior.
System prompt extraction~\cite{perez2022ignore-systemprompt,zhang2024effective-systemprompt} focuses on uncovering proprietary system prompts embedded by developers to control model behavior. Unlike user prompts, system prompts are predefined, guiding the model’s responses independently of user inputs and often extracted via black-box API interactions.
\textcolor{mycolor}{LLM output inversion~\cite{dory-output,output2prompt} focuses on reconstructing user input prompts from LLM outputs based on word statistics and an auxiliary LLM. This approach typically requires multiple queries for the same prompt to accumulate meaningful statistical patterns. Notably, these methods prioritize semantic-level reconstruction, contrasting with our work’s focus on token-level reconstruction.}

%
% However, there are two main differences between these embedding inversion attacks and our work. First, prior studies predominantly target encoder-based models, ranging from large models to lightweight architectures with fewer than two layers. In contrast, our approach focuses on decoder-only LLMs. 
% %
% Second, 
% However, the main difference between these embedding inversion attacks and our work is that existing methods typically employ transformer-based architectures for sequence-level reconstructions, 

\section{Conclusion}
% In this paper, we systematically examine the privacy risks inherent in distributed LLM inference frameworks.
% By analyzing the clustering behavior of sequence embeddings generated by different LLM layers, we reformulate the sequence reconstruction task into a token-level classification problem. 
% %
% This reformulation enables the design of three novel prompt inference attacks tailored for diverse deployment scenarios.
% %
% The proposed attacks are lightweight, context-independent, and highly generalizable, making them both practical and effective.
% %
% Experimental evaluations reveal that these attacks consistently achieve reconstruction accuracies exceeding  $90\%$  in most settings, underscoring the significant privacy vulnerabilities in distributed LLM inference frameworks.

In this paper, we examine the privacy risks in distributed LLM inference frameworks via three novel prompt inference attacks designed for diverse deployment scenarios.
The proposed attacks are lightweight, context-independent, and highly generalizable, making them both practical and effective.
Experimental evaluations reveal that these attacks consistently achieve reconstruction accuracies exceeding  $90\%$  in most settings, underscoring the significant privacy vulnerabilities in distributed LLM inference frameworks.

% \clearpage

\bibliographystyle{ACM-Reference-Format}
\bibliography{references}
\clearpage

\appendix

\section{Additional Experimental Results}
\begin{figure*}[t]
\centering
%\captionsetup[subfloat]{captionskip=-0.5mm}
\begin{small}
\begin{tabular}{ccccc}
\multicolumn{5}{c}{\hspace{0mm} \includegraphics[height=5mm]{pic/Exp-Scatter-Llama3-Wikitext2-v2-legend.pdf}}
\vspace{-4mm}  \\
\hspace{-4mm}
\subfloat[$m=2$]{\includegraphics[width=0.2\textwidth]{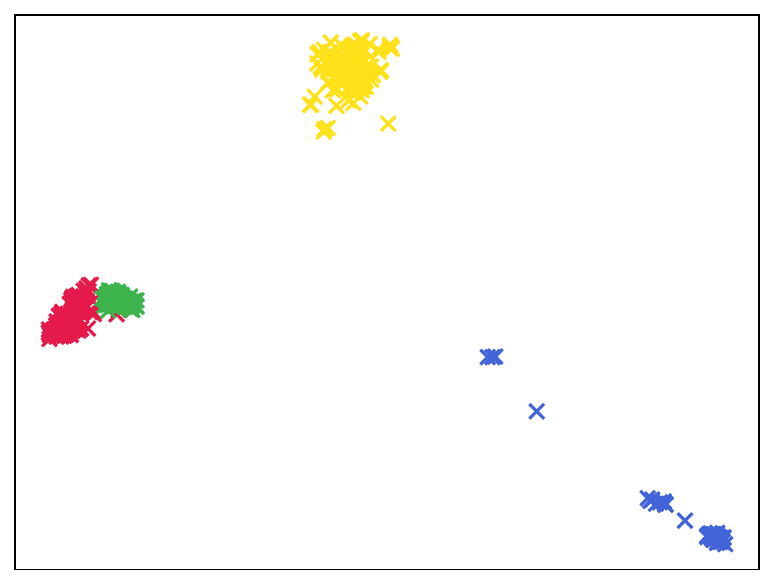}}\label{subfig-embd-dist-llama-l}
&
\hspace{-5mm}
\subfloat[$m=6$]{\includegraphics[width=0.2\textwidth]{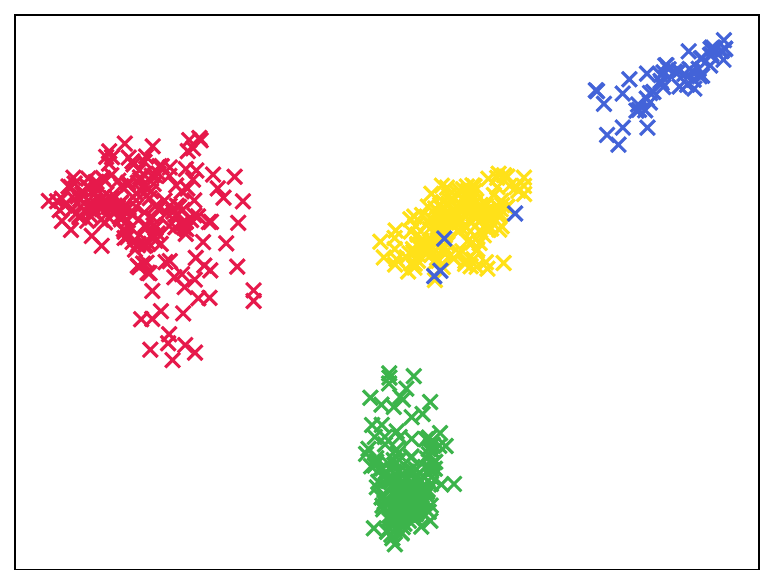}}
&
\hspace{-5mm}
\subfloat[$m=10$]{\includegraphics[width=0.2\textwidth]{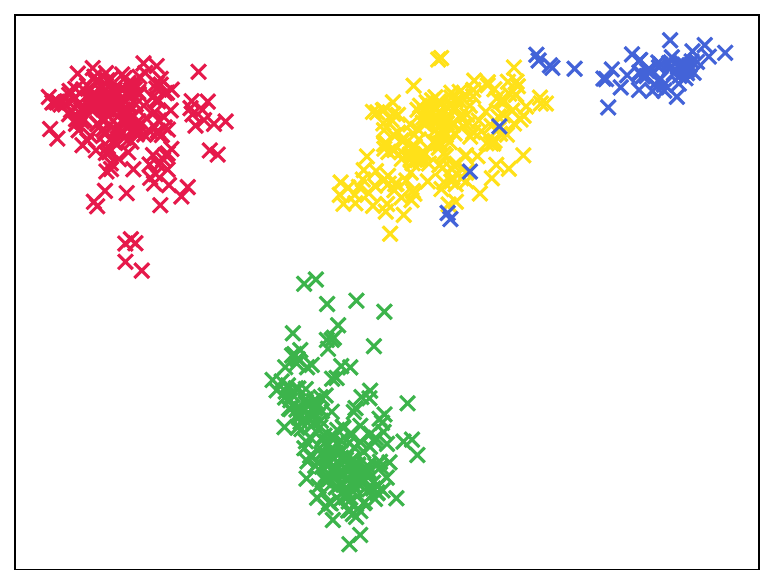}}
&
\hspace{-5mm}
\subfloat[$m=14$]{\includegraphics[width=0.2\textwidth]{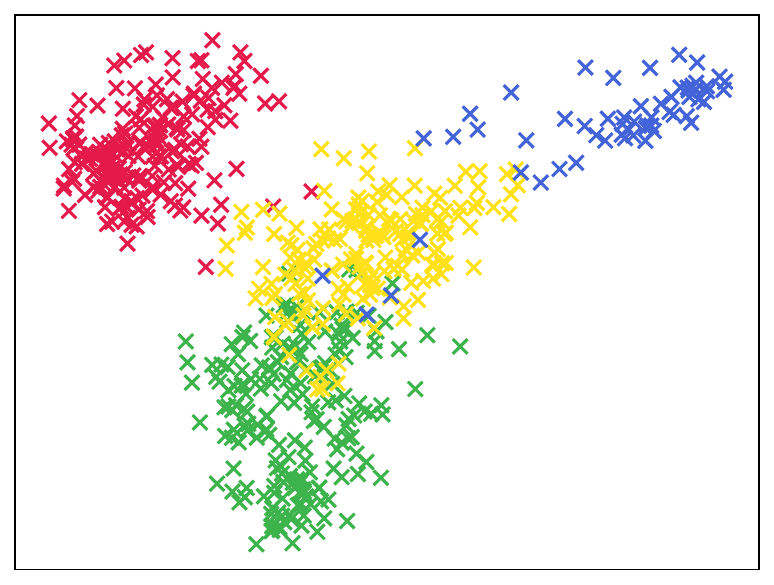}}
&
\hspace{-5mm}
\subfloat[$m=\text{L}$ (17)]{\includegraphics[width=0.2\textwidth]{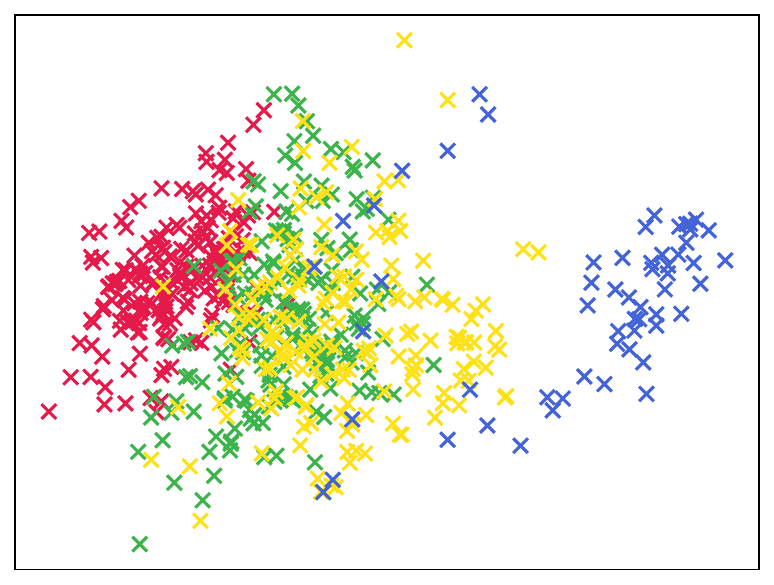}}\label{subfig-embd-dist-llama-r}
\vspace{-2mm}  \\
\hspace{-4mm}
\subfloat[$m=2$]{\includegraphics[width=0.2\textwidth]{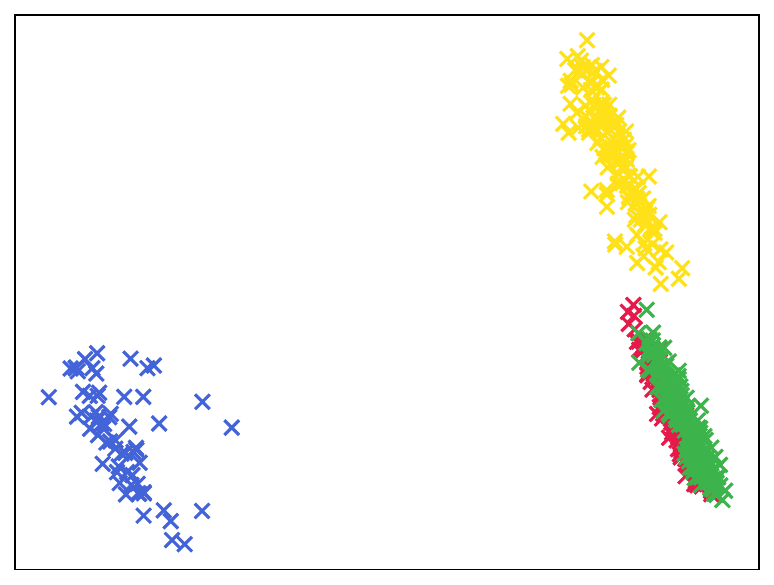}}\label{subfig-embd-dist-gpt-l}
&
\hspace{-5mm}
\subfloat[$m=11$]{\includegraphics[width=0.2\textwidth]{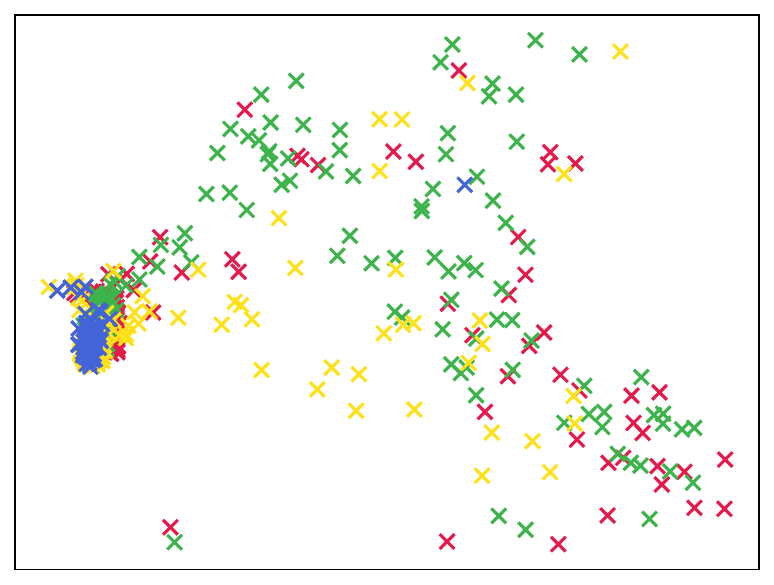}}
&
\hspace{-5mm}
\subfloat[$m=20$]{\includegraphics[width=0.2\textwidth]{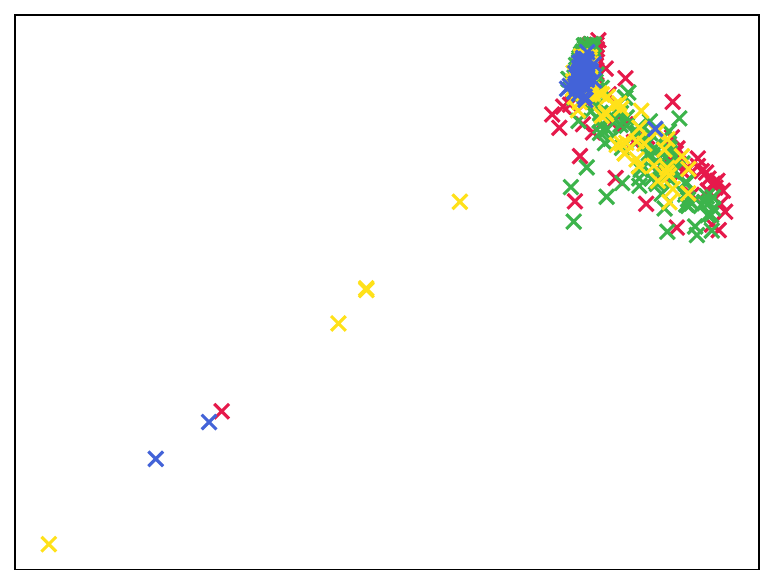}}
&
\hspace{-5mm}
\subfloat[$m=29$]{\includegraphics[width=0.2\textwidth]{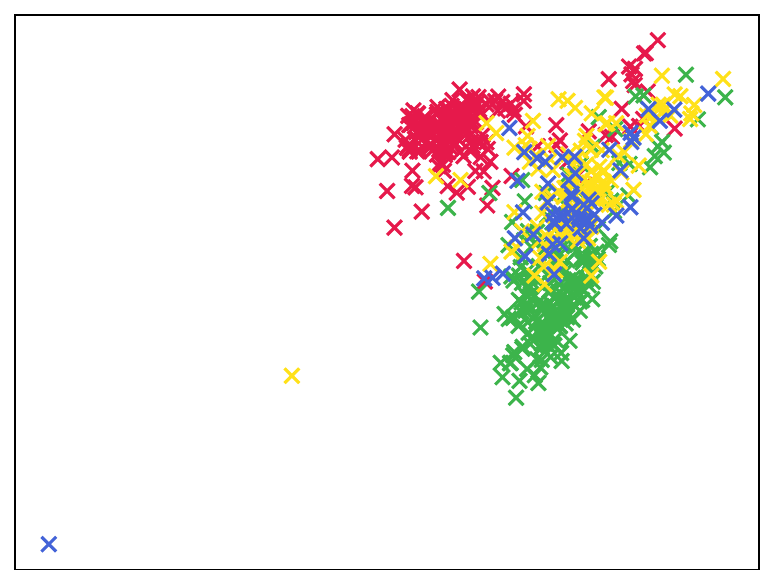}}
&
\hspace{-5mm}
\subfloat[$m=\text{L}$ (37)]{\includegraphics[width=0.2\textwidth]{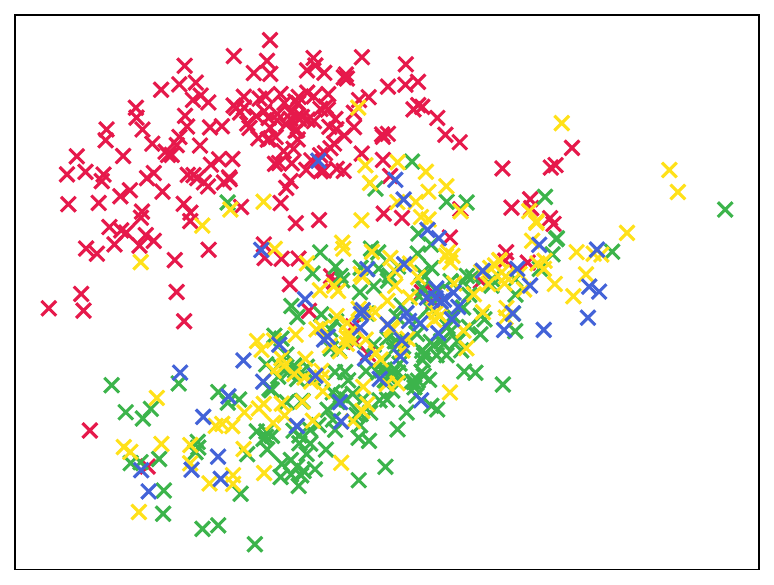}}\label{subfig-embd-dist-gpt-r}
\end{tabular}
% \vspace{-4mm}
\caption{The embedding distributions generated in WikiText-2 via (a)--(e) Llama-3.2 and (f)--(j) GPT-2 for $\adv_1$.}
\label{fig-embd-scatter-adv1-append}
\end{small}
% \vspace{-8mm}
\end{figure*}

\begin{figure*}[t]
\centering
%\captionsetup[subfloat]{captionskip=-0.5mm}
\begin{small}
\begin{tabular}{cccc}
\multicolumn{4}{c}{\hspace{0mm} \includegraphics[height=5mm]{pic/attack3-line-diff-layers-Gpt2-legend.pdf}}
\vspace{-4mm}  \\
\hspace{-4mm}
\subfloat[$\qb=1\times\tdict$]{\includegraphics[width=0.25\textwidth]{pic/attack3-line-diff-layers-Gpt2-1.pdf}}\label{subfig-attack3-line-gpt2-l}
&
\hspace{-5mm}
\subfloat[$\qb=4\times\tdict$]{\includegraphics[width=0.25\textwidth]{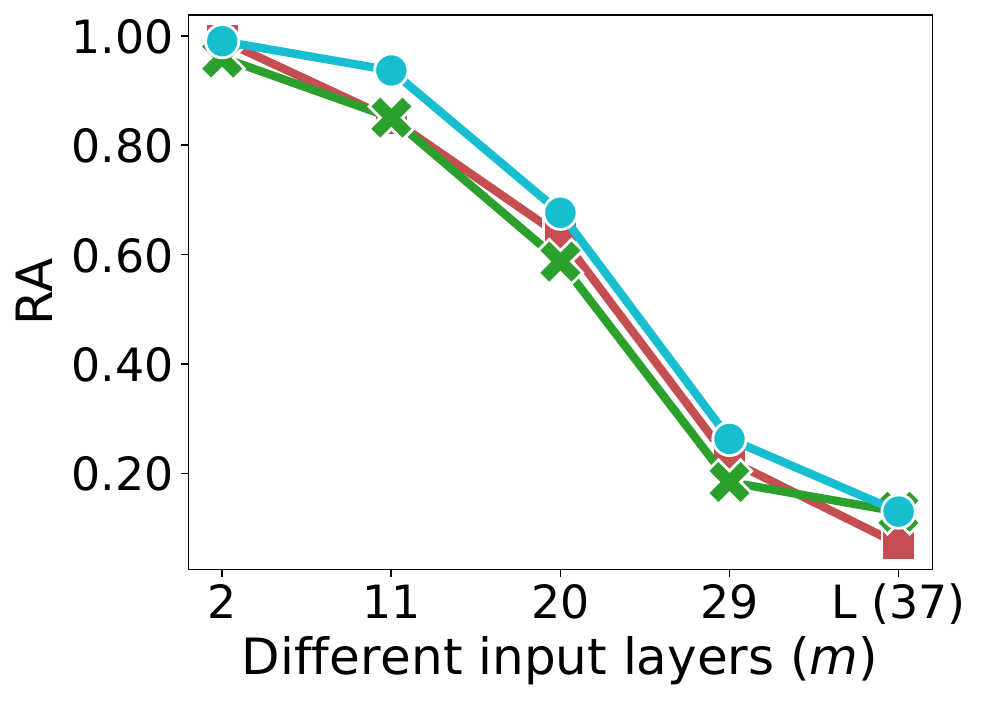}}
&
\hspace{-5mm}
\subfloat[$\qb=8\times\tdict$]{\includegraphics[width=0.25\textwidth]{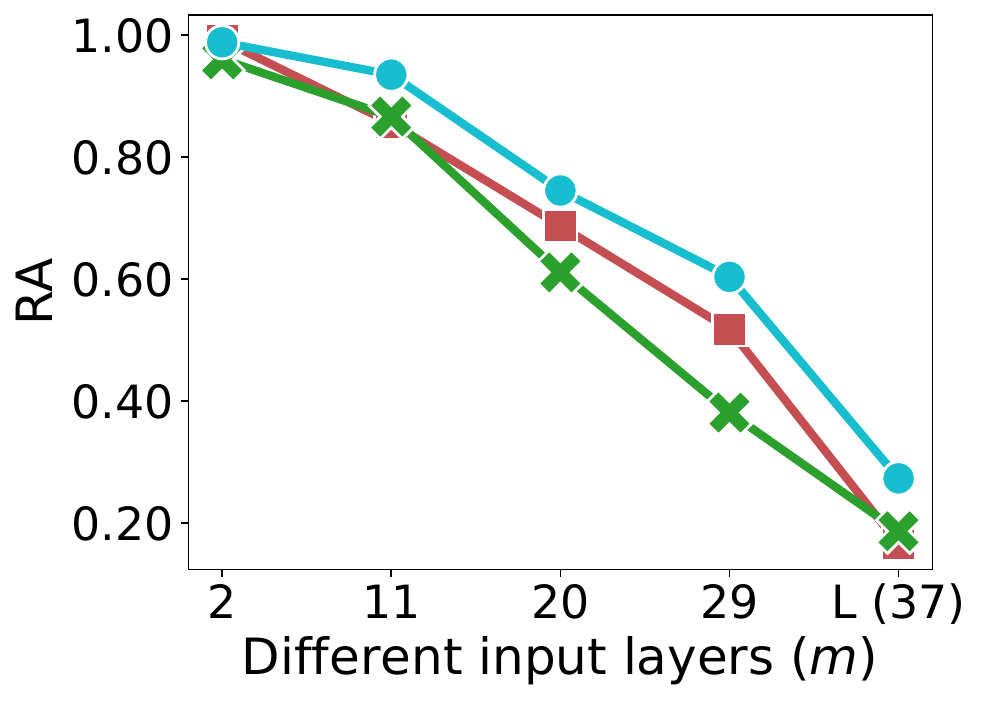}}
&
\hspace{-5mm}
\subfloat[$\qb=16\times\tdict$]{\includegraphics[width=0.25\textwidth]{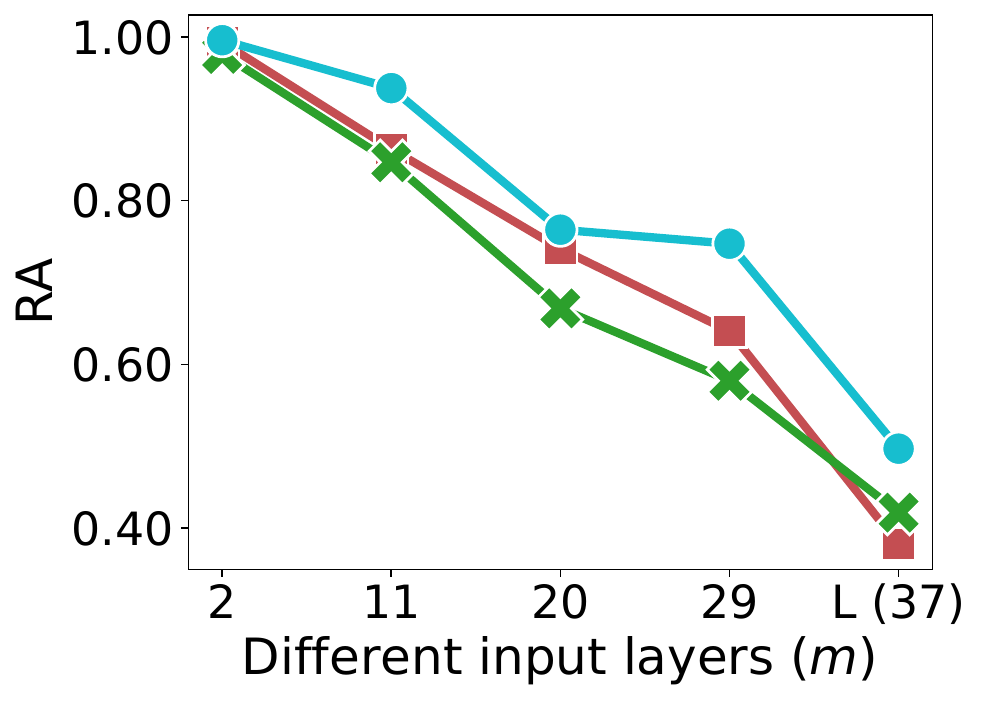}}\label{subfig-attack3-line-gpt2-r}
\vspace{-2mm}  \\
\hspace{-4mm}
\subfloat[$\qb=1\times\tdict$]{\includegraphics[width=0.25\textwidth]{pic/attack3-line-diff-layers-Bert-1.pdf}}\label{subfig-attack3-line-bert-l}
&
\hspace{-5mm}
\subfloat[$\qb=4\times\tdict$]{\includegraphics[width=0.25\textwidth]{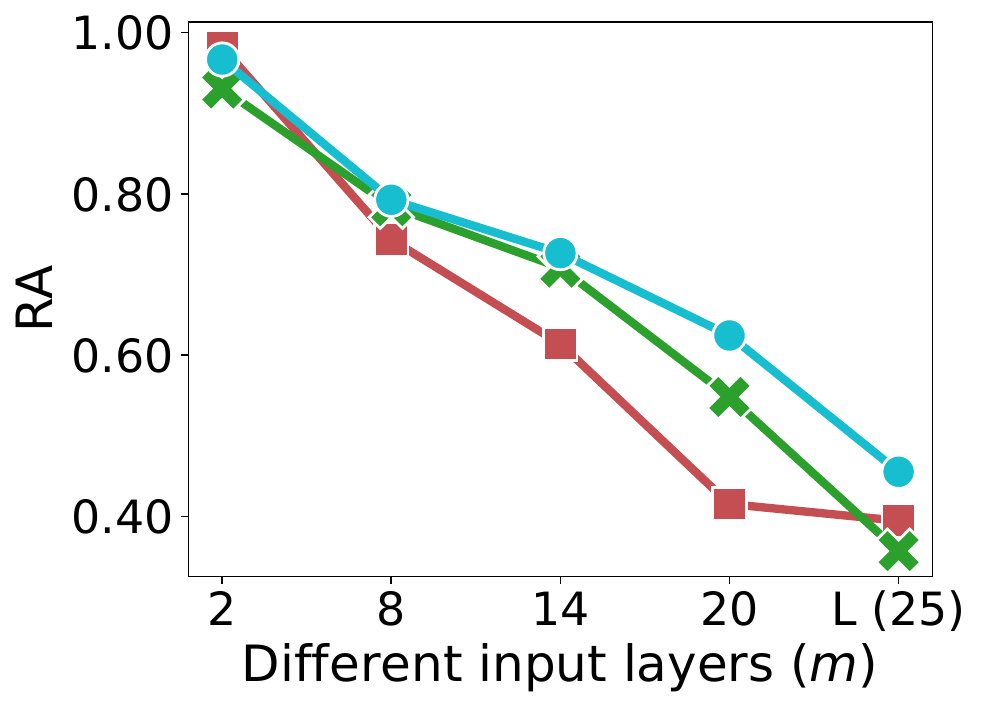}}
&
\hspace{-5mm}
\subfloat[$\qb=8\times\tdict$]{\includegraphics[width=0.25\textwidth]{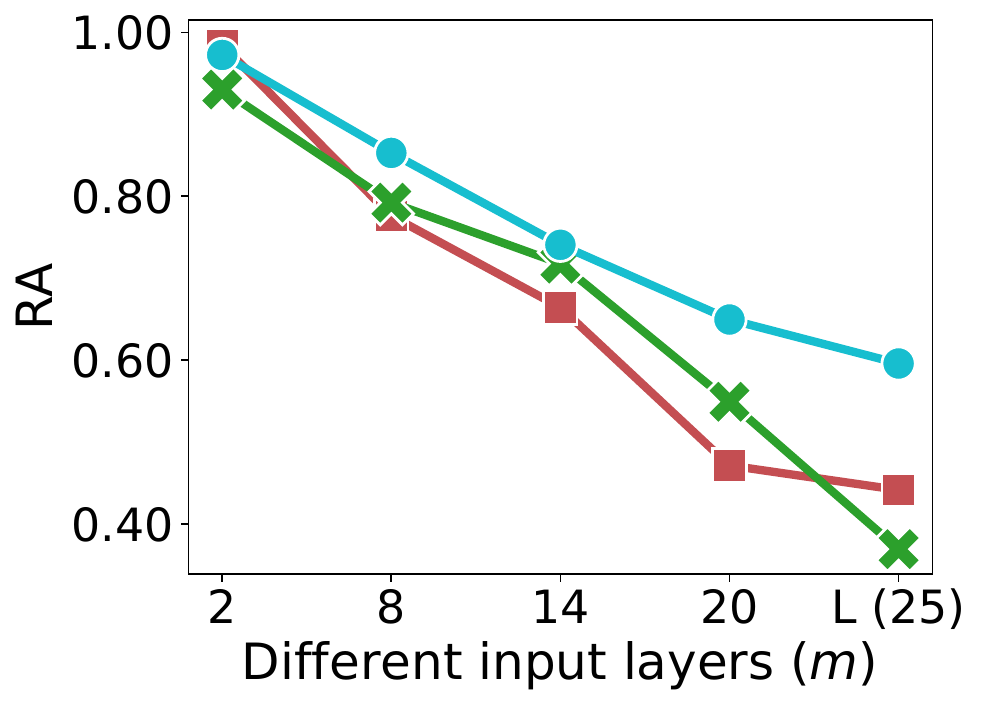}}
&
\hspace{-5mm}
\subfloat[$\qb=16\times\tdict$]{\includegraphics[width=0.25\textwidth]{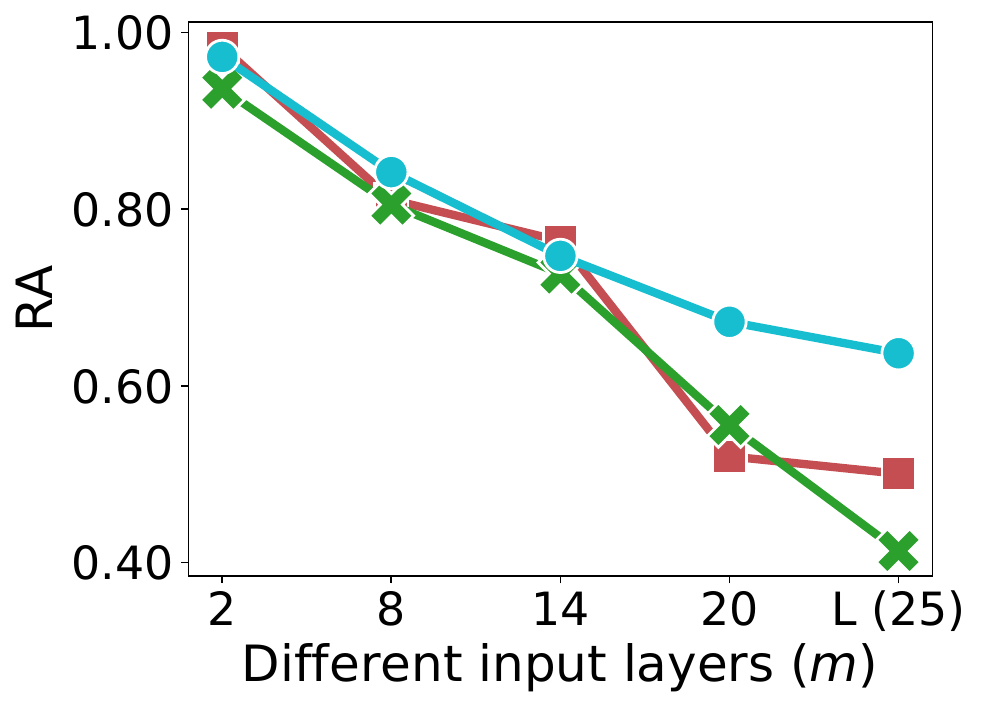}}\label{subfig-attack3-line-bert-r}
\end{tabular}
% \vspace{-4mm}
\caption{The reconstruction accuracies of $\adv_3$ performed on (a)--(d) GPT-2 and (e)--(h) BERT with different query budgets.}
\label{fig-attack3-accur-append}
\end{small}
% \vspace{-8mm}
\end{figure*}

Fig.~\ref{fig-embd-scatter-adv1-append} shows the embedding distributions generated by Llama-3.2 and GPT-2 for $\adv_1$, serving as a complement to Fig.~\ref{fig-embd-scatter-adv1} in Section~\ref{subsec-exp-adv1}.
Fig.~\ref{fig-attack3-accur-append} presents the complete results of $\adv_3$ on GPT-2 and BERT \textit{w.r.t.} different query budgets and layers, serving as a complement to Fig.~\ref{fig-attack3-accur} in Section~\ref{subsec-exp-adv3}.
Table~\ref{tb-exp-adv1-append}, \ref{tb-exp-adv2-append}, and \ref{tb-exp-adv3-in-append-m2}--\ref{tb-exp-adv3-in-append-m37} provide examples of reconstructed prompts for $\adv_1$, $\adv_2$, and $\adv_3$, respectively.

\subsection{\color{mycolor}Linear Layer-based Baseline Attack}\label{append-subsec-bll}
{\color{mycolor}
A simple baseline for prompt reconstruction is connecting the linear layer of LLMs to intermediate embeddings at each layer and interpreting the linear outputs as token reconstructions. To evaluate its effectiveness, we test this baseline attack on Phi-3.5 using WikiText-2, with results summarized in Table~\ref{tb-bll}.

As shown in the second row, the baseline performs poorly on current-token reconstruction, which aligns with the fact that the LLM’s linear layer is optimized for next-token prediction rather than recovering the current token. This motivates a follow-up evaluation: how well does the baseline perform on next-token reconstruction? Results in the third row reveal that its performance improves in deeper layers and even surpasses $\adv_3$ at the final layer. Hence, when query budgets are limited, this baseline can serve as a practical complement to $\adv_3$ for the last transformer layer.

It is important to note that this baseline requires access to both the linear layer and intermediate embeddings --- resources often unavailable for most participants in distributed LLM frameworks --- thereby limiting its usability for most framework participants. }

\begin{table}[t]
\color{mycolor}
\setlength{\tabcolsep}{5.6pt}
\caption{\color{mycolor}RA results of the linear layer-based baseline (B-LL) attack, with Phi-3.5 and WikiText-2 as the testing LLM and dataset.}\label{tb-bll}
% \vspace{-3mm}
\center
\small
\begin{tabular}{c|ccccc}
\hline

Method  &  2 & 10 & 18 & 26 & L (33)   \\
\hline 
\hline
$\adv_3$   & 0.9539 &	0.8156 &	0.7599 &	0.6134 &	0.3928\\
\hline
B-LL (Current Token)    & 0 &	0.0010 &	0.0010 &	0.0010 &	0.0072 \\
\hline
B-LL (Next Token)    & 0.0070 &	0.0171 &	0.0615 &	0.2356 &	0.4431 \\
\hline
\end{tabular}
% \vspace{-3mm}
\end{table}

\clearpage
\onecolumn
% \begin{table*}[t]
% \setlength{\tabcolsep}{7pt}
% \caption{Reconstructed examples in $\adv_1$. All prompts are reconstructed based on the embeddings fed into the linear layer, i.e., $m=\text{L}$. }\label{tb-exp-adv1}
% \vspace{-3mm}
% \center
% \small
\begin{longtable}{c|c|c|c}
\caption{Reconstructed examples in $\adv_1$. All prompts are reconstructed based on the embeddings fed into the linear layer, i.e., $m=\text{L}$. }\label{tb-exp-adv1-append}\\
\hline
Dataset & Ground Truth Prompt & Model & Reconstructed Prompt \\
\hline 
\hline
\multirow{4}{*}{WikiText-2}   &  \multirow{4}{*}{\parbox{\aonegtTW\textwidth}{\texttt{Jordan is also known for his product endorsements. He fueled the success of Nike 's Air Jordan <unk>, which were introduced in 1985 and remain popular today. Jordan also starred in the 1996 feature film Space Jam as himself. In 2006, he became part @-@ owner and head of basketball operations for the then @-@ Charlotte Bobcats, buying a controlling interest in 2010. In 2015, as a result of the increase in value of NBA franchises, Jordan became the first billionaire NBA player in history and the world 's second @-@ richest African @-@ American.} }} & Phi-3.5 &	\parbox{\aonereTW\textwidth}{\texttt{\textbf{Jordan is also known for his product endors}ations.he \textbf{fueled the success of N}ime's \textbf{Air Jordan <unk>, which were introduced in 1985 and remain popular today}" \textbf{Jordan also starred in the 1996 feature film Space} Io \textbf{as himself}" \textbf{In 2006}, \textbf{he became part @-@ owner and head of basketball operations for the }today \textbf{@-@ Charlotte Bobc}ars, \textbf{buying a controlling interest in 2010}" \textbf{In 2015}, \textbf{as a result of the increase in value of NBA franch}ies, \textbf{Jordan became the first billionaire NBA player in history and the world 's second @-@ richest African @-@ American}"} } \\
\cline{3-4}
                                 &                            & Llama-3.2 &	\parbox{\aonereTW\textwidth}{\texttt{\textbf{Jordan is also known for his product} nominations. \textbf{He} heated \textbf{the success} for \textbf{Nike 's Air Jordan <unk}, \textbf{which were introduced in 1985 and remain popular today. Jordan also starred in the 199}4 \textbf{feature film Space Jam as himself. In 2006}, \textbf{he became part @-@ owner and head of basketball operations for the then @-@} Stuart \textbf{Bob} Bulldogs, buy aalling advantage \textbf{in 2010. In 2015}, \textbf{as a result of the increase in value of NBA} franchise, \textbf{Jordan became the first} millionaire \textbf{NBA player in history and the world 's} third \textbf{@-@ richest African @-@ American}. } } \\
\cline{3-4}                                 
                                 &                            & GPT-2 &	\parbox{\aonereTW\textwidth}{\texttt{\textbf{Jordan is also known for his product} affiliation " \textbf{He} helped \textbf{the success of Nike 's Air Jordan <unk}, that \textbf{were introduced in 1985} that become \textbf{popular} life " \textbf{Jordan also} starring \textbf{in the} 2016 \textbf{feature} filmmakers \textbf{Space} Jelly \textbf{as himself} " \textbf{In} 2005, \textbf{he became part @-@ owner and head of basketball} coordination of \textbf{the then @-@ Charlotte Bob} Carolina — \textbf{buying a controlling interest} of \textbf{2010} " \textbf{In 2015} when \textbf{as a result of the increase} of \textbf{value of NBA franchises}, \textbf{Jordan} is \textbf{the first billionaire NBA player in history and the world 's second @-@ richest African @-@ American} ' } } \\
\cline{3-4}                                 
                                 &                            & BERT &	\parbox{\aonereTW\textwidth}{\texttt{Alabama \textbf{is also known for his product endorsements. He fueled}, \textbf{success of Nike's Air Jordan < unk >, which were introduced in 1985 and remain popular today. Jordan also starred in the 1996 feature film Space Jam as himself. In 2006, he became part @ - @ owner and head, basketball operations for the then @ - @ Charlotte Bobcats, buying}, dominating story\textbf{ in 2010. In 2015, as}, \textbf{result}, \textbf{the increase in value of NBA franchises}, \textbf{Jordan became the first billionaire NBA player in history and the world's second @-@ richest African @-@ American}.} } \\  
\hline
\multirow{4}{*}{SQuAD 2.0}    &  \multirow{4}{*}{\parbox{\aonegtTW\textwidth}{\texttt{Who was the name of the Soviet double agent who was the correspondent for The Times in Spain in the late 1930s?} }} & Phi-3.5 &	\parbox{\aonereTW\textwidth}{\texttt{\textbf{Who was the name of the Soviet double agent who was the correspond}ist \textbf{for The Times in Spain in the late 1930s?}} } \\
\cline{3-4}
                                 &                            & Llama-3.2 &	\parbox{\aonereTW\textwidth}{\texttt{\textbf{Who was the name of the Soviet double} editor\textbf{ who was the} journalist \textbf{for The Times in Spain in the late 1930s?}} } \\
\cline{3-4}                                 
                                 &                            & GPT-2 &	\parbox{\aonereTW\textwidth}{\texttt{\textbf{Who was the name of the Soviet double agent who was the} ambassador on \textbf{The Times in Spain in the late 1930s?}} } \\
\cline{3-4}                                 
                                 &                            & BERT &	\parbox{\aonereTW\textwidth}{\texttt{\textbf{What was the name of the Soviet} vice \textbf{agent who was the correspondent for The Times in Spain in the late 1930s?}} } \\              
\hline
\multirow{4}{*}{\makecell{Midjourney\\ prompts}}    &  \multirow{4}{*}{\parbox{\aonegtTW\textwidth}{\texttt{why did you leave?   many elements,  hidden meanings, representational, realism, mod, maximalist, highly detailed 1971 album cover} }} & Phi-3.5 &	\parbox{\aonereTW\textwidth}{\texttt{\textbf{why did you leave?} \textbf{many elements, hidden} hide\textbf{ings, representational, realism, mod, maximalist, highly detailed 1971 album cover}} } \\
\cline{3-4}
                                 &                            & Llama-3.2 &	\parbox{\aonereTW\textwidth}{\texttt{\textbf{why did you} off?   \textbf{many elements,  hidden meanings}, transforms\textbf{ational, realism, mod, maximalist, highly detailed 1971 album cover}} } \\
\cline{3-4}                                 
                                 &                            & GPT-2 &	\parbox{\aonereTW\textwidth}{\texttt{\textbf{why} when \textbf{you} outside?  \textbf{ many elements,  hidden} defines, \textbf{represent}ative, \textbf{realism, mod, maximalist, highly detailed 1971 album cover}} } \\
\cline{3-4}                                 
                                 &                            & BERT &	\parbox{\aonereTW\textwidth}{\texttt{r \textbf{did you leave? many elements, hidden meanings, representational, realism,} alod, al\textbf{ximalist, highly detailed 1971 album cover}} } \\    
\hline
\end{longtable}
% \vspace{-3mm}
% \end{table*}

\clearpage
% \begin{table*}[t]
% \setlength{\tabcolsep}{7pt}
% \caption{Reconstructed examples in $\adv_2$. WikiText-2 is used as the auxiliary dataset. All prompts are reconstructed based on the embeddings fed into the linear layer, i.e., $m=\text{L}$. }\label{tb-exp-adv2}
% \vspace{-3mm}
% \center
% \small
% \begin{tabular}{c|c|c|c|c}
\begin{longtable}{c|c|c|c|c}
\caption{Reconstructed examples in $\adv_2$. WikiText-2 is used as the auxiliary dataset. All prompts are reconstructed based on the embeddings fed into the linear layer, i.e., $m=\text{L}$. }\label{tb-exp-adv2-append}\\
\hline
Dataset & Ground Truth Prompt & Model & Reconstructed Prompt Without $\adv_2$ & Reconstructed Prompt With $\adv_2$\\
\hline 
\hline
\multirow{12}{*}{SQuAD 2.0}    &  \multirow{4}{*}{\parbox{\atwogtTW\textwidth}{\texttt{If your computer runs Vista, what's the maximum number of start and end rules you can have for each time zone setting?} }} & Phi-3.5 &	\parbox{\atworeTW\textwidth}{\texttt{let \textbf{your computer runs} Vami, whose0s \textbf{the} minimum \textbf{number of start and} start \textbf{rules you can be for each time} domain \textbf{setting} :} } &	\parbox{\atworeTW\textwidth}{\texttt{Before \textbf{your computer runs} Vptions, \textbf{what's the maximum number of start and end rules you can have for each time Zone setting})?} } \\
\cline{3-5}
                                 &                            & Llama-3.2 &	\parbox{\atworeTW\textwidth}{\texttt{\textbf{If} our \textbf{computer runs} Meteor, \textbf{what} has \textbf{the} potential \textbf{number of} starts \textbf{and end rules you can} want \textbf{for each time} distance \textbf{setting} :} } &	\parbox{\atworeTW\textwidth}{\texttt{\textbf{If your computer runs} Mercedes, \textbf{what} is \textbf{the maximum number of start and end rules you can} desires \textbf{for each time zone setting?}} }\\
\cline{3-5}                                 
                                 &                            & GPT-2 &	\parbox{\atworeTW\textwidth}{\texttt{\textbf{If} \textbf{your computer} ran \textbf{Vista} win What is \textbf{the maximum number of start and end rules you can have for each time zone setting} (} } &	\parbox{\atworeTW\textwidth}{\texttt{\textbf{If your computer} ran Suns> \textbf{what} is \textbf{the maximum number of start and end rules} they must\textbf{ have for each time zone setting?}} }\\
\cline{3-5}                                 
                                 &                            & BERT &	\parbox{\atworeTW\textwidth}{\texttt{As \textbf{your computer runs}é, \textbf{what} of s of \textbf{maximum number of start and end rules you can have for each time zone setting?}.} } &	\parbox{\atworeTW\textwidth}{\texttt{The \textbf{your computer runs Vista, what's}'\textbf{maximum number of start and end rules you can have for each time zone setting?}} }\\  
                                 \cline{2-5}
                                &  \multirow{4}{*}{\parbox{\atwogtTW\textwidth}{\texttt{In terms of profession, what were people like Philip Hardwick, William Adams Nicholson and Thomas de Grey?} }} & Phi-3.5 &	\parbox{\atworeTW\textwidth}{\texttt{\textbf{In terms of profession, what were people like Philip Hard}ford, \textbf{William Adams Nicho}lman \textbf{and Thomas de} Harris "} } &	\parbox{\atworeTW\textwidth}{\texttt{\textbf{In terms of profession, what were people like Philip Hard}mann, Robert \textbf{Adams Nichol}ison \textbf{and Thomas de} Pam} } \\
\cline{3-5}                                
                                 &                            & Llama-3.2 &	\parbox{\atworeTW\textwidth}{\texttt{\textbf{In terms} on \textbf{profession}, \textbf{what were people like Philip Hard} Edward, \textbf{William Adams} Ingram \textbf{and Thomas de}bury "} } &	\parbox{\atworeTW\textwidth}{\texttt{\textbf{In terms of profession, what were people like Philip Hard}ford, \textbf{William Adams Nicholson and Thomas de}ley!} }\\
\cline{3-5}                                 
                                 &                            & GPT-2 &	\parbox{\atworeTW\textwidth}{\texttt{\textbf{In terms of profession} that What \textbf{were people like Philip Hard}etter to \textbf{William Adams Nicholson and Thomas de} themselves :} } &	\parbox{\atworeTW\textwidth}{\texttt{\textbf{In terms} for \textbf{profession} and \textbf{what were people} of \textbf{Philip Hard}ered, \textbf{William Adams Nicholson and Thomas de}re.} }\\
\cline{3-5}                                 
                                 &                            & BERT &	\parbox{\atworeTW\textwidth}{\texttt{\textbf{In} lengths \textbf{of profession, what were people like} Benjamin of\textbf{wick, William Adams Nicholson and Thomas de Grey?}} } &	\parbox{\atworeTW\textwidth}{\texttt{The \textbf{terms of} retirement, \textbf{what were people like Philip} '\textbf{wick, William Adams Nicholson and Thomas de Grey} said} }\\   
                                 \cline{2-5}
                                &  \multirow{4}{*}{\parbox{\atwogtTW\textwidth}{\texttt{What did Jefferson's concept of 'separation of church and state" became part of what jurisprudence?} }} & Phi-3.5 &	\parbox{\atworeTW\textwidth}{\texttt{sweet \textbf{did} Michigan\textbf{'s concept of}'separ \textbf{separation of church and state " become part} to that \textbf{juris}laedacy "} } &	\parbox{\atworeTW\textwidth}{\texttt{\textbf{What} done \textbf{Jefferson's concept of}'separ \textbf{separation of church and state " become part to what juris}breurgance?} }\\
\cline{3-5}                                
                                 &                            & Llama-3.2 &	\parbox{\atworeTW\textwidth}{\texttt{\textbf{What did Jefferson}s \textbf{concept of '} se \textbf{separation of church and} rule " \textbf{became part} into that isgrance "} } &	\parbox{\atworeTW\textwidth}{\texttt{\textbf{What} had \textbf{Jefferson}s \textbf{concept of} "es \textbf{separation of church and}ines" \textbf{became part} it that \textbf{juris}ityjustice?} }\\
\cline{3-5}                                 
                                 &                            & GPT-2 &	\parbox{\atworeTW\textwidth}{\texttt{\textbf{What did Jefferson}s \textbf{concept of} " separated \textbf{separation} both \textbf{church and}ments " \textbf{became part} into that Juris custom legislation "} } &	\parbox{\atworeTW\textwidth}{\texttt{\textbf{What did Jefferson}s \textbf{concept of} "separ \textbf{separation} both \textbf{church} man inches ' \textbf{became part of what juris} Prology.} }\\
\cline{3-5}                                 
                                 &                            & BERT &	\parbox{\atworeTW\textwidth}{\texttt{\textbf{What did Jefferson} " \textbf{s concept of'separation of church and state " become part} " \textbf{what} juan " discrimination?} } &	\parbox{\atworeTW\textwidth}{\texttt{\textbf{What did Jefferson} of \textbf{s concept of'separation of church and state "} posed \textbf{part of what jurisprudence?}} }\\                   
\hline
\multirow{12}{*}{\makecell{Midjourney\\ prompts}}    &  \multirow{4}{*}{\parbox{\atwogtTW\textwidth}{\texttt{hand-painted oil painting 4k post-processing highly detailed soldier in a sandstorm, oil painting concept art artstation --ar 16:8} }} & Phi-3.5 &	\parbox{\atworeTW\textwidth}{\texttt{stand-p paint \textbf{painted oil painting 4}g \textbf{post- processing highly detailed soldier in a sands}ttain, \textbf{oil painting concept art artstation -ar 16} /\textbf{8}} } &	\parbox{\atworeTW\textwidth}{\texttt{\textbf{hand painted oil painting 4}w \textbf{post-processing highly detailed soldier in a sandstorm, oil painting concept art artstation} \textbf{-ar 16:8}} }\\
\cline{3-5}
                                 &                            & Llama-3.2 &	\parbox{\atworeTW\textwidth}{\texttt{ hand sp \textbf{painted oil painting 4}w \textbf{post} edited \textbf{highly} detail \textbf{soldier in a sand} hairstyle, powder \textbf{painting concept} picture \textbf{art} setting —\textbf{ar 16} 6} } &	\parbox{\atworeTW\textwidth}{\texttt{should-p \textbf{painted oil painting 4}hk \textbf{post-processing highly detailed soldier in a sand}undra of \textbf{oil painting concept} painting \textbf{art} documentary ::\textbf{ar 16:}5} }\\
\cline{3-5}                                 
                                 &                            & GPT-2 &	\parbox{\atworeTW\textwidth}{\texttt{ Hand Hand \textbf{painted oil painting 4k post} Postulation \textbf{highly detailed soldier in a sand}land, \textbf{oil painting concept art art}and —\textbf{ar 16} :\textbf{8}} } &	\parbox{\atworeTW\textwidth}{\texttt{\textbf{hand-painted oil painting 4k post} Post\textbf{processing highly}esque \textbf{soldier in a sandstorm, oil painting concept art artstation —ar 16:8}} }\\
\cline{3-5}                                 
                                 &                            & BERT &	\parbox{\atworeTW\textwidth}{\texttt{s a \textbf{painted oil painting 4k post} a \textbf{processing highly detailed soldier in a sand} Storm, \textbf{oil painting concept art arts} sounding - - \textbf{ar 16:8}} } &	\parbox{\atworeTW\textwidth}{\texttt{The of \textbf{painted oil painting 4k post-processing highly detailed soldier in a sandstorm}s, \textbf{oil painting concept art artst}ruction - \textbf{-ar 16:8}} }\\    
                                 \cline{2-5}
                                &  \multirow{4}{*}{\parbox{\atwogtTW\textwidth}{\texttt{in a dystopian future, a city has been forgotten in the jungle, turned to ruins, gloomy atmosphere, rainy and dark, photorealistic} }} & Phi-3.5 &	\parbox{\atworeTW\textwidth}{\texttt{\textbf{in a dystopian future, a city has been forgotten in the jungle, turned to ruins,} glorible \textbf{atmosphere}, rafare \textbf{and dark}, \textbf{photore} moralative} } &	\parbox{\atworeTW\textwidth}{\texttt{\textbf{in a dystopian future, a city has been forgotten in the jungle, turned to ruins}, gloisy \textbf{atmosphere, rainy and dark, photorealistic}} }\\
\cline{3-5}                                
                                 &                            & Llama-3.2 &	\parbox{\atworeTW\textwidth}{\texttt{\textbf{in a dystopian future, a city has been forgotten in the jungle, turned to ruins}, lo buried \textbf{atmosphere},rels \textbf{and dark}, \textbf{phot} realistic nostalgic} } &	\parbox{\atworeTW\textwidth}{\texttt{\textbf{in a dyst} dementia \textbf{future} of \textbf{a city has been forgotten in the jungle, turned to ruins, gloomy atmosphere, rainy and dark,} bot réal visualization} }\\
\cline{3-5}                                 
                                 &                            & GPT-2 &	\parbox{\atworeTW\textwidth}{\texttt{\textbf{in a dystopian future, a city has been forgotten in the jungle, turned to} destroying, desolate \textbf{atmosphere}, rainfall \textbf{and dark,} microsc virtualize} } &	\parbox{\atworeTW\textwidth}{\texttt{\textbf{in a dystopian future, a city has been forgotten in the jungle, turned to} destroyed, \textbf{gloomy} environment,\textbf{ rainy and dark, photoreal}etrical} }\\
\cline{3-5}                                 
                                 &                            & BERT &	\parbox{\atworeTW\textwidth}{\texttt{" aocy\textbf{stopian future, a city has been forgotten in the jungle, turned to ruins},ocbrandoc \textbf{atmosphere, rainy and dark, photorealistic}} } &	\parbox{\atworeTW\textwidth}{\texttt{\textbf{in a dystopian future, a city has been forgotten in the jungle, turned to ruins}, of dusk of \textbf{atmosphere, rainy and dark, photorealistic}} }\\   
                                 \cline{2-5}
                                &  \multirow{4}{*}{\parbox{\atwogtTW\textwidth}{\texttt{Rick Caruso campaign event on stage in front of a sell out crowd photographed in the 1980s at the Great Western Forum in Inglewood} }} & Phi-3.5 &	\parbox{\atworeTW\textwidth}{\texttt{irk Caro \textbf{campaign event on stage in front of a} listenown \textbf{crowd photographed in the 1980s at the Great Western} Hall \textbf{in} Inville\textbf{wood}} } &	\parbox{\atworeTW\textwidth}{\texttt{D\textbf{ick Caruso campaign event on stage in front of a sell out crowd}raphed \textbf{in the 1980s at the Great Western}position \textbf{in Inglewood}} }\\
\cline{3-5}                                
                                 &                            & Llama-3.2 &	\parbox{\atworeTW\textwidth}{\texttt{\textbf{Rick Car}oso \textbf{campaign event on stage in front of a sell out crowd photographed in the 1980s at the Great Western} Memorial \textbf{in In}lyington} } &	\parbox{\atworeTW\textwidth}{\texttt{\textbf{Rick Car}ura \textbf{campaign event on stage in front of a sell} head \textbf{crowd photographed in the 1980s at the Great Western}um \textbf{in In}ale California} }\\
\cline{3-5}                                 
                                 &                            & GPT-2 &	\parbox{\atworeTW\textwidth}{\texttt{\textbf{Rick Caruso campaign event on} room \textbf{in front of a sell out crowd photographed in the 1980} 1968 \textbf{at the Great Western} Pavilion \textbf{in Ingle}illo} } &	\parbox{\atworeTW\textwidth}{\texttt{\textbf{Rick Caruso campaign event on stage in front} by \textbf{a sell out crowd photographed in the 1980} " \textbf{at the Great Western Forum in Inglewood}} }\\
\cline{3-5}                                 
                                 &                            & BERT &	\parbox{\atworeTW\textwidth}{\texttt{Alabama \textbf{Caruso campaign event on stage in front of a sell}'\textbf{crowd photographed in the 1980s at the Great Western} actor \textbf{in In}ise\textbf{wood}} } &	\parbox{\atworeTW\textwidth}{\texttt{\textbf{Rick Caruso campaign event on stage in front of a} drive, \textbf{crowd photographed in the 1980s at the Great Western} Liga \textbf{in Inglewood}} }\\           
\hline
\end{longtable}
% \vspace{-3mm}
% \end{table*}

\clearpage

% \begin{table*}[t]
% \setlength{\tabcolsep}{7pt}
% \caption{Reconstructed examples in $\adv_3$. Experiments are performed on GPT-2. All prompts are reconstructed from the embeddings fed into the layer $m=2$. }\label{tb-exp-adv3-in-append-m2}
% \vspace{-3mm}
% \center
% \small
\begin{longtable}{c|c|c|c}
\caption{Reconstructed examples in $\adv_3$. Experiments are performed on GPT-2. All prompts are reconstructed from the embeddings fed into the layer $m=2$. Tokens in [] denote the tokens reconstructed by beam search.}\label{tb-exp-adv3-in-append-m2}\\
\hline
Dataset & Ground Truth Prompt & \makecell{Query\\ Budgets\\ ($\times\tdict$)} & Reconstructed Prompt \\
\hline 
\hline
\multirow{4}{*}{WikiText-2}   &  \multirow{4}{*}{\parbox{\athreegtTW\textwidth}{\texttt{Critical reception for the episode was mixed ; certain critics believed the episode was not an " instant classic " in contrast to the other episodes of the season but called it " memorable " and " brilliant " nevertheless, while others regarded it as the black sheep of the season. The episode caused controversy in Canada for the episode's final gag, in which Peter states that " Canada sucks. "} }} & 1 &	\parbox{\athreereTW\textwidth}{\texttt{\textbf{Critical reception for episode was mixed ; certain critics believed episode was not " instant classic " in contrast other episodes season but called it " memorable " " brilliant " nevertheless, while others regarded it as black sheep season.The episode caused controversy in Canada for episode 's final gag, in which Peter states that " Canada sucks. " }} } \\
\cline{3-4}
                                 &                            & 4 &	\parbox{\athreereTW\textwidth}{\texttt{\textbf{Critical reception for episode was mixed ; certain critics believed the episode was not " instant classic " in contrast the other episodes the season but called it " memorable " " brilliant " nevertheless, while others regarded it as the black sheep of the season. The episode caused controversy in Canada for the episode 's final gag, in which Peter states that " Canada sucks. "} } } \\
\cline{3-4}                                 
                                 &                            & 8 &	\parbox{\athreereTW\textwidth}{\texttt{\textbf{Critical reception for the episode was mixed ; certain critics believed the episode was not " instant classic " in contrast the other episodes the season but called it " memorable " " brilliant " nevertheless, while others regarded it as the black sheep the season. The episode caused controversy in Canada for the episode 's final gag, in which Peter states that " Canada sucks. "} } } \\
\cline{3-4}                                 
                                 &                            & 16 &	\parbox{\athreereTW\textwidth}{\texttt{\textbf{Critical reception for the episode was mixed ; certain critics believed the episode was not}  a \textbf{" instant classic " in contrast to the other episodes the season but called it " memorable " and " brilliant " nevertheless, while others regarded it as the black sheep of the season. The episode caused controversy in Canada for the episode 's final gag, in which Peter states that " Canada sucks. " }} } \\          
\hline
\multirow{4}{*}{SQuAD 2.0}    &  \multirow{4}{*}{\parbox{\athreegtTW\textwidth}{\texttt{In dirhams, what was the yearly cost of the Iraqi postal service when Yusuf bin Umar was governor?} }} & 1 &	\parbox{\athreereTW\textwidth}{\texttt{\textbf{In dirhams, what was} The \textbf{yearly cost} for The \textbf{Iraqi postal service when Yusuf bin Umar was governor?}} } \\
\cline{3-4}
                                 &                            & 4 &	\parbox{\athreereTW\textwidth}{\texttt{\textbf{In dirhams, what was the yearly cost of the Iraqi postal service when Yusuf bin Umar was governor?}} } \\
\cline{3-4}                                 
                                 &                            & 8 &	\parbox{\athreereTW\textwidth}{\texttt{\textbf{In dirhams, what was the yearly cost of the Iraqi postal service when Yusuf bin Umar was governor?}} } \\
\cline{3-4}                                 
                                 &                            & 16 &	\parbox{\athreereTW\textwidth}{\texttt{\textbf{In dirhams} in \textbf{what was the yearly cost of the Iraqi postal service when Yusuf bin Umar was governor?}} } \\        
\hline
\multirow{4}{*}{\makecell{Midjourney\\ prompts}}    &  \multirow{4}{*}{\parbox{\athreegtTW\textwidth}{\texttt{a flying drone tied to the ground by fluorescent coloured seat belts on a simple concrete background with studio lighting, octane, --ar 16:9} }} & 1 &	\parbox{\athreereTW\textwidth}{\texttt{\textbf{a flying drone tied} for [the] \textbf{ground by fluorescent coloured seat belts on} \textbf{simple concrete background with studio lighting} \textbf{octane --ar 16:9}} } \\
\cline{3-4}
                                 &                            & 4 &	\parbox{\athreereTW\textwidth}{\texttt{\textbf{a flying drone tied ground by fluorescent coloured seat belts on a simple concrete background with studio lighting, octane, --ar 16:9}} } \\
\cline{3-4}                                 
                                 &                            & 8 &	\parbox{\athreereTW\textwidth}{\texttt{\textbf{a flying drone tied the ground by fluorescent coloured seat belts on a simple concrete background with studio lighting, octane, --ar 16:9}} } \\
\cline{3-4}                                 
                                 &                            & 16 &	\parbox{\athreereTW\textwidth}{\texttt{\textbf{a flying drone tied to the ground by fluorescent coloured seat belts on a simple concrete background with studio lighting, octane, --ar 16:9}} } \\         
\hline
\end{longtable}
% \vspace{-3mm}
% \end{table*}

% \begin{table*}[t]
% \setlength{\tabcolsep}{7pt}
% \caption{Reconstructed examples in $\adv_3$. Experiments are performed on GPT-2. All prompts are reconstructed from the embeddings fed into the layer $m=20$. }\label{tb-exp-adv3-in-append-m20}
% \vspace{-3mm}
% \center
% \small
\begin{longtable}{c|c|c|c}
\caption{Reconstructed examples in $\adv_3$. Experiments are performed on GPT-2. All prompts are reconstructed from the embeddings fed into the layer $m=20$. Tokens in [] denote the tokens reconstructed by beam search.}\label{tb-exp-adv3-in-append-m20}\\
\hline
Dataset & Ground Truth Prompt & \makecell{Query\\ Budgets\\ ($\times\tdict$)} & Reconstructed Prompt \\
\hline 
\hline
\multirow{4}{*}{WikiText-2}   &  \multirow{4}{*}{\parbox{\athreegtTW\textwidth}{\texttt{A music video was filmed in support of the song. It was directed by British film maker Dominic <unk>, and shot at <unk>'<unk> Building in London. It features the band performing the song, with a laser show, in front of a staged audience, mostly local college students. Stage effects and blue @-@ red light transitions give the video a surreal feel, while a stoic crowd make up the audience.} }} & 1 &	\parbox{\athreereTW\textwidth}{\texttt{ a. \textbf{video}s is \textbf{filmed} for \textbf{support}s \textbf{of} The \textbf{song}). \textbf{It was directed by British} Film \textbf{maker Dominic <unk>, and}[ shoot] \textbf{at} <ritchiece'." <ritch. \textbf{Building} have \textbf{London. It features} The \textbf{band performing} The \textbf{song}, \textbf{With a Laser Show, In front of a staged audience}, largely \textbf{local College Student}.[ stage] \textbf{effects and blue @-@ red Light  transition}ing gave The \textbf{video a surreal feel} a \textbf{while a sto}[etic] \textbf{crowd} Make prise The \textbf{audience}. } } \\
\cline{3-4}
                                 &                            & 4 &	\parbox{\athreereTW\textwidth}{\texttt{ a \textbf{was filmed} \textbf{Support} \textbf{of The song}!.\textbf{ It was} directs \textbf{by British film maker Dominic <unk>}, And \textbf{shot at <unk>}"\textbf{<unk>} \textbf{building in London}\%. \textbf{It featur}ing The \textbf{band performing} The \textbf{song}, With \textbf{a laser show}, \textbf{In front of a staged audience, mostly local College Students}!. \textbf{Stage effects and blue @-@ red Light transitions give} The \textbf{video a surreal feel, while a stoic crowd make up} Thegoers.). 
} } \\
\cline{3-4}                                 
                                 &                            & 8 &	\parbox{\athreereTW\textwidth}{\texttt{ a \textbf{was filmed in support of the song}). \textbf{It was directed by British film maker Dominic <unk>}, And \textbf{shot at <unk>}" \textbf{<unk>} \textbf{building in London. It featur}ing  The \textbf{band performing the song}, \textbf{with a laser show, in front of a staged audience, mostly local college Students}. \textbf{Stage effects and blue @-@ red light transitions gives The video a surreal feel, while a sto}etic \textbf{crowd make up} The \textbf{audience}!.
} } \\
\cline{3-4}                                 
                                 &                            & 16 &	\parbox{\athreereTW\textwidth}{\texttt{ a \textbf{was filmed in support of the song}). \textbf{It was} directs \textbf{by British film maker Dominic}\textbf{ <unk>, and shot at <unk>}" \textbf{<unk>} \textbf{Building in London. It features the band performing the song, with a laser show, in front of a staged audience, mostly local college students.} \textbf{stage effects and Blue @-@ red light transitions give}s \textbf{the video a surreal feel, while a sto}aic \textbf{crowd make up the audience.} } } \\          
\hline
\multirow{4}{*}{SQuAD 2.0}    &  \multirow{4}{*}{\parbox{\athreegtTW\textwidth}{\texttt{What movie did the American Film Institute rank as \#1 on their 10 Greatest American Films ever Made list in 2007?} }} & 1 &	\parbox{\athreereTW\textwidth}{\texttt{ kidding Movie \textbf{did the American film Institute ranks as \#} 6 \textbf{on their 10 Greatest American Films} never \textbf{made lists In 2007?}} } \\
\cline{3-4}
                                 &                            & 4 &	\parbox{\athreereTW\textwidth}{\texttt{ unless \textbf{movie did the American}[ Film] \textbf{institute rank}s \textbf{as \#} \textbf{on their 10 Greatest American Films ever made list in}[ 2008]?} } \\
\cline{3-4}                                 
                                 &                            & 8 &	\parbox{\athreereTW\textwidth}{\texttt{63 \textbf{movie did the American Film Institute rank}s \textbf{as \#} \textbf{on their 10 Greatest American Films ever made list in 2007?}
} } \\
\cline{3-4}                                 
                                 &                            & 16 &	\parbox{\athreereTW\textwidth}{\texttt{[What] \textbf{movie did the American Film Institute rank}ed \textbf{as \#1 on their}[ ten] \textbf{Greatest American films ever Made list in 2007?}} } \\        
\hline
\multirow{4}{*}{\makecell{Midjourney\\ prompts}}    &  \multirow{4}{*}{\parbox{\athreegtTW\textwidth}{\texttt{an impressionist oil painting of a vase of small white flowers in the style of Vincent Van Gogh, matte painting, mint green background color} }} & 1 &	\parbox{\athreereTW\textwidth}{\texttt{ctic \textbf{impression}sinian \textbf{oil painting of a} viage \textbf{of Small white flowers} a \textbf{The style}s \textbf{of Vincent Van Go}gers. \textbf{matte painting}. [Mint][ Green] \textbf{background color}} } \\
\cline{3-4}
                                 &                            & 4 &	\parbox{\athreereTW\textwidth}{\texttt{ \textbf{impression}s \textbf{oil painting of a v} \textbf{of small white flowers in} \textbf{The Style of Vincent Van Gogh} \textbf{matte painting} \textbf{mint green background color}} } \\
\cline{3-4}                                 
                                 &                            & 8 &	\parbox{\athreereTW\textwidth}{\texttt{ists \textbf{oil Painting of a} v \textbf{of small white flowers in the style of Vincent Van Go} \textbf{matte Painting, mint green background color}} } \\
\cline{3-4}                                 
                                 &                            & 16 &	\parbox{\athreereTW\textwidth}{\texttt{\textbf{an impressionist oil Painting of a} v \textbf{of small White flowers in the style of Vincent Van Go, matte Painting, mint green background color}} } \\         
\hline
\end{longtable}
% \vspace{-3mm}
% \end{table*}

% \begin{table*}[t]
% \setlength{\tabcolsep}{7pt}
% \caption{Reconstructed examples in $\adv_3$. Experiments are performed on GPT-2. All prompts are reconstructed from the embeddings fed into the layer $m=\text{L}$ (37). }\label{tb-exp-adv3-in-append-m37}
% \vspace{-3mm}
% \center
% \small
\begin{longtable}{c|c|c|c}
\caption{Reconstructed examples in $\adv_3$. Experiments are performed on GPT-2. All prompts are reconstructed from the embeddings fed into the layer $m=\text{L}$ (37). Tokens in [] denote the tokens reconstructed by beam search.}\label{tb-exp-adv3-in-append-m37}\\
\hline
Dataset & Ground Truth Prompt & \makecell{Query\\ Budgets\\ ($\times\tdict$)} & Reconstructed Prompt \\
\hline 
\hline
\multirow{4}{*}{WikiText-2}   &  \multirow{4}{*}{\parbox{\athreegtTW\textwidth}{\texttt{In 1903, a senior Kodokan instructor named Yamashita <unk> traveled to the United States at the request of the Seattle businessman Sam Hill. In Washington, DC, Yamashita's students included Theodore Roosevelt and other prominent Americans. At Roosevelt's request, Yamashita also taught judo at the US Naval Academy. <unk> on the publicity, the Japanese <unk> in the USA asked the Kodokan to send more judo teachers to America, providing continuity to Yamashita's work. Tomita <unk> accepted the task ; Maeda and Satake embraced the opportunity. } }} & 1 &	\parbox{\athreereTW\textwidth}{\texttt{\textbf{In} 263[ Inn] young young young young young young young young young young young young young young[ \textbf{to}][ Leafs][ Leafs][ Leafs][ Giants][ Serbian][ Serbian][ incarnation][ Serbian][ Serbian][ incarnation][ Hornets][ Hornets][ Hornets][ Hornets][st][st][st][st][ners][st][st][ fifth][ fifth][ Fellow][ Fellow][ Fellow][ conferred][ whose][ inclusion][ initiatives][ initiatives].[ Charles][ Stanton][ Fellow][ Fellow][ Fellow][ Fellow][ Fellow][ Fellow][ Mama][ once][ once][ retiring][ once] Same[ ABC][ ABC][ ABC][book].[ Fellow][ Fellow][y][ Fellow][ hall][ hall][ hall][ hall][ enic][ hall][the][ atre][ atre][ atre][ atre][arding][ atre][the][ atre][the][ atre][arding][atre][ners][oka][oka][oka][ institute][ periodic][ periodic][ periodic][ visit][ periodic][ periodic][ visit][ periodic][ simple][ periodic][[ peasant]][ periodic][ periodic] peasant[ periodic][ periodic][ periodic][ dissolved][ hill]racted[ Les][ier][ dance][ periodic][ periodic][ periodic][ endangered][ endangered][ endangered][ couldn]
} } \\
\cline{3-4}
                                 &                            & 4 &	\parbox{\athreereTW\textwidth}{\texttt{\textbf{In} In Mits Fundamental Fundamental[ Fundamental][ Fundamental][ lashes][ lashes][ lashes][ Dil][ Dil][ sou][anton][ Dil][ sou][bourne] \textbf{to}[ the][ carnage][ carnage] carnage carnage[ Leafs] carnage carnage[ Leafs][ carnage][ carnage][ Leafs] Used[ Leafs][ Shaun] JFK[ JFK] JFK[ JFK][ Ec][ash] "[onel][onel][onel][onel][onel] gentlemen perilous perilous perilous. \textbf{At}[ \textbf{Roosevelt}] "[onel][onel] JFK[UL][UL][UL] perilous perilous[ flagged][ansen][anton] carnage[ carnage][ Devil][ Awards] Used JFK Restore Restore \textbf{on}[ the][ fairness] carnage carnage carnage carnageigh carnagearate carnage carnage[ carnage] carnage[ DH][ DH][ DH] carnage[ DH] carnage[ sympt][omatic][ Equipment] to[ Equipment] JFK[ optimal][ Path] \textbf{to}[ Join][ Join][ Join][ Join][ Join][TP] Used[ Tom][ Gap] Restoreigh Restore[TP] carnage[ DH] carnage[ DH][ DH] gentlemen[ contemplate][ Rory][ drove] carnage[ DH] Used 
} } \\
\cline{3-4}                                 
                                 &                            & 8 &	\parbox{\athreereTW\textwidth}{\texttt{ \textbf{In} In Fernandez Fundamental Fundamental[ Fundamental][ Fundamental][ lashes][ lashes][ lashes] \textbf{Yamash}[ghan][ Gap][ Gap][ Gap][ Gap] \textbf{to} carnage[ carnage][ carnage] carnage carnage[ carnage] TIME carnage[ carnage][ carnage] \textbf{Sam}[idates] Used[ carnage] \textbf{Washington} JFK \textbf{DC} JFK JFK\textbf{ash}[UL][UL][UL][onel][onel][onel][obby] perilous[[ perilous perilous perilous Used \textbf{At}[iration]bit[ Latest][ Statement] JFK[UL]\textbf{ash}[bourne] perilous perilous \textbf{jud}[do] on \textbf{the}[ carnage] \textbf{Naval Academy} Used \textbf{<unk>} \textbf{on}[ \textbf{the}][ carnage] carnage carnage[ Print] \textbf{<unk} calfarate carnage[ Print][ carnage] carnage \textbf{Kod}[ Equipment][ Equipment] \textbf{to}[ Equipment] few jud[orum][ Equipment] \textbf{to}[ Equipment] JFK \textbf{providing}[ Path] \textbf{to Yam}[nesia][ Gap][[ Statement]][TP] \textbf{work} Used \textbf{Tom}[TP] \textbf{<unk>}[TP] carnage[ DH] merciless[ DH][ DH] gentlemen[ Eb][anton][anton] carnage[ fairness] Used } } \\
\cline{3-4}                                 
                                 &                            & 16 &	\parbox{\athreereTW\textwidth}{\texttt{ In Inonel Fundamental Fundamental[ Fundamental][ Fundamental]an[ Fundamental][ Dil] \textbf{Yam}ak[ Dil] \textbf{<unk>}[ Gap] \textbf{to}[ \textbf{the}] \textbf{United States}[ Types][ \textbf{the}] \textbf{request}[ Types][omatic] \textbf{Seattle businessman Sam Hill} Used[ Shaun] \textbf{Washington} JFK Northwest JFK \textbf{Yamash}[UL] "s[keye] \textbf{included Theodore}[ \textbf{Roosevelt}] perilous perilous perilous perilous Used \textbf{At Roosevelt}['][ Latest][ Statement] JFK \textbf{Yamash}[ thanked] perilous perilous \textbf{jud} perilous \textbf{at}[ perilous] \textbf{US Naval Academy}[[ Used]] \textbf{<unk> on}[orum] \textbf{publicity}[ DH][ DH] \textbf{Japanese} \textbf{<unk>}arate[ carnage] \textbf{USA} told[ DH] \textbf{Kodokan to} ask[ DH] \textbf{jud}[orum] \textbf{teachers to America} JFK \textbf{providing continuity to Yamash}[imura] "\textbf{s work} Used \textbf{Tom}[ Gap] \textbf{<unk> accepted}[ Print] \textbf{task}[ Print] \textbf{Ma}[edi] gentlemen \textbf{Satake}[ contemplate][ DH] \textbf{opportunity}[ DH] } } \\          
\hline
\multirow{4}{*}{SQuAD 2.0}    &  \multirow{4}{*}{\parbox{\athreegtTW\textwidth}{\texttt{In what year did Microsoft imply that they would be making changes to support RealTimeIsUniversal in a step towards compatibility with UTC?} }} & 1 &	\parbox{\athreereTW\textwidth}{\texttt{\textbf{In} domain Missouri[ football][ football][ indicating] \textbf{that}[ \textbf{they}][ that]['re][ \textbf{making}][Why][ sell][ Characters][ Characters][ Characters][ Characters][ Characters][ planned][Her][ Characters][Though][ Characters][ Characters][ Characters]'?
} } \\
\cline{3-4}
                                 &                            & 4 &	\parbox{\athreereTW\textwidth}{\texttt{\textbf{In}[ \textbf{what}] \textbf{year}[[ Nathan]][ Nathan][ \textbf{imply}] \textbf{that}[ Characters][ Characters][ Characters][ Characters][ Characters][ Characters][ Characters][ Characters][ Characters][ Characters][ Characters][ Characters][ Characters][ Characters][ Characters][ Characters][ Characters][PI].?} } \\
\cline{3-4}                                 
                                 &                            & 8 &	\parbox{\athreereTW\textwidth}{\texttt{\textbf{In}[ \textbf{what}] Cort Cort[ NA][ \textbf{imply}][ \textbf{that}][ \textbf{they}][ \textbf{will}][ Characters][ Characters][ Characters][ Characters][ Characters][ Characters][External]\textbf{Is}[\textbf{Universal}][Characters][ Characters][ Characters][ Characters][ Characters][ Characters][ Characters][?]} } \\
\cline{3-4}                                 
                                 &                            & 16 &	\parbox{\athreereTW\textwidth}{\texttt{\textbf{In}[ \textbf{what}] \textbf{year} Cort \textbf{Microsoft imply}[ \textbf{that}] you[ \textbf{would}][Characters]\textbf{ making changes}[ for] \textbf{support RealTimeIsUniversal}[Characters][Characters]\textbf{ step towards}[Characters] \textbf{with}[PI][?]
} } \\        
\hline
\multirow{4}{*}{\makecell{Midjourney\\ prompts}}    &  \multirow{4}{*}{\parbox{\athreegtTW\textwidth}{\texttt{trailcam footage livestream video of the minions playing in a cursed crop circle glitch grain noise datamosh corrupted horror deformed scary} }} & 1 &	\parbox{\athreereTW\textwidth}{\texttt{\textbf{tra}[\textbf{il}][\textbf{cam}][cam][cam][ shoot][ earthquake][ letter][ letter][ letter][ letter][ suburban][one][ Earth][ orbit][ Earth][ orbit][ era][ era][never][never][ tsunami][ era][never][never][ tsunami]} } \\
\cline{3-4}
                                 &                            & 4 &	\parbox{\athreereTW\textwidth}{\texttt{\textbf{tra}[\textbf{il}][\textbf{cam}][ Ham][ Ham][ Ham][ 1998][ Colossus][ Colossus][ Colossus][ Hyde][ conversation][ conversation][ Earth][ conversation][ Earth][ Earth][ conversation][lich][lich][lich][lich][ humor][lich][lich][racist]} } \\
\cline{3-4}                                 
                                 &                            & 8 &	\parbox{\athreereTW\textwidth}{\texttt{\textbf{tra}[\textbf{il}][\textbf{cam}][ activate][ \textbf{livestream}] sympathetic corrupted tribes fighting Sith corrupted corrupted corrupted corrupted corrupted corrupted corrupted corrupted corrupted corrupted corrupted \textbf{corrupted}[ \textbf{horror}][ racist] corrupted
} } \\
\cline{3-4}                                 
                                 &                            & 16 &	\parbox{\athreereTW\textwidth}{\texttt{\textbf{tra}\textbf{il}\textbf{cam} livestream \textbf{livestream} \textbf{video} [ Luna][ visitors] \textbf{minions}[ visitors] \textbf{in} close \textbf{cursed crop circle} blast \textbf{grain noise datam}ap \textbf{corrupted horror deformed}[lich]} } \\         
\hline
\end{longtable}
% \vspace{-3mm}
% \end{table*}

\end{document}